\documentclass[11pt,a4paper,twoside]{article}
\usepackage{t1enc}
\usepackage[latin1]{inputenc}
\usepackage[english]{babel}
\usepackage{yfonts}
\usepackage{bbm}
\usepackage{bm}
\usepackage{amssymb}
\usepackage{amsmath}
\usepackage{amsfonts}
\usepackage{fancyhdr}
\usepackage[all,cmtip,line]{xy}
\usepackage{hyperref}
\usepackage{cite}
\hypersetup{final=true}

\newcommand{\bra}[1]{\langle #1|}
\newcommand{\ket}[1]{| #1\rangle}
\newcommand{\braket}[2]{\langle #1|#2\rangle}
\newcommand{\sod}[0]{\hat{\mathfrak{so}}(d)_1}
\newcommand{\so}[0]{\hat{\mathfrak{so}}}

\newcommand{\g}[0]{\mathfrak{g}}
\newcommand{\h}[0]{\mathfrak{h}}

\newcommand{\gk}[0]{\hat{\mathfrak{g}}_k}
\newcommand{\ak}[0]{\hat{\mathfrak{a}}_{\kappa}}
\newcommand{\gkt}[0]{\hat{\mathfrak{g}}_{\tilde{k}}}
\newcommand{\bJ}[0]{\mathsf{J}}
\newcommand{\pJ}[0]{\mathcal{J}}

\numberwithin{equation}{section}

\begin{document}

\title{\textsc{Kondo flow invariants, twisted K-theory and Ramond-Ramond charges}}
\author{\\ \textsc{Samuel Monnier\footnote{samuel.monnier@math.unige.ch}} \\ \\ \small Université de Genève, Section de Mathématiques, \\ \small 2-4 Rue du Lièvre, Case postale 64, 1211 Genève 4, Switzerland}
\maketitle

\upshape

\begin{abstract}
We take a worldsheet point of view on the relation between Ramond-Ramond charges, invariants of boundary renormalization group flows and K-theory. In compact super Wess-Zumino-Witten models, we show how to associate invariants of the generalized Kondo renormalization group flows to a given supersymmetric boundary state. The procedure involved is reminiscent of the way one can probe the Ramond-Ramond charge carried by a D-brane in conformal field theory, and the set of these invariants is isomorphic to the twisted K-theory of the Lie group. We construct various supersymmetric boundary states, and we compute the charges of the corresponding D-branes, disproving two conjectures on this subject. We find a complete agreement between our algebraic charges and the geometry of the D-branes.
\end{abstract}

{\small
\newpage
\tableofcontents
}

\section{Introduction}

It has been known for a decade that Ramond-Ramond charges of D-branes, invariants of boundary renormalization group flows and K-theory are intimately linked. The first steady indication of the link between Ramond-Ramond charges \cite{Polchinski:1995mt} and K-theory came from a detailed analysis of the coupling of the gauge and scalar fields supported on the D-brane worldvolume to the Ramond-Ramond fields living in the bulk \cite{Minasian:1997mm}. Along separate line of development, it became clear that Sen's conjectures \cite{Sen:2004nf} implied that the set of D-brane configurations modulo the action of the boundary renormalization group flows is given by an appropriate version of the K-theory of the target space \cite{Witten:1998cd, Horava:1998jy}. (See \cite{Olsen:1999xx, Witten:2000cn, Evslin:2006cj} for a complete overview of K-theory applied to strings.)

The boundary renormalization group flows are triggered by boundary perturbations of the conformal field theory describing the open string modes living on the D-branes considered. They generically map a D-brane configuration onto another (possibly empty) D-brane configuration plus closed string radiation. As closed strings do not carry any charge with respect to the Ramond-Ramond gauge fields, it is natural to expect that the Ramond-Ramond charges of the two D-brane configurations coincide. This property has been checked in \cite{Sen:2002an} for the flat space case, in \cite{Gaberdiel:1999ch} for the case of orbifolds of toroidal compactifications, and in \cite{SchaferNameki:2003aj, SchaferNameki:2004yr} for Kazama-Suzuki coset models. Ramond-Ramond charges therefore provide natural invariants of the boundary renormalization group flows. 

The aim of this paper is to construct invariants of generalized Kondo flows \cite{Affleck:1990by, Alekseev:2000fd, Fredenhagen:2002qn, Monnier:2005jt, Alekseev:2007in} in super Wess-Zumino-Witten (sWZW) models on a compact simply connected Lie group, using a procedure that is formally a measurement of the Ramond-Ramond charge of the boundary state by a Ramond-Ramond test state (see for instance \cite{DiVecchia:1999rh}, section 8). We used ``formally'' in the last sentence because sWZW models do not contain massless Ramond-Ramond gauge fields, so the concept of Ramond-Ramond charge is ill-defined in this setting. The ``charges'' constructed in this paper are well-defined as invariants of the boundary renormalization group flows, however.

\vspace{.5cm}

WZW models are conformally invariant sigma models with a Lie group $G$ as target space. When $G$ is compact, the corresponding conformal field theory is rational and can be solved exactly. The set of brane configurations modulo boundary renormalization group flows is conjectured to be isomorphic to the twisted K-theory of the Lie group $G$, where the twist is provided by the class of the NS-NS 3-form $H$ (this class is determined by the level $k$ of the WZW model). The twisted K-theory for $G$ compact, simple and simply connected has been computed in \cite{Braun:2003rd}, and is given by the direct product of $2^{r-1}$ copies of $\mathbb{Z}/M\mathbb{Z}$ :
\begin{equation}
\label{K-theoryG}
K_H(G) = \left ( \mathbb{Z}/M\mathbb{Z} \right )^{(2^{r-1})},
\end{equation}
where $r$ is the rank of the group $G$ and $M$ is an integer depending $G$ and on the level $k$. (The explicit expression for $M$ is not very enlightening and can be found in \cite{Braun:2003rd}.) One remarkable feature of these K-groups is that they have torsion (they are direct products of finite cyclic groups), so that the corresponding D-brane charge is conserved only modulo the integer $M$. In particular, a stack of $M$ identical D-branes can decay into closed string radiation. The fact that WZW models are well-known CFTs, that their target spaces admit a non-trivial (and non-torsion) $H$ field and that their K-theory groups are cyclic makes them interesting examples on which one can test the K-theory conjecture.

Comparison between the Kondo flow action on WZW D-brane and the K-theory of the underlying Lie groups have been performed in several papers \cite{Alekseev:2000jx, Fredenhagen:2000ei, Maldacena:2001xj, Bouwknegt:2002bq, Gaberdiel:2003kv}. The general idea underlying these tests of the K-theory conjecture is the following. One associates to the known D-branes a $\mathbb{Z}/M\mathbb{Z}$ valued charge, and consider the constraints imposed by the maximally symmetric Kondo boundary flows on these charges. These constraints can be solved, and impose that the value of $M$ coincides with the result of the K-theory computation. Given an arbitrary D-brane, one can generate by Kondo flows other D-branes in the same $\mathbb{Z}/M\mathbb{Z}$ factor, and the universality \cite{Graham:2003nc, Bachas:2004sy} of the Kondo renormalization group flows explains why the integer $M$ is the same in each factor of \eqref{K-theoryG}. However, it is impossible to test the number of $\mathbb{Z}/M\mathbb{Z}$ factors in \eqref{K-theoryG} with these techniques. It is also impossible to know whether two boundary states carry the same type of charge (ie. lie in the same $\mathbb{Z}/M\mathbb{Z}$ factor) if they are not linked by a boundary renormalization group flow of the Kondo type. For instance, in the two papers \cite{Gaberdiel:2004hs, Gaberdiel:2004za}, the authors claim to display boundary states carrying charges from each of the factors of \eqref{K-theoryG}, but no steady argument supports these conjectures. We will see that actually both of them prove wrong.

This situation makes it desirable to have a well-defined prescription to assign to a given boundary state an element of the K-theory group \eqref{K-theoryG}. In this paper, we will describe a procedure that allows to associate to a worldsheet supersymmetric boundary state of the super Wess-Zumino-Witten model a quantity invariant under the generalized Kondo renormalization group flows. We will check that these invariants form a group isomorphic to the twisted K-theory of the Lie group $G$.

The paper is organized as follows. In the next section 
we review the classification of flat space type IIB D-branes by K-theory, and show how this K-theory charge can be probed by computing a coupling of the boundary state with a Ramond-Ramond test state. We then review briefly and roughly our construction. We also summarize the main results about the Kondo flow invariants that we will build. In section \ref{SecBasNot}, we review the super Wess-Zumino-Witten model, as well as the relation between quantum Wilson operators and the Kondo flows in the bosonic WZW model. 

In section \ref{sWZWBranes}, we construct a large class of worldsheet supersymmetric boundary states for the sWZW model, using well-known constructions available for the bosonic WZW model. In section \ref{SupSymKondPert}, we show how Wilson operators can be used to describe the supersymmetric Kondo flows of the sWZW model. In section \ref{TS-CSupC}, we compute the cohomology of the worldsheet supercharge, which determines the space of Ramond-Ramond test states. We also compute the action of Wilson operators on these test states. Section \ref{SecChargeBoundSt} is devoted to the construction of  the charges. At the end of this section we compute in full generality the charges of the simplest boundary states constructed in section \ref{sWZWBranes} (namely the maximally symmetric, coset and twisted boundary states). Several specific examples of more elaborate boundary states in $SU(4)$ are treated in section \ref{SecExamples}, as an illustration of our formalism. In section \ref{SecRelHom}, we use the so-called Kostant conjecture to connect our procedure with the familiar picture of the classification of brane charges by homology. We check that in the examples considered in section \ref{SecExamples}, the charges that we found coincide with the geometric intuition. We end with some concluding remarks in section \ref{SecConcl}.

\section{An overview}

\label{SecOverview}

The aim of this section is to convey a general feeling of the ideas that will be relevant to the understanding of the construction elaborated in the sections \ref{TS-CSupC} and \ref{SecChargeBoundSt}.

\subsection{The flat space case}

First, it is instructive to have a look at the flat space case. A more detailed account of the K-theory classification of D-branes in flat space can be found in \cite{Olsen:1999xx}.

So let us consider type IIB string theory on $\mathbb{R}^{10}$, and recall that the type IIB D-branes are located along even dimensional hyperplanes (in spacetime). According to \cite{Witten:1998cd}, any D-brane in this background can be constructed from a tachyon field configuration on a stack of an equal number (say $n$) of D9 and anti-D9 branes. This stack carries a Chan-Patton bundle with structure group $U(n) \times U(n)$, and the tachyon field $T$ is a complex Lorentz scalar transforming in the bifundamental representation of $U(n) \times U(n)$. (It is the field corresponding to open strings stretched between a brane and an anti-brane.) In this way, a D$p$-brane supported on a $p+\!1$-dimensional hyperplane of $\mathbb{R}^{10}$ is identified with a solitonic configuration of $T$ vanishing on this hyperplane, and constant in the directions parallel to it. Such a tachyon profile can be built out of the Atiyah-Bott-Shapiro construction \cite{Atiyah1964}, and these tachyonic configurations are classified by the K-theory with compact support in the directions normal to the D$p$-brane \cite{Olsen:1999xx}. After compactification of $\mathbb{R}^{9-p}$ by the addition of its sphere at infinity, compactly supported K-theory is isomorphic to the relative K-theory $K(B^{9-p}, S^{8-p}) = \mathbb{Z}$ ($p$ odd), where $B^m$ and $S^m$ denote the $m$-dimensional ball and sphere. 
This K-theory classifies the charges of all of the brane configurations which are trivial along a prescribed $p+\!1$-dimensional hyperplane (in the sense that any two sections of such configurations orthogonal to the hyperplane are homotopic). We consider here only branes belonging to this family.

According to the philosophy exposed in \cite{Schomerus:2004ds}, any object that can be constructed from target space concepts should have a counterpart on the worldsheet, in the conformal field theory formalism. So let us see what is the analogue of the K-group $K(B^{9-p}, S^{8-p}) = \mathbb{Z}$ from the CFT point of view. Closed type IIB string theory in $\mathbb{R}^{10}$ is described by the conformal field theory (CFT) consisting of ten free bosons and fermions tensored with a ghost CFT, and subject to the appropriate GSO projection. In the Ramond-Ramond sector, zero modes $\psi^\mu_0$ of the holomorphic and antiholomorphic fermions form two anticommuting copies of the Clifford algebra $\mbox{Cl}(\mathbb{R}^{9+1})$. They satisfy :
$$
\{\psi^\mu_0,\psi^\nu_0\} = \eta_{\mu\nu}\;, \qquad \{\bar{\psi}^\mu_0,\bar{\psi}^\nu_0\} = \eta_{\mu\nu}\;, \qquad \{\psi^\mu_0,\bar{\psi}^\nu_0\} = 0 \;,
$$
where $\eta_{\mu\nu}$ is the standard Minkowski metric. The GSO projection keeps only operators formed by an even number of fermionic zero modes. The massless Ramond-Ramond sector of this theory (at zero momentum) is given by the even part of a ($\mathbb{Z}/2\mathbb{Z}$ graded) tensor product of two Clifford modules. The following representation of this vector space is useful. Pick an orthonormal basis $\{e^\mu\}$ such that the first $p+1$ vectors are tangent to the worldvolume of $p+1$ dimensional hyperplane considered above, define $\psi^\mu_{\pm} = \frac{1}{\sqrt{2}}(\psi^\mu_0 \pm i \bar{\psi}^\mu_0)$, and let $\ket{1}$ be the state of unit norm such that $\psi^\mu_{+}\ket{1} = 0$ for all $\mu$. Then the massless Ramond-Ramond sector is generated by the states :
$$
\ket{e^{\mu_1} \wedge ... \wedge e^{\mu_{p+1}}} := \psi^{\mu_1}_{-} ... \; \psi^{\mu_{p+1}}_{-}\ket{1}
$$
with odd $p$. We will call these states ``test states'', for the following reason. Up to a normalization factor, the component on the Ramond-Ramond ground states of the D-branes in our family is given (in the light cone gauge) by $\ket{e^{2} \wedge ... \wedge e^{p+1}} + \ket{e^{p+2} \wedge ... \wedge e^{9}}$. The Ramond-Ramond charge of a given D-brane can be probed by computing the overlap $\braket{e^{2} \wedge ... \wedge e^{p+1}}{\mbox{D}p}$ (see for instance section 8 of \cite{DiVecchia:1999rh} for more details).
Now it is well-known that boundary states form a lattice (ie. one can stack only an integer number of D-branes), so the overlap with the test states are quantized for all of the boundary states. After a suitable normalization, the linear form $\bra{e^{2} \wedge ... \wedge e^{p+1}}$ can be seen as a map from the set of boundary states into $K(B^{9-p}, S^{8-p}) = \mathbb{Z}$, which assigns to each D-brane its K-theory charge. \cite{Sen:2002an} provided good indications that these overlaps between boundary states and test states are invariant under renormalization group flows. Indeed, the D9 and anti-D9 configuration with a non-trivial tachyon field configuration seems to carry a Ramond-Ramond charge identical to the charge of the equivalent D$p$-brane. Therefore on the worldsheet, the map associating to a D-brane its K-theory charge is realized by taking the overlap of the corresponding boundary state with a test state in the Ramond-Ramond vacuum. Note that the study of the coupling of D-branes to test states is completely equivalent to the study of their intersection form, and yields their charges more straightforwardly.

We will only aim at probing the Ramond-Ramond charges of worldsheet supersymmetric boundary states, that is, boundary states $\ket{B}$ satisfying $D_- \ket{B} = 0$. $D_-$ is a worldsheet supercharge, and we define $D_+ = (D_-)^\dagger$. In non-trivial conformal field theories, some of the charges obtained by taking the scalar product of supersymmetric boundary state with Ramond-Ramond test states may be linearly dependent. Indeed, any test state $\ket{RR}$ of the form $\ket{RR} = D_+ \ket{RR'}$ has vanishing scalar product with $\ket{B}$. There may also exist massless Ramond-Ramond states which are not supersymmetric. We will ignore such states and restrict ourselves to test states $\ket{RR}$ which lie in the kernel of the adjoint of the worldsheet supercharge \nolinebreak: $D_+\ket{RR} = 0$. Therefore the set of Ramond-Ramond test states we propose to consider is provided by a basis of representatives of the cohomology of the supercharge $D_+$ on the set of massless Ramond-Ramond states. Note that cohomology appears here exactly for the same reason as in the BRST formalism : we are restricting ourselves to states lying in the kernel of a nilpotent operator. In the flat space case, $D_+$ vanishes on the set of Ramond-Ramond ground states, which therefore coincides with the cohomology. However the worldsheet supercharge of the super Wess-Zumino-Witten model does have a non-trivial cohomology, and it is necessary to take this fact into account to obtain agreement with the prediction of K-theory on the charge group. 


\subsection{Summary of the construction}

When trying to apply the ``test-state'' procedure reviewed above to the super Wess-Zumino-Witten model, one faces at least three major difficulties :
\begin{itemize}
	\item The physical state space of super WZW models does not contain any massless Ramond-Ramond state \cite{Fuchs:1988gm}. This shows that the very concept of Ramond-Ramond charge is ill-defined, as there are no massless Ramond-Ramond gauge fields in the target space theory. (It is the reason why one should think of the charges to be found below only as invariants of the boundary renormalization group flow.)
	\item When one tries to use the (massive) Ramond-Ramond ground states to probe D-branes, the charge obtained is not quantized, and moreover it is modified by the action of the Kondo flows \cite{Bachas:2000ik}.
	\item Finally, to match the K-theory prediction \eqref{K-theoryG}, the charge must be identified modulo $M$. It is unclear how one could possibly get a periodically identified charge from a scalar product between the boundary state and a test state.
\end{itemize}
There is however a way of solving these three problems all at once : we have to look for truly massless Ramond-Ramond test states outside the physical state space of the super WZW model. 

Let us explain this idea. Recall that the state space of the super Wess-Zumino-Witten model on a simply connected Lie group in the Ramond-Ramond sector is a direct sum of irreducible modules of the form : 
$$
V_\lambda = H^\mathfrak{g}_\lambda \otimes \bar{H}^\mathfrak{g}_{\lambda^\ast} \otimes F_{R} \otimes F_{R} \;,
$$
where $H^\mathfrak{g}_\lambda$ and $\bar{H}^\mathfrak{g}_{\lambda^\ast}$ are integrable modules for a Kac-Moody algebra $\gk$ associated to the group $G$, and $F_{R}$ are Fock modules for $d = \mbox{dim}G$ free fermions in the Ramond sector. The condition that the Ramond-Ramond ground states be massless reads (see section \ref{SectCohomSupCharge}) :
$$
(\lambda,\lambda + 2\rho) + \frac{h^\vee d}{12} = 0\;,
$$
where $h^\vee$ is the dual Coxeter number of $G$ and $\rho$ its Weyl vector (half the sum of the positive roots). As any integrable module for $\gk$ satisfies $(\lambda,\rho) \geq 0$, the physical sector of the sWZW model does not contain any massless Ramond-Ramond state. However, the ansatz $\lambda = -\rho$ satisfies the equation and is the unique solution (by Freudenthal-de Vries strange formula). While the non-integrable module $V_{-\rho}$ does not appear in the physical state space of the theory, it does carry a representation of the spectrum generating algebra of the sWZW model, so one can in particular study the cohomology of the supercharge. It turns out that this cohomology has dimension $2^r$, where $r$ is the rank of $G$, and is supported on a highest grade subspace of $V_{-\rho}$, for some appropriate grading. This space will be our space of Ramond-Ramond test states.

To pursue the procedure sketched above in the case of flat space, one would like to measure the charge of a given boundary state by taking its scalar product with a given Ramond-Ramond test state. This is not readily possible because the boundary state does not have any component along $V_{-\rho}$. One should therefore complete the boundary state in the virtual sector $V_{-\rho}$. This completion can be performed in a consistent way as follows. Recall that boundary states are linear combinations of Ishibashi states, which are themselves a set of linearly independent solutions to the gluing conditions imposed on the boundary state. These gluing conditions specify how the bulk fields of the theory are reflected at the boundary, and they can also be solved in $V_{-\rho}$\footnote{To be precise, we will have to solve them in a bundle which fibers are composed of highest weight modules $V_{-\rho}$ with twisted action of the chiral algebra of the model. This is necessary to preserve the global $G \times G$ symmetries of the model, see section \ref{GenTestStates}.}. While there may be several linearly independent solutions to the gluing conditions in $V_{-\rho}$, only one of them intersects the representatives of the cohomology, and does so along a one-dimensional subspace, so the gluing conditions select an element of the cohomology up to normalization. 

The latter can be fixed by considering the action of Wilson operators \cite{Alekseev:2007in}. Roughly speaking, Wilson operators encode the reflection coefficients of the boundary state (the prefactors of the Ishibashi states). But they are also normal-ordered series in the Kac-Moody current, which have a well-defined action on any highest weight module. This important property can be used to perform an appropriate continuation of the components of the boundary states from their value in the physical state space to $V_{-\rho}$. 

The charge of a boundary state are then measured by taking the scalar product of a representative of the cohomology with the completed boundary state. 

We will show that :
\begin{itemize}
	\item The resulting charges are quantized (that is, integer up to normalization). (Section \ref{SubSecNormPer})
	\item The action of Wilson operators naturally imposes that these charges are periodic, with the right periodicity $M$, so they can actually be taken to lie in $\mathbb{Z}/M\mathbb{Z}$. (Section \ref{SubSecNormPer})
	\item Whenever a generalized Kondo flow sends a brane configuration onto another one, the charges of the two configurations are equal. These charges are therefore the invariants we are looking for. (Section \ref{SubSecInvKondFlows})
	\item Half of the linearly independent test states cannot couple to any boundary state, so the number of independent charges is $2^{r-1}$. This yields a charge group of the form : $\left ( \mathbb{Z}/M\mathbb{Z} \right )^{(2^{r-1})}$, which coincides with the twisted K-theory group \eqref{K-theoryG} of $G$. (Section \ref{SecRelHom})
	\item There is a distinguished basis of the representative of the cohomology of the supercharge that can be naturally identified with the generators of the homology of the Lie group. This provides the link between our algebraic picture of the charges and the more familiar geometric picture, in term of homology classes. We will check in several examples that the algebraic charge coincides with the geometric one. (Section \ref{SecRelHom})
\end{itemize}

We will now make these statements precise.

\section{Basic notions}

\label{SecBasNot}

\subsection{The super Wess-Zumino-Witten model}

We start by reviewing the super Wess-Zumino-Witten (sWZW) model \cite{DiVecchia:1984ep, Nemeschansky:1984mr, Kiritsis:1986wx, 1987PhLB..187..340N, 1987NuPhB.286..455F, Fuchs:1988gm}.

\label{ChirAlg}

\subsubsection*{The chiral algebra}

We describe here the chiral algebra of the sWZW model. To simplify the notations in this paper, we will not distinguish typographically the various infinite dimensional Lie algebras from their respective vertex algebras.

Let $G$ be a compact, simple and simply connected Lie group of dimension $d$. Let $\mathfrak{g} = \mbox{Lie}(G)$ be its Lie algebra, and $\hat{\mathfrak{g}}_{\tilde{k}}$ the corresponding Kac-Moody algebra at level $\tilde{k} = k+h^\vee$, $k>0$, where $h^\vee$ is the dual Coxeter number of $\mathfrak{g}$. Let us choose an orthonormal basis $\{e^a\}_{a = 1}^d$ of $\mathfrak{g}$ with respect to the Killing form, let $\{J^a(z)\}$ be the components of the Kac-Moody current on this basis, and $\{J^a_n\}$ ($n\in \mathbb{Z}$) their Laurent modes. All the sums on the Lie algebra indices $a,b,c,...$ will be implicit.

Let $\hat{\mathfrak{f}}$ be the Lie superalgebra generated by $d$ free fermions $\{\psi^a(z)\}$ with antiperiodic (Neveu-Schwarz) or periodic (Ramond) boundary conditions, and $\psi^a_r$ their Laurent modes. Here $r\in \mathbb{Z}+\frac{1}{2}$ in the Neveu-Schwarz sector, and $r\in \mathbb{Z}$ in the Ramond sector. We will adopt this convention throughout the rest of this paper. We will sometimes see the $d$ free fermions $\{\psi^a\}$ as a single $\mathfrak{g}$-valued fermionic field $\psi$. The chiral algebra $\hat{\mathfrak{c}}$ of the level $\tilde{k}$ super Wess-Zumino-Witten (sWZW) model is given by the semidirect product $\hat{\mathfrak{c}} = \hat{\mathfrak{g}}_{\tilde{k}} \ltimes \hat{\mathfrak{f}}$, where the action of $\hat{\mathfrak{g}}_{\tilde{k}}$ on $\hat{\mathfrak{f}}$ is given by (\ref{actionGsurF}) below. Explicitly, the generators satisfy the following commutation relations \nolinebreak:
$$
[J^a_n, J^b_m] = f_{abc} J^c_{m+n} + \tilde{k}n\delta_{ab} \delta_{n+m,0} \;, \mbox{\hspace{1cm}} \{\psi^a_r, \psi^b_s\} = \delta_{ab} \delta_{r+s,0} \;,
$$
\begin{equation}
\label{actionGsurF}
[J^a_n, \psi^b_r] = f_{abc} \psi^c_{n+r} \;,
\end{equation}
where $f_{abc}$ are the structure constants of $\mathfrak{g}$.

\subsubsection*{Subalgebras}

The chiral algebra $\hat{\mathfrak{c}}$ defined above contains several important subalgebras (in the sense of vertex algebras).

First, $\hat{\mathfrak{f}}$ contains a subalgebra isomorphic to $\hat{\mathfrak{g}}_{h^\vee}$, the Kac-Moody algebra based on $\g$ at level $h^\vee$. The generators for this subalgebra are :
\begin{equation}
\label{FermCurrent}
(J_\psi)^a_n =  -\frac{1}{2}f_{abc} \sum_{r} \psi^b_r \psi^c_{n-r}\;.
\end{equation}
One can then define the ``bosonic'' current $\bJ$ :
\begin{equation}
\label{BosCurrent}
\bJ = J - J_\psi \;.
\end{equation}
It generates a Kac-Moody subalgebra $\hat{\mathfrak{g}}_{k}$ of $\hat{\mathfrak{c}}$ which has the crucial property of commuting with $\hat{\mathfrak{f}}$ :
$$
[\bJ^a_n, \psi^b_r] = 0 \;.
$$
This shows that $\hat{\mathfrak{c}}$ is in fact isomorphic to $\hat{\mathfrak{g}}_{k} \oplus \hat{\mathfrak{f}}$.

$\hat{\mathfrak{c}}$ also contains a copy of the $N = 1$ superconformal algebra, with generators given by :
\begin{align}
\label{ConfGen}
L_n =& \frac{1}{2\tilde{k}} \sum_m :\bJ^a_m \bJ^a_{n-m}: + \frac{1}{2} \sum_r r :\psi^a_{n-r} \psi^a_r: \left ( + \frac{d}{16}\delta_{n,0} \right ) ,\\
\label{SupConfGen}
G_r =& -\frac{1}{\sqrt{\tilde{k}}} \left ( \sum_m \bJ^a_m \psi^a_{r-m} - \frac{1}{6}f_{abc} \sum_{s,t} \psi^a_s \psi^b_t \psi^c_{r-s-t} \right ) ,
\end{align}
where the indices $r$, $s$ and $t$ are summed over $\mathbb{Z}$ or $\mathbb{Z}+\frac{1}{2}$ in the Ramond and Neveu-Schwarz sector, and $m$ and $n$ are always summed over $\mathbb{Z}$. The term between parenthesis in the definition of $L_0$ is present only in the Ramond sector. They satisfy :
\begin{equation}
\label{SupConfAlg}
\begin{split}
[L_m,L_n] =& \; (m-n)L_{m+n} + \frac{c}{12}(m^3-m)\delta_{m+n,0} \;, \\
\{G_r,G_s\} =& \; 2L_{r+s} + \frac{c}{12}(4r^2-1)\delta_{r+s,0} \;, \\
[L_m,G_r] =& \; \frac{m-2r}{2}G_{m+r} \;,
\end{split}
\end{equation}
where $c = \frac{3k+h^\vee}{2(k+h^\vee)}d$ is the central charge. Their action on the generators of $\hat{\mathfrak{c}}$ is given by :
$$
[L_m,J^a_n] = -nJ^a_{m+n}\;, \mbox{\hspace{1cm}} [L_m,\psi^a_n] = -\frac{2n+m}{2}\psi^a_{m+n}\;, 
$$
$$
[G_m,J^a_n] = \sqrt{\tilde{k}}n\psi^a_{m+n}\;,  \mbox{\hspace{1cm}} \{G_m,\psi^a_n\} = -\frac{1}{\sqrt{\tilde{k}}}J^a_{m+n}\;.
$$

Finally, we note that the bosonic part of $\hat{\mathfrak{f}}$ is isomorphic to the Kac-Moody algebra $\hat{\mathfrak{so}}(d)_1$ at level one, with generators :
\begin{equation}
\label{SOCurrent}
J_{\mathfrak{so}}^i(z) = \frac{1}{2}t^i_{ab}\psi^a(z) \psi^b (z) \;, \mbox{\hspace{1cm}} i = 1, ..., \frac{1}{2}(d^2-d) \;,
\end{equation}
where $t^i_{ab}$ are the matrix elements of the generators $\{t^i\}$ of $\mathfrak{so}(d)$ in the defining representation. The bosonic part of the chiral algebra is therefore given by $\hat{\mathfrak{g}}_{k} \oplus \hat{\mathfrak{so}}(d)_1$.

The full spectrum generating algebra is the direct sum $\hat{\mathfrak{c}} \oplus \hat{\mathfrak{c}}$ of a holomorphic and antiholomorphic copy of $\hat{\mathfrak{c}}$. Here we understand the direct sum in a $\mathbb{Z}/2\mathbb{Z}$-graded sense, so that holomorphic and antiholomorphic elements having both odd fermion number anticommute. The fields in the holomorphic sector will be denoted as above, and the ones in the antiholomorphic sector will carry a bar ($\bar{J}$, $\bar{\psi}$, ...).

\subsubsection*{The state space}

We start by describing highest weight modules for $\hat{\mathfrak{g}}_{k}$ and  $\hat{\mathfrak{f}}$. 

The highest weight modules for $\hat{\mathfrak{g}}_{k}$ that will be relevant to the construction of the state space of the sWZW model are the integrable modules. They are indexed by the (finite) set of integrable dominant weights at level $k$, $P^+_k \subset \mathfrak{h}^\ast$ where $\mathfrak{h}$ is the Cartan subalgebra of $\mathfrak{g}$. Given an integrable weight $\lambda \in P^+_k$, we will denote the corresponding integrable highest weight module by $H^{\mathfrak{g}}_\lambda$, with a superscript indicating the corresponding Kac-Moody algebra in the situations when an ambiguity might occur. Integrable highest weight modules carry a hermitian invariant bilinear form. The elements of the compact form of the Kac-Moody algebra are anti-self-adjoint with respect to this form, and the adjoints of the Kac-Moody generators are given by $(J^a_n)^\dagger = -J^a_{-n}$.

There are only two irreducible highest weight modules for $\hat{\mathfrak{f}}$, one in the Neveu-Schwarz sector, and one in the Ramond sector, and we denotes them respectively by $F_{NS}$ and $F_{R}$. 

All of these modules inherit a grading $\mbox{gr}_{L_0}$ from the adjoint action of $L_0$ and their components with negative grade are trivial. Note that the grade zero subspace $(F_{NS})_0$ of $F_{NS}$ is one dimensional, whereas $(F_{R})_0$ is an irreducible Clifford module for the $d$-dimensional Clifford algebra $\mbox{Cl}(d)$ generated by the zero modes $\psi_0$ of the fermions in the Ramond sector. 

We want now to construct the state space of the sWZW model. This state space can be fixed by the requirement that the torus partition function be modular invariant when the mapping class group $SL(2,\mathbb{Z})$ of the torus acts. This modular invariance condition forces us to impose a GSO projection that breaks the spectrum generating algebra $\hat{\mathfrak{c}} \oplus \hat{\mathfrak{c}}$ to its bosonic part. 

We are working with a compact simple Lie group of arbitrary dimension, so generically only the type 0 GSO projection is available. The type 0 GSO projector is given by $P_{GSO} = \frac{1}{2}(1+(-1)^{(\mathcal{F}+\bar{\mathcal{F}})})$, where $\mathcal{F}$ and $\bar{\mathcal{F}}$ denote the fermion numbers in the holomorphic and antiholomorphic sectors, ie. the $\mathbb{Z}/2\mathbb{Z}$ gradings coming from the superalgebra structures of the left and right factor of $\hat{\mathfrak{c}} \oplus \hat{\mathfrak{c}}$. The operators of total grade zero are the only ones which survive the projection, so the remaining chiral algebra is given by $\gk \oplus \sod \oplus \gk \oplus \sod$. Note that the superconformal generators $G$ and $\bar{G}$ are projected out.

It turns out that the problem of constructing a partition function for $P_{GSO}(\hat{\mathfrak{c}} \oplus \hat{\mathfrak{c}})$ which has the required modular invariance properties boils down to constructing a modular invariant partition function for the bosonic WZW model based on the chiral algebra $\gk \oplus \sod$. We will construct these partition functions, and then comment on why the underlying module is a module for $P_{GSO}(\hat{\mathfrak{c}} \oplus \hat{\mathfrak{c}})$.

The state space of the $\gk \oplus \sod$ WZW model factorizes into $\hat{\mathfrak{g}}_{k}$-modules and $\hat{\mathfrak{so}}(d)_1$-modules\footnote{This is not necessarily the case for non simply connected groups, see \cite{Braun:2004qg, Fredenhagen:2004xp}.} \nolinebreak:
\begin{equation}
\label{StateSpacesWZW}
\mathcal{H} = \mathcal{H}_{\mathfrak{g}} \otimes \mathcal{H}^X_{\mathfrak{so}}\;,
\end{equation}
where $X =$ 0A, 0B or 0 indexes different possible choices, see below. We will choose the charge conjugation modular invariant for the $\hat{\mathfrak{g}}_{k}$ theory :
$$
\mathcal{H}_{\mathfrak{g}} = \bigoplus_{\lambda \in P^+_k} H_\lambda \otimes H_{\lambda^\ast} \;,
$$
where $\lambda^\ast$ is the weight conjugate to $\lambda$. In the $k \rightarrow \infty$ limit, the WZW model with charge conjugation modular invariant describes a string evolving in the simply connected Lie group $G$ constructed from $\g$. The extension of our results to models with different modular invariants may be non trivial.

$\sod$ has four integrable representations when $d$ is even, namely the ones corresponding to the trivial ($t$), the defining (or fundamental) ($f$), and the two spinorial ($s$ and $s'$) representations of $\mathfrak{so}(d)$. For odd $d$ the trivial and defining representations are still present, but there is a single spinorial representation, that we denote by $s$. In even $d$, $\mathfrak{so}(d)$ admits an outer automorphism, so we can construct a modular partition function using either this outer automorphism or the trivial one. The two modular invariants obtained correspond respectively the type 0A and 0B GSO projections of string theories. In odd dimension, there is a unique type 0 modular invariant. The state space for the $\sod$ part therefore reads :
\begin{align}
\label{StateSpaceSod}
\begin{split}
\mathcal{H}^0_{\mathfrak{so}} =&\; (H_t \otimes H_t) \oplus (H_f \otimes H_f) \oplus (H_s \otimes H_s)\;, \qquad \qquad \qquad \quad \, d \mbox{ odd,} \\
\mathcal{H}^{0A}_{\mathfrak{so}} =&\; (H_t \otimes H_t) \oplus (H_f \otimes H_f) \oplus (H_{s'} \otimes H_s) \oplus (H_s \otimes H_{s'})\;, \quad d \mbox{ even,} \\
\mathcal{H}^{0B}_{\mathfrak{so}} =&\; (H_t \otimes H_t) \oplus (H_f \otimes H_f) \oplus (H_s \otimes H_s) \oplus (H_{s'} \otimes H_{s'})\;, \quad d \mbox{ even.}
\end{split}
\end{align}

By standard CFT arguments (for instance \cite{CFT1997}, chapter 17), the state spaces $\mathcal{H}_{\mathfrak{g}}$, $\mathcal{H}^0_{\mathfrak{so}}$, $\mathcal{H}^{0A}_{\mathfrak{so}}$, $\mathcal{H}^{0B}_{\mathfrak{so}}$ all yield modular invariant partitions functions. Therefore $\mathcal{H}$ as defined in (\ref{StateSpacesWZW}) yields a modular invariant partition function in each case.

To see that these state spaces are really modules for our spectrum generating algebra $P_{GSO}(\hat{\mathfrak{c}} \oplus \hat{\mathfrak{c}})$, note how the highest weight modules for $\hat{\mathfrak{f}}$ decompose into $\sod$ modules :
$$
F_{NS} \rightarrow H^{\mathfrak{so}}_t \oplus H^{\mathfrak{so}}_f \;,
$$
$$
F_{R} \rightarrow H^{\mathfrak{so}}_s \oplus H^{\mathfrak{so}}_{s'} \;\;(d \mbox{ even), \hspace{1cm}} F_{R} \rightarrow H^{\mathfrak{so}}_s \;\;(d \mbox{ odd}).
$$
In the decomposition of $F_{NS}$, $H^{\mathfrak{so}}_t$ appears at grade zero, while $H^{\mathfrak{so}}_f$ appears at grade $\frac{1}{2}$. The two spinorial $\sod$-modules are already present at grade zero in the decomposition of $F_{R}$ in even dimension. One way of checking these relations is to compute the dimensions of the grade 0 and $\frac{1}{2}$ subspaces of the relevant modules, and then recall that any product of an even number of fermionic generators can be expressed in term of the currents of $\sod$. Under the operator-state mapping of the vertex algebra $\hat{\mathfrak{f}}$, $H^{\mathfrak{so}}_t$ corresponds to operators with even fermion number, while $H^{\mathfrak{so}}_f$ corresponds to operators with odd fermion number. For even $d$, one has the same picture in the Ramond sector, where the two $\sod$ spinorial modules correspond to even and odd fermion number operators (which is which depends on how we choose the grading on the grade zero subspace). For odd $d$, however, $F_{R}$ does not carry any $\mathbb{Z}/2\mathbb{Z}$ grading. We have the same picture in the antiholomorphic sector.

Now by the remark above and \eqref{StateSpaceSod}, we see that the postulated state spaces (\ref{StateSpacesWZW}) are modules for $P_{GSO}(\hat{\mathfrak{c}} \oplus \hat{\mathfrak{c}})$. Operators with $\mathcal{F} = \bar{\mathcal{F}} = 0$ preserve the $\sod$ modules, while those with $\mathcal{F} = \bar{\mathcal{F}} = 1$ permute the summands in (\ref{StateSpaceSod}).

\subsubsection*{Supersymmetric states}

It will be crucial for us to have a notion of ``supersymmetric state''. The problem is that the superconformal generators $G$ and $\bar{G}$ defined in \eqref{SupConfGen} do not act on the state space of the sWZW model. They have odd fermion number and are projected out by the GSO projection. 

However, we will be able to define supersymmetric states if we construct a $\mathbb{Z}/2\mathbb{Z}$-graded module $\mathcal{H}'$ for $\hat{\mathfrak{c}} \oplus \hat{\mathfrak{c}}$ such that the state space of the sWZW model coincides with the even part of $\mathcal{H}'$ : $\mathcal{H} = (\mathcal{H}')_0$. Denote by $i_{\mathcal{H}}$ this embedding. Then the supersymmetry generators map $(\mathcal{H}')_0$ to $(\mathcal{H}')_1$, and we can define a state $\ket{X} \in \mathcal{H}$ to be supersymmetric if :
$$
(G_r - i \epsilon  \bar{G}_{-r})i_{\mathcal{H}}(\ket{X}) = 0 \quad \mbox{in} \; \mathcal{H}'.
$$
$\epsilon = \pm 1$ depends on the supercharge chosen to be preserved.

The module $\mathcal{H}'$ is straightforward to construct when $d$ is even. One can take in this case :
\begin{equation}
\label{SupSymStateSpace}
\mathcal{H}' = \mathcal{H}_{\mathfrak{g}} \otimes (F_{NS} \otimes F_{NS} \oplus F_{R} \otimes F_{R})\;,
\end{equation}
where it is understood that in the fermionic modules of the form $F \otimes F$, the holomorphic modes $\psi^a_n$ act on the first factor through the action of $\hat{\mathfrak{f}}$, and the anti-holomorphic modes acts on the first factor by $(-1)^{\mathcal{F}}$ and on the second through the usual action of $\hat{\mathfrak{f}}$. The non-trivial action of the anti-holomorphic modes on the first factor is necessary to make them anticommute with the holomorphic modes, rather than commute. Depending on the choice of $\mathbb{Z}/2\mathbb{Z}$-grading on the two copies of $F_{R}$ we get an extension of the state space of the 0A or 0B sWZW model.

$\mathcal{H}'$ is less easy to construct when $d$ is odd. By what was said above, we see readily that a construction analogous to the one in the even case is impossible, as $F_R$ does not admit any $\mathbb{Z}/2\mathbb{Z}$-grading in the odd case. However, as long as we do not impose the reality condition $z^\ast = \bar{z}$ on the worldsheet coordinates, the holomorphic and antiholomorphic modes of the fermions form a Lie algebra isomorphic to the Lie algebra of the modes of $2d$ chiral fermions. A Ramond module $F_{R}^{2d}$ for $2d$ chiral fermions admits a $\mathbb{Z}/2\mathbb{Z}$-grading, namely the fermion number. Therefore, the following is a module for $\hat{\mathfrak{c}} \oplus \hat{\mathfrak{c}}$ :
$$
\mathcal{H}' = \mathcal{H}_{\mathfrak{g}} \otimes (F_{NS} \otimes F_{NS} \oplus F_{R}^{2d})\;.
$$
The action of $\hat{\mathfrak{c}} \oplus \hat{\mathfrak{c}}$ does not split into a holomorphic and antiholomorphic module in the Ramond sector, as was the case in the even $d$ case. But after the GSO projection, the even part of $F_{R}^{2d}$ becomes a spinorial module $H^{2d}_s$ for $\hat{\mathfrak{so}}(2d)_1$, which is isomorphic to $H_s \otimes H_s$ as a $\sod \oplus \sod$-module (as can be seen by a computation of the dimensions of the grade zero subspaces, for instance). So the holomorphic/antiholomorphic splitting is recovered after the GSO projection. 

Actually, the same construction can be applied in the even case, and is equivalent to the one we used because for $d$ even, $F_{R}^{2d} \simeq F_{R} \otimes F_{R}$ as $\hat{\mathfrak{f}} \oplus \hat{\mathfrak{f}}$-modules.

We have now a well-defined notion of a supersymmetric state in $\mathcal{H}$. In the remaining of this paper, we will omit to write explicitly the map $i_{\mathcal{H}}$ to avoid cluttering the notation too much. But it should be understood each time an odd operator acts on a state of $\mathcal{H}$.

\vspace{.5cm}

We now describe a useful parametrization of the grade zero subspace of $\mathcal{H}'$ generated by the zero modes of the holomorphic and antiholomorphic fermions in the Ramond-Ramond sector. These zero modes satisfy the following relations :
$$
\{\psi^a_0,\psi^b_0\} = \delta_{ab}\;, \qquad \{\bar{\psi}^a_0,\bar{\psi}^b_0\} = \delta_{ab}\;, \qquad \{\psi^a_0,\bar{\psi}^b_0\} = 0\;.
$$
From the discussion above, they generate a Clifford module $F^{2d}_0$ for the Clifford algebra $\mbox{Cl}(2d)$ in dimension $2d$. We can make the following change of basis :
\begin{equation}
\label{DefFermPol}
\psi^a_{0+} = \frac{1}{\sqrt{2}}(\psi^a_0 + i\bar{\psi}^a_0) \;, \qquad \psi^a_{0-} = \frac{1}{\sqrt{2}}(\psi^a_0 - i\bar{\psi}^a_0)\;.
\end{equation}
The new generators satisfy :
$$
\{\psi^a_{0+},\psi^b_{0+}\} = 0 \;, \qquad \{\psi^a_{0+},\psi^b_{0-}\} = \delta_{ab} \;, \qquad \{\psi^a_{0-},\psi^b_{0-}\} = 0 \;.
$$
Defining $\ket{1}$ to be the state with unit norm such that $\psi^a_{0+}\ket{1} = 0$ for all $a = 1,...,d$, the Clifford module $F^{2d}_0$ is freely generated from $\ket{1}$ by the set $\{\psi^a_{0-}\}$. We can therefore parametrize the vectors in $F^{2d}_0$ by elements of the exterior algebra $\bigwedge \mathfrak{g}$ :
\begin{equation}
\label{BaseCliff}
e^{a_1} \wedge ... \wedge e^{a_p} \mapsto \ket{e^{a_1} \wedge ... \wedge e^{a_p}} := \psi^{a_1}_{0-}...\psi^{a_p}_{0-} \ket{1} \;,
\end{equation}
where we denoted the product in the exterior algebra by $\wedge$.

Finally, let us note that, by definition, the state $\ket{e^{a_1} \wedge ... \wedge e^{a_p}}$ satisfies the following relations \nolinebreak:
\begin{equation}
\label{GlueCondVecBasFermMod}
(\psi^a_0 + i\epsilon\bar{\psi}^a_0)\ket{e^{a_1} \wedge ... \wedge e^{a_p}} = 0 \;,
\end{equation}
with $\epsilon = -1$ si $a \in \{a_1,...,a_p\}$ and $\epsilon = 1$ else.

\subsection{Wilson loops and Kondo renormalization group flows}

\label{WilsLoopKondRGF}

We define here the Kondo perturbation in the bosonic case and recall how the fixed points of the induced boundary renormalization group flow can be identified by mean of quantized Wilson operators.

\subsubsection*{The Kondo perturbation}

Consider a purely bosonic WZW model with holomorphic current $J \in \hat{\mathfrak{g}}_k$, defined on a surface (possibly with boundaries) $\Sigma$, with an embedded time-like cycle $C$. Let $A : \mathfrak{g} \rightarrow \mathbb{C}^n$ a $n$-dimensional representation of $\mathfrak{g}$, and $A^a = A(e^a)$. Let us tensor the state space $\mathcal{H}_{\mathfrak{g}}$ of the WZW model with $\mathbb{C}^n$. One can perturb the WZW action with the following term, acting on $\mathcal{H}_{\mathfrak{g}} \otimes \mathbb{C}^n$ \nolinebreak:
\begin{equation}
\label{KondoPert}
\Delta S = l \int_C d\sigma J^a(\sigma)A^a \;,
\end{equation}
where $l$ is a coupling, and $\sigma$ a parametrization of $C$. One can see this perturbation as a point-like charged defect with worldline $C$ and spin $A$, which interacts minimally with the current $J$. 

From the string theory point of view, the Kondo perturbation has a different interpretation. Consider an open string cylinder amplitude between two D-branes. We have $\Sigma = S^1 \times [0,1]$, with worldsheet time running along $S^1$. Let us choose $C = S^1 \times \{0\}$, so that the perturbation is supported on one of the boundaries of the cylinder. $\mathcal{H}_\g \otimes \mathbb{C}^n$ can now be interpreted as the state space for open strings stretched between a stack of $n$ identical D-branes at $S^1 \times \{0\}$ and a given D-brane at $S^1 \times \{1\}$. This perturbation amounts to turning on a constant field $A$ on the stack of D-branes \cite{Alekseev:2000fd}. For generic $l$, this perturbation breaks the superconformal symmetry of the model, and one can study the boundary renormalization group flow that it triggers. The IR fixed point of this flow is described by the Affleck-Ludwig prescription \cite{Affleck:1990by}, and is again a WZW model, with a new boundary condition on the boundary initially perturbed. When the perturbed stack of D-brane is composed of $n$ maximally symmetric branes of label $\lambda \in P^+_k$, the final D-brane configuration is given by a set of $\mathcal{N}_{\lambda\mu}^{\;\;\nu}$ maximally symmetric branes of label $\nu$, where $\mu$ is the highest weight of the $\mathfrak{g}$-representation $A$ and $\mathcal{N}_{\lambda\mu}^{\;\;\nu}$ are the fusion coefficients of $\hat{\mathfrak{g}}_k$. A rigorous justification of the Affleck-Ludwig principle can be found in \cite{Alekseev:2007in}, section 5.

It is however instructive to look at Kondo perturbations from the worldsheet dual theory. 

\subsubsection*{Quantum Wilson loops}

By open-closed string duality, one can consider the same setting, but now with worldsheet time running along the non-periodic direction of the cylinder. This amplitude has now the interpretation of a closed string exchange between the branes sitting at each end of the cylinder. The cycle $C$ is spacelike, and the perturbation can be seen as a defect supported on $C$. 

Classically, this defect is a Wilson loop having the following expression :
\begin{equation}
\label{ClassWilsLoop}
w(\mu,l) = \mbox{Tr}_{\mathbb{C}^n} \mbox{P} \exp \left ( il \int_C d\sigma j^a(\sigma) A^a \right )\;,
\end{equation}
where $P$ denotes the path-ordered exponential, $\mu$ is the highest weight of the representation $A$ of $\g$ on $\mathbbm{C}^n$ and $j^a(\sigma)$ are the components of the classical current $j$. These classical observables are topological : they depend only on the homotopy class of $C$. When $l = \frac{1}{k}$, $w(\mu,l)$ even preserves the full symmetry of the WZW model. Indeed, it has vanishing Poisson bracket with the classical current $j$.

To understand the Kondo perturbation from the closed string point of view, one needs a quantized version of the classical Wilson loop. This quantization was performed in \cite{Alekseev:2007in} in the case $l = \frac{1}{k}$. The quantized Wilson loop $W_\mu$ corresponding to the classical Wilson loop $w(\mu,\frac{1}{k})$ is a normal-ordered series in the quantum Kac-Moody current $J$. The special symmetries of $w(\mu,\frac{1}{k})$ are preserved by this quantization procedure, which means that $W_\mu$ commutes with every element of $\hat{\mathfrak{g}}_k$. Hence it acts by scalar multiplication on any irreducible $\hat{\mathfrak{g}}_k$-module. The power of the quantization procedure of \cite{Alekseev:2007in} is that the spectrum of $W_\mu$ is obtained explicitly. Let $\eta$ be any weight at level $k > -h^\vee$, and $M_\eta$ is the Verma module of highest weight $\eta$. Then \nolinebreak:
\begin{equation}
\label{EigValWilsLoop}
W_\mu = \chi_\mu \left ( -\frac{2\pi i}{k+h^\vee} (\eta + \rho) \right ) \mathbbm{1} \mbox{ \hspace{.5cm} on } M_\eta \;,
\end{equation}
where $\chi_\mu$ is the $\mathfrak{g}$-character of the representation with highest weight $\mu$. On the integrable highest weight module $H_\lambda$ and for integrable $\mu$, the eigenvalue can be written as :
$$
W_\mu = 
 \frac{S_{\mu\lambda}}{S_{0\lambda}} \mathbbm{1} \mbox{ \hspace{.5cm} on } H_\lambda \;,
$$
where $S_{\mu\lambda}$ is the modular $S$-matrix of $\hat{\mathfrak{g}}_k$. However, the fact that the action of $W_\mu$ is well-defined on any highest weight module at level $k$ will be crucial to our argument. 

We have now a well-defined expression for the quantized Wilson operator at the special coupling value $l = \frac{1}{k}$. This value corresponds to the (classical) IR fixed point of the renormalization group flow equation starting from the UV fixed point $l = 0$. Moreover, as $W_\mu$ commutes with $\hat{\mathfrak{g}}_k$, it also commutes with the associated Virasoro algebra. Therefore the theory defined on the cylinder in which $W_\mu$ is inserted in all the amplitudes is still a conformal field theory, which still has a $\hat{\mathfrak{g}}_k \oplus \hat{\mathfrak{g}}_k$ symmetry. As was shown in \cite{Alekseev:2007in}, it is actually the Affleck-Ludwig fixed point of the Kondo flow. Put differently, when the Wilson loop $W_\mu$ acts on a boundary state $\ket{B}$, it yields the infrared fixed point of the RG flow triggered by the corresponding Kondo perturbation on $d_\mu\ket{B}$. Because the spectrum of $W_\mu$ is completely explicit, this provides a very simple and efficient way of investigating Kondo flows. We will repeat this argument in detail below in section \ref{SupSymKondPert}, in the case of supersymmetric Kondo perturbations.

The discussion above is not restricted to maximally symmetric boundary states, because the construction of the Wilson operator was completely independent from the boundary conditions imposed at the ends of the cylinder. We can see the Kondo flow as acting on defect operators as $d_\mu\mathbbm{1} \mapsto W_\mu$. This flow on defect operators turns into a boundary flow when we let these two operators act on a given boundary state. Wilson operators therefore provide a generalization of the Affleck-Ludwig prescription : they describes ``universal'' Kondo flows starting from any D-brane. (See \cite{Bachas:2004sy} for a deeper discussion of the universal properties of these flows.)

\vspace{.5cm}

Let us add here an important remark. The Wilson operators described above form a ring isomorphic to the representation ring of $\mathfrak{g}$, because their eigenvalues \eqref{EigValWilsLoop} are given by characters of $\mathfrak{g}$. However, the physical state space $\mathcal{H}_{\mathfrak{g}}$ of the sWZW model decomposes into a direct sum of integrable highest weight $\gk$-modules. Let $w$ be an element of the affine Weyl group of $\gk$, and $\epsilon(w)$ its sign. The Wilson operators $W^\mathfrak{g}_\mu$ and $\epsilon(w)W^\mathfrak{g}_{w(\mu)}$ have an identical action on every integrable module, because their eigenvalues are the same. One way to see that the eigenvalues \eqref{EigValWilsLoop} coincide is to use an argument similar to the one leading to the Kac-Walton formula (see \cite{CFT1997}, §16.2.1). We have therefore the equality : $W^\mathfrak{g}_\mu|_{\mathcal{H}_{\mathfrak{g}}} = \epsilon(w)W^\mathfrak{g}_{w(\mu)}|_{\mathcal{H}_{\mathfrak{g}}}$. We can define an equivalence relation on the ring generated by Wilson operators, where two elements are equivalent if they coincide on $\mathcal{H}_{\mathfrak{g}}$. Denoting the equivalence class of $W^\mathfrak{g}_\mu$ by $[W^\mathfrak{g}_\mu]$, we have :
\begin{equation}
\label{RingClassWilsOp}
[W^\mathfrak{g}_\lambda] [W^\mathfrak{g}_\mu] = \mathcal{N}_{\lambda\mu}^{\;\;\nu} [W^\mathfrak{g}_\nu] \;,
\end{equation}
where $\mathcal{N}_{\lambda\mu}^{\;\;\nu}$ are the fusion coefficients of $\gk$. Indeed, on every highest weight module $H_\lambda$, $[W^\mathfrak{g}_\mu]$ acts by scalar multiplication by $\frac{S_{\mu\lambda}}{S_{0\lambda}}$, so \eqref{RingClassWilsOp} is equivalent to Verlinde's formula.

\vspace{.5cm}

The formalism described above allows to treat a more general set of boundary perturbations \cite{Monnier:2005jt, Alekseev:2007in}. Choose a semi-simple subalgebra $\mathfrak{a} \subset \mathfrak{g}$, with embedding index $x$. Then we have an induced embedding of Kac-Moody algebras $\hat{\mathfrak{a}}_{xk} \subset \hat{\mathfrak{g}}_k$. Suppose that $P_\mathfrak{a}$ is the orthogonal projection (with respect to the Killing form) of $\mathfrak{g}$ on $\mathfrak{a}$, and let $A' : \mathfrak{a} \rightarrow \mathbb{C}^n$ be a representation of $\mathfrak{a}$ with highest weight $\tau$. One can then choose $A = A' \circ P$. (One therefore has $A^a = 0$ whenever $e^a \in \mathfrak{g}/\mathfrak{a}$.) 

The generalized Kondo perturbation (\ref{KondoPert}) also triggers a boundary renormalization group flow. One can repeat the quantization procedure, and get a quantized Wilson loop $W^\mathfrak{a}_\tau$. $W^\mathfrak{a}_\tau$ does not commute with all of $\hat{\mathfrak{g}}_k$, but rather with $\hat{\mathfrak{a}}_{xk} \oplus \hat{\mathfrak{g}}_k/\hat{\mathfrak{a}}_{xk}$, where the coset algebra $\hat{\mathfrak{g}}_k/\hat{\mathfrak{a}}_{xk}$ is the algebra formed by all the operators in the vertex algebra associated to $\hat{\mathfrak{g}}_k$ which commute with the subalgebra $\hat{\mathfrak{a}}_{xk}$. The spectrum of $W^\mathfrak{a}_\tau$ is also obtained in an explicit way :
\begin{equation}
\label{SymBreakWilLoop}
W^\mathfrak{a}_\tau = \chi^\mathfrak{a}_\tau \left ( -\frac{2\pi i}{xk+h^\vee(\mathfrak{a})} (\upsilon + \rho_\mathfrak{a}) \right ) \mathbbm{1} \mbox{ \hspace{.5cm} on } M^\mathfrak{a}_\upsilon \;,
\end{equation}
where $\chi^\mathfrak{a}$ is the $\mathfrak{a}$-character, $h^\vee(\mathfrak{a})$ and $\rho_{\mathfrak{a}}$ the dual Coxeter number and the Weyl vector of $\mathfrak{a}$, and $M^\mathfrak{a}_\upsilon$ the Verma $\hat{\mathfrak{a}}_{xk}$-module of highest weight $\upsilon$. To determine the action of $W^\mathfrak{a}_\tau$ on a $\hat{\mathfrak{g}}_k$ module, one first decomposes it into $\hat{\mathfrak{a}}_{xk}$-modules, on which the action is given by (\ref{SymBreakWilLoop}).

Again, one can use these Wilson operators to find the infrared fixed point boundary states of the generalized Kondo perturbations. These states preserve only $\hat{\mathfrak{a}}_{xk} \oplus \hat{\mathfrak{g}}_k/\hat{\mathfrak{a}}_{xk} \subset \hat{\mathfrak{g}}_k$, and were described in \cite{Quella:2002ct}.

\vspace{.5cm}

Wilson operators also provide a very convenient way of building boundary states. Whenever a Wilson operator commutes with a given subalgebra of $\hat{\mathfrak{a}} \subset \hat{\mathfrak{g}}_k$, its action on a boundary state preserving $\hat{\mathfrak{a}}$ automatically yields another boundary state preserving $\hat{\mathfrak{a}}$. This property will be used below in the construction of supersymmetric boundary states for the sWZW model.

\section[Boundary states in the sWZW model]{Boundary states in the sWZW model\protect\footnote{Many thanks to Stefan Fredenhagen for very useful discussions on this point.}}
\label{sWZWBranes}

In this section, we consider various well-known D-branes of the bosonic Wess-Zumino-Witten model, and show how to construct their supersymmetric counterparts in the sWZW model. The supersymmetric D-branes that we will construct are based on :
\begin{itemize}
	\item the maximally symmetric D-branes \cite{Cardy:1989ir},
	\item the twisted D-branes \cite{Birke:1999ik, Gaberdiel:2002qa, Petkova:2002yj},
	\item the ``coset'' D-branes \cite{Maldacena:2001ky, Quella:2002ct, Quella:2002ns},
	\item the ``twisted coset'' D-branes \cite{Maldacena:2001ky, Quella:2002ct, Quella:2002ns, Gaberdiel:2004hs, Gaberdiel:2004za}.
\end{itemize}
We will determine the charge of the first three families of D-branes in section \ref{SecChargeBoundSt} and the charge of some members of the fourth family in section \ref{SecExamples}. For some pedagogical introduction to the treatment of D-branes in conformal field theory, see \cite{Schomerus:2002dc, Gaberdiel:2002my, Zuber:2000ia}.

A D-brane is fully specified once its couplings with all of the closed string states are given. Therefore, it can be pictured as a functional on the state space of the conformal field theory. It is often convenient to see this functional as a ``boundary state'' in a completion of the state space of the model (these states are not normalizable in general). We will construct the D-branes mentioned above by exhibiting their corresponding boundary states.

Let us recall that a boundary state will be called ``supersymmetric'' if its components in the R-R and NS-NS sector satisfy the equations :
$$
(G_r - i \epsilon  \bar{G}_{-r})i_{\mathcal{H}}(\ket{B}) = 0 \;,
$$
where $i_{\mathcal{H}}$ is the map defined in section \ref{ChirAlg}.

\subsection{Maximally symmetric D-branes}

In the WZW model based on $\hat{\mathfrak{g}}_k$, the maximally symmetric D-branes, as their name indicates, preserve the maximal amount of the bulk symmetry, namely the diagonal $\hat{\mathfrak{g}}_k$ subalgebra of the spectrum generating algebra $\hat{\mathfrak{g}}_k \oplus \hat{\mathfrak{g}}_k$. 

We would like our supersymmetric maximally symmetric D-branes of the sWZW model to have the same property. The latter can be implemented in the gluing conditions satisfied by the corresponding boundary state $\ket{B}$ :
\begin{equation}
\label{MaxSymCond}
(J^a_n  + \bar{J}^a_{-n})\ket{B} = 0 \;.
\end{equation}
We want this boundary state to be conformal, which imposes :
\begin{equation}
\label{ConfCond}
(L_n - \bar{L}_{-n}) \ket{B} = 0 \;.
\end{equation}
Moreover, we also want it to be supersymmetric :
\begin{equation}
\label{GlobSupSymCond}
(G_r - i \epsilon \bar{G}_r) \ket{B} = 0 \;,
\end{equation}
with $\epsilon = \pm 1$. 

Given (\ref{MaxSymCond}) and the explicit form of the superconformal generators (\ref{SupConfGen}), of the bosonic current (\ref{BosCurrent}) and of the $\hat{\mathfrak{so}}(d)_1$ current (\ref{SOCurrent}), we see that imposing :
\begin{equation}
\label{FermCond}
(\psi^a_r + i\epsilon \bar{\psi}^a_{-r}) \ket{B} = 0
\end{equation}
implies :
\begin{equation}
\label{MaxSymCondSuSo}
(\bJ^a_n  + \bar{\bJ}^a_{-n})\ket{B} = 0 \;, \qquad ((J_{\mathfrak{so}})^i_n +  (\bar{J}_{\mathfrak{so}})^i_{-n})\ket{B} = 0 \;,
\end{equation}
and hence (\ref{ConfCond}) and (\ref{GlobSupSymCond}). Conversely, (\ref{MaxSymCondSuSo}) implies (\ref{MaxSymCond}) and (\ref{FermCond}) for some $\epsilon = \pm 1$.

(\ref{MaxSymCondSuSo}) are exactly the maximally symmetric gluing conditions of the bosonic WZW model based on $\hat{\mathfrak{g}}_{k} \oplus \hat{\mathfrak{so}}(d)_1$, which state space coincides with the sWZW model after GSO projection. Therefore we see that the supersymmetric maximally symmetric boundary states for the sWZW model are the maximally symmetric boundary states of the $\hat{\mathfrak{g}}_{k} \oplus \hat{\mathfrak{so}}(d)_1$ WZW model.

These states are tensor products of boundary states for $\hat{\mathfrak{g}}_{k}$ and $\hat{\mathfrak{so}}(d)_1$. For $\hat{\mathfrak{g}}_{k}$, the first set of gluing conditions in (\ref{MaxSymCondSuSo}) is relevant. There is one linearly independent solution (so-called Ishibashi state) $\ket{\lambda} \! \rangle$ in each $\hat{\mathfrak{g}}_k \oplus \hat{\mathfrak{g}}_k$-module $H_\lambda \otimes H_{\lambda^\ast} \subset \mathcal{H}_{\mathfrak{g}}$ with conjugated weights. It is convenient to rescale it so that we have :
$$
\langle \! \bra{\mu} q^{\frac{1}{2} \left ( L_0 + \bar{L}_0 -\frac{kd}{12(k+h^\vee)} \right )} \ket{\lambda} \! \rangle = \delta_{\mu\lambda} S_{0\lambda} \chi_\lambda(q) \;,
$$
where $q$ is a formal variable, $S$ the modular $S$ matrix of $\hat{\mathfrak{g}}_k$ and $\chi_\mu(q)$ is the specialized character of $H_\mu$. (Our notation does not distinguish the Virasoro zero modes of the sWZW model from the ones associated to $\hat{\mathfrak{g}}_k$ and $\hat{\mathfrak{so}}(d)_1$. It should be clear from the context which one should be used.) The elementary maximally symmetric boundary states are indexed by integrable highest $\hat{\mathfrak{g}}_k$-weights, and they are given by :
$$
\ket{B_{bos}, \mu} = \sum_{\mu \in P^+_k} \frac{S_{\mu\lambda}}{S_{0\lambda}} \ket{\lambda} \! \rangle \;.
$$

To construct boundary states for $\hat{\mathfrak{so}}(d)_1$, we have to distinguish several cases :
\begin{enumerate}
	\item {\bf ${\bm d}$ odd} : If $d$ is odd, we define $I = \{t,f,s\}$.
	\item $\bm{d = 0 \mod 4}$ : In this case the spinorial representations are self-conjugate. Therefore we see that it will be possible to solve the second gluing condition of (\ref{MaxSymCondSuSo}) in the Ramond-Ramond sector only when choosing the 0B GSO projection. So in this case we will consider for now only the 0B GSO projection, and set $I = \{t,f,s,s'\}$.
	\item $\bm{d = 2 \mod 4}$ : The spinorial representations are exchanged by charge conjugation, so the gluing condition can be solved in the Ramond-Ramond sector only when choosing the 0A GSO projection. So we consider the 0A GSO projection in this case and set $I = \{t,f,s,s'\}$.
	\item The cases $d = 0 \mod 4$ with 0A GSO projection, and $d = 2 \mod 4$ with 0B GSO projection will be treated further below.
\end{enumerate}	
We therefore have one Ishibashi state for each element of the set $I$ : $\ket{x} \! \rangle \in H_x \otimes H_x$, $\forall x \in I$. We normalize them as :
\begin{equation}
\label{NormIshSod}
 \langle \! \bra{x} q^{\frac{1}{2} \left ( L_0 + \bar{L}_0 -\frac{d}{24} \right )} \ket{y} \! \rangle = \delta_{xy} S^{\mathfrak{so}}_{tx} \chi_x(q) \;.
\end{equation}
where again, $S^{\mathfrak{so}}$ the modular $S$ matrix of $\hat{\mathfrak{so}}(d)_1$ and $\chi_x(q)$ is the specialized character of $H_x$, $x \in I$. From these Ishabashi states we can construct the following elementary boundary states \nolinebreak:
$$
\ket{B_{ferm}, x} = \sum_{y\in I} \frac{S^\mathfrak{so}_{xy}}{S^\mathfrak{so}_{ty}} \ket{y} \! \rangle \;.
$$

The maximally symmetric supersymmetric boundary states for the $\hat{\mathfrak{g}}_{\tilde{k}}$ are therefore given by :
$$
\ket{B,\mu,x} = \ket{B_{bos}, \mu} \otimes \ket{B_{ferm}, x} \;.
$$
It is obvious that these boundary states satisfy the Cardy condition, because the $\ket{B_\mu, \hat{\mathfrak{g}}_k}$ and $\ket{B_x, \hat{\mathfrak{so}}(d)_1}$ satisfy it separately.

It is instructive to compute explicitly the ratio $\frac{S^\mathfrak{so}_{xy}}{S^\mathfrak{so}_{ty}}$ appearing in the expression for the $\hat{\mathfrak{so}}(d)_1$ boundary states \nolinebreak:
$$
\begin{array}{ll}
\begin{array}{c|ccc}
y \backslash x & t & f & s  \\
\hline  t    & 1 & 1  & \sqrt{2}   \\
f     & 1 & 1 & -\sqrt{2}  \\
s     & 1 & -1 & 0 
\end{array} & \mbox{d odd} \vspace{.5cm}\\
\begin{array}{c|cccc}
y \backslash x & t & f & s & s'  \\
\hline  t    & 1 & 1  & 1  & 1  \\
f     & 1 & 1 & -1 & -1 \\
s & 1 & -1 & (-1)^{d/4} & (-1)^{d/4+1} \\
s'     & 1 & -1 & (-1)^{d/4+1} & (-1)^{d/4}
\end{array} & \mbox{d even}
\end{array} 
$$
We see that for odd $d$, $\ket{B,\mu,f}$ describes the anti-brane of $\ket{B,\mu,t}$ (the sign of the Ramond-Ramond component of the boundary state is reversed). $\ket{B,\mu,s}$ describes a brane which does not couple to the closed string Ramond-Ramond sector, therefore we do not expect it to carry any conserved charge. 

For even $d$, we see that for each $\mu \in P^+_k$, we have two boundary states $\ket{B,\mu,t}$ and $\ket{B,\mu,s}$, together with their respective anti-branes $\ket{B,\mu,f}$ and $\ket{B,\mu,s'}$. Note that $\ket{B,\mu,t}$ and $\ket{B,\mu,f}$ satisfy the gluing conditions (\ref{FermCond}) with $\epsilon = +1$, while $\ket{B,\mu,s}$ and $\ket{B,\mu,s'}$ satisfy the gluing conditions with $\epsilon = -1$. Therefore they preserve different supersymmetries. Interestingly, seeing the type 0 GSO projected free fermions as a $\hat{\mathfrak{so}}(d)_1$ WZW model provides straightforwardly the consistent fermionic boundary states. When building them directly from the free fermion theory, one obtains a bigger set of branes, which has to be reduced to the set found above by considering the consistency of the open string CFTs between these branes (see for instance \cite{Gaberdiel:2000jr}, section 2.3).

We still have to consider the cases when $d = 0 \mod 4$ with 0A GSO projection or $d = 2 \mod 4$ with 0B GSO projection. In these cases, one finds two $\hat{\mathfrak{so}}(d)_1$ Ishibasi states $\ket{t} \! \rangle$ and $\ket{a} \! \rangle$ in the NS sector, but there is no way to solve the gluing condition in the Ramond-Ramond sector. Still, the states $2\ket{t} \! \rangle \pm 2\ket{f} \! \rangle$ are admissible boundary states. Tensoring them with a maximally symmetric boundary states $\ket{B_\mu, \hat{\mathfrak{g}}_k}$ yields consistent boundary states for the sWZW model. Indeed, these states are nothing but the states $\ket{B,\mu,t}+\ket{B,\mu,f}$ and $\ket{B,\mu,s}+\ket{B,\mu,s'}$, which are (non-elementary) maximally symmetric boundary states for the sWZW model with opposite GSO projection. They lead to consistent open string partition function, because the two GSO projections coincide in the NS-NS sector. Just like for the third elementary boundary state in the odd $d$ case, we do not expect them to carry any charge, due to the fact that they do not couple to the Ramond-Ramond sector.

Finally, we remark that we have :
$$
\ket{B,\mu,x} = W_\mu^{\mathfrak{g}} W_x^\mathfrak{so} \ket{B,0,t}\;,
$$
where $W_\mu^{\mathfrak{g}}$ and $W_x^\mathfrak{so}$ are Wilson operators associated respectively to $\hat{\mathfrak{g}}_k$ and $\hat{\mathfrak{so}}(d)_1$, as defined in the previous section. $W_\mu^{\mathfrak{g}}$ is a normal ordered series in the current $\bJ$, which commutes with $\bJ$. As $\bJ$ commutes with $\psi$ and as the superconformal generators can be expressed in term of $\bJ$ and $\psi$, we deduce that $W_\mu^{\mathfrak{g}}$ commutes with the superconformal algebra. Similarly, $W_x^\mathfrak{so}$ is a normal ordered series in the current $J_{\mathfrak{so}}$ which commutes with $J_{\mathfrak{so}}$. Its eigenvalues can be found in the two tables above, and one checks that $W_x^\mathfrak{so}$ commutes with $\psi$ when $x = t,f$ (actually $W_t^\mathfrak{so}$ is always the identity operator), while it anticommutes with $\psi$ when $x = s,s'$. As it obviously commutes with $\bJ$, $W_x^\mathfrak{so}$ commutes or anticommutes with the superconformal generators, depending on $x$. We deduce from these consideration that given a supersymmetric boundary state $\ket{B}$, $W_\mu^{\mathfrak{g}}$ will map it onto another boundary state preserving the same supercharge, $W_a^\mathfrak{so}$ will map it onto its anti-brane, while $W_s^\mathfrak{so}$ and $W_{s'}^\mathfrak{so}$ will map it onto the corresponding brane and anti-brane preserving the opposite supercharge. (It will reverse the sign of $\epsilon$ in the gluing condition (\ref{GlobSupSymCond}).)

\subsection{Twisted D-branes}

In the bosonic case, the twisted D-branes satisfy the following gluing conditions on the Kac-Moody currents :
\begin{equation}
\label{TwistCond}
(J^a_n + \Omega(\bar{J}^a_{-n}))\ket{B} = 0 \;.
\end{equation}
where $\Omega$ is the outer automorphism of $\hat{\mathfrak{g}}_k$ induced by an outer automorphism of the corresponding finite Lie algebra $\mathfrak{g}$. (We will denote the latter by the same symbol $\Omega$.) We want to construct supersymmetric boundary states in the WZW model such that the condition \eqref{TwistCond} is satisfied by the full current $J$ generating $\gkt$.

The fermionic field $\psi(z)$ is $\mathfrak{g}$-valued, hence $\Omega$ acts naturally on it. This defines an automorphism of the chiral algebra $\hat{\mathfrak{c}}$. Using the fact that $\Omega$ is an orthogonal transformation preserving the Killing form on $\g$ and an automorphism of the Lie bracket, one can check that $\Omega$ leaves invariant both $L(z)$ and $G(z)$. The action of $\Omega$ on $\psi(z)$ extends to an action on the $\sod$ current $J_\mathfrak{so}$. This action is given by ad$_\Omega$, seeing $\Omega$ as an element of $O(d)$.

It follows from these considerations that we can readily apply the familiar construction of twisted boundary states \cite{Birke:1999ik, Gaberdiel:2003kv} to the bosonic WZW model based on $\gk \oplus \sod$, with automorphism given by $\Omega \times \mbox{ad}_\Omega$. The resulting boundary states are products of twisted boundary states for $\gk$ and for $\sod$, because the automorphism factorizes. 

The twisted boundary states for $\gk$ are indexed by the weights $\dot{\mu} \in P^+_{\Omega,k}$ of $\Omega$-twisted representations of $\gk$. 

The fermionic boundary states are indexed by ad$_\Omega$ twisted representations of $\sod$. We have to distinguish several cases again :
\begin{itemize}
	\item {\bf det$\,\bm{\Omega = 1}$} : When $\Omega$ belongs to $SO(d)$, the twist is inner in the fermionic sector. The twisted boundary state is given by the action of $\Omega$ on the maximally symmetric boundary states $\ket{B_{ferm}, x}$ constructed in the previous section. For even $d$, they have a non-trivial component in the Ramond-Ramond sector when we choose the 0B (0A) GSO projection for $d=0 \mod 4$ ($d=2 \mod 4$).
	\item {\bf det$\,\bm{\Omega = -1}$ and ${\bm d}$ odd} : $\Omega \notin SO(d)$, but it can be written as $-\Omega'$ where $\Omega' \in SO(d)$. Therefore the twist on $\sod$ is inner, given by ad$_{\Omega'}$. The twisted boundary state is given by $\Omega'\ket{B_{ferm}, x}$. Because of the relative factor of $-1$ between $\Omega$ and $\Omega'$, the resulting twisted boundary states $\ket{B_{ferm},\Omega, x}$ preserve the opposite supercharge (with $\epsilon = -1$ in \eqref{GlobSupSymCond}).
	\item {\bf det$\,\bm{\Omega = -1}$ and ${\bm d}$ even} : Then up to some inner automorphism, $\Omega$ is the outer automorphism exchanging the two fundamental weights associated with the spinorial representations. So these boundary states can have a non-trivial component in the Ramond-Ramond sector only when $d = 0 \mod 4$ in the 0A case, and when $d = 2 \mod 4$ in the 0B case (the opposite as for maximally symmetric boundary states). They are labeled by the integrable weights $\{t,f\}$ of the twisted Kac-Moody algebra $D^{(2)}_{d/2}$ at level one.
\end{itemize}

All in all, provided we choose the right GSO projection, we get the following set of twisted supersymmetric boundary states :
$$
\ket{B_\Omega, \dot{\mu}, x}\;, \qquad \dot{\mu} \in P^+_{\Omega,k} \;,
$$
where $x \in \{t,f,(s,s')\}$, $s'$ appearing only in the case where $\det \Omega = 1$ and $d$ is even, and $s$ being absent when $\det \Omega = -1$ and $d$ is even. $\ket{B_\Omega, \dot{\mu}, t}$ and $\ket{B_\Omega, \dot{\mu}, f}$ are each other's anti-brane.

It follows from the general theory of twisted boundary states \cite{Birke:1999ik, Gaberdiel:2003kv} that these states satisfy Cardy's condition. They are supersymmetric because they satisfy \eqref{TwistCond} as well as the corresponding gluing condition on the fermionic modes :
\begin{equation}
\label{TwistCondFerm}
(\psi^a_n + i\epsilon\Omega(\bar{\psi}^a_{-n})) \ket{B_\Omega, \dot{\mu}, x} = 0 \;.
\end{equation}
This implies the gluing condition :
$$
(G_r - i \epsilon \Omega(\bar{G}_r)) \ket{B_\Omega, \dot{\mu}, x} = 0 \;,
$$
which implies \eqref{GlobSupSymCond} because $G$ is invariant under $\Omega$.

Remark that just like for supersymmetric maximally symmetric boundary states, products of Wilson loop operators like $W_\mu^{\mathfrak{g}} W_x^\mathfrak{so}$ act on these states.

\subsection{Coset D-branes}

\label{ConstrCosDbranes}

Let $\hat{\mathfrak{a}}_{k'} \subset \gk$ be a Kac-Moody subalgebra generated by an embedding $\mathfrak{a} \subset \mathfrak{g}$ of reductive finite Lie algebras. In \cite{Maldacena:2001ky, Quella:2002ct, Quella:2002ns}, the authors considered the bosonic WZW model based on $\gk$ and constructed branes preserving only $\hat{\mathfrak{a}}_{k'} \oplus  \gk/\hat{\mathfrak{a}}_{k'} \subset \gk$, where $\gk/\hat{\mathfrak{a}}_{k'}$ denotes the coset vertex algebra, ie. all the normal ordered products of generators of $\gk$ commuting with the elements of $\hat{\mathfrak{a}}_{k'}$.

Again, we want to find the corresponding supersymmetric boundary states for the sWZW model. We will see that the construction of \cite{Quella:2002ct} can be used without any further modification, once the proper Kac-Moody subalgebra $\ak \subset \gk \oplus \sod$ has been found. Supersymmetry of the resulting boundary states will follow from a simple reasoning using Wilson operators.

So consider an embedding of a semi-simple finite Lie algebra $\mathfrak{a}$ in $\mathfrak{g}$, with embedding index $x$. It will be useful to choose the orthonormal basis $\{e^a\}$ of $\mathfrak{g}$ so that the first $d_\mathfrak{a}$ vectors generate $\mathfrak{a}$. Capital indices $A, B, C, ...$ will run from $1$ to $d_\mathfrak{a}$. 

Let us define the ``partial'' current \cite{1989NuPhB.321..232K} :
\begin{equation}
\label{PartCurrents}
\mathcal{J}_n^A = J^A_n + \frac{1}{2} f_{ABC} \sum_r \psi_r^B \psi_{n-r}^C \;,
\end{equation}
where as always the sum on $r$ is on half-integers in the Neveu-Schwarz sector and on integers in the Ramond sector. Note that the partial current differs from the restriction $\mathsf{J}|_\mathfrak{a}$ of the bosonic current $\mathsf{J}$ to $\mathfrak{a}$, because the sum on the Lie algebra indices $B$ and $C$ on the right hand size is restricted to $\mathfrak{a}$. It generates a Kac-Moody subalgebra $\hat{\mathfrak{a}}_{\kappa} \subset \gk \oplus \sod$, where $\kappa = x\tilde{k} - h^\vee_{\mathfrak{a}}$, $h^\vee_{\mathfrak{a}}$ being the dual Coxeter number of $\mathfrak{a}$. This current satisfies :
$$
[\mathcal{J}_n^A, \psi^B_r] = 0 \;,
$$
but in general it does not commute with $\psi|_{\mathfrak{a}^\perp}$, the component of the fermionic field associated with the orthogonal complement of $\mathfrak{a}$ in $\mathfrak{g}$. Define furthermore :
\begin{equation}
\label{PartStressSupCh}
\begin{split}
G^\mathfrak{a}_r =&\; -\frac{1}{\sqrt{\tilde{k}}} \left ( \sum_m \pJ^A_m \psi^A_{r-m} - \frac{1}{6}f_{ABC} \sum_{s,t} \psi^A_s \psi^B_t \psi^C_{r-s-t} \right ) \; ,\\
G^{\mathfrak{g}/\mathfrak{a}}_r =&\; G_r - (G_\mathfrak{a})_r \; ,\\
L^\mathfrak{a}_n = &\; \frac{1}{2\tilde{k}} \sum_m :\pJ^a_m \pJ^a_{n-m}: + \frac{1}{2} \sum_r r :\psi^A_{n-r} \psi^A_r: \left ( + \frac{d_{\mathfrak{a}}}{16}\delta_{n,0} \right ) \; ,\\
L^{\mathfrak{g}/\mathfrak{a}}_n =&\; L_n - (L_\mathfrak{a})_n \; ,
\end{split}
\end{equation}
where the term in parenthesis on the third line should be added in the Ramond sector only.

The crucial properties of $G^{\mathfrak{g}/\mathfrak{a}}$ and $L^{\mathfrak{g}/\mathfrak{a}}$ is that they commute with $\mathcal{J}$ and $\psi|_{\mathfrak{a}}$ \cite{1989NuPhB.321..232K}. Therefore they also commute with $G^\mathfrak{a}$ and $L^\mathfrak{a}$.

Let $P^+_{\mathfrak{a},\kappa}$ denotes the integrable highest weights for $\hat{\mathfrak{a}}_{\kappa}$, and $S^\mathfrak{a}_{\sigma\tau}$ be the modular $S$ matrix of $\hat{\mathfrak{a}}_{\kappa}$. Now consider a Wilson loop operator $W^\mathfrak{a}_\sigma$, $\sigma \in P^+_{\mathfrak{a},\kappa}$, built from the classical current corresponding to $\pJ$.  Then $W^\mathfrak{a}_\sigma$ commutes with the current $\pJ$, and it acts by scalar multiplication by $\frac{S^\mathfrak{a}_{\sigma\tau}}{S^\mathfrak{a}_{0\tau}}$ on any highest weight $\hat{\mathfrak{a}}_{\kappa}$-module appearing in $\mathcal{H}$. Moreover, it can be expressed as a normal-ordered series in $\pJ$, so it also commutes with $\psi|_{\mathfrak{a}}$, with $G^{\mathfrak{g}/\mathfrak{a}}$ and with $L^{\mathfrak{g}/\mathfrak{a}}$. Because $G^\mathfrak{a}$ and $L^\mathfrak{a}$ are expressed in term of $\pJ$ and $\psi|_{\mathfrak{a}}$, $W^\mathfrak{a}_\sigma$ commutes with them, so it commutes with $L$ and $G$. Therefore it maps a conformal supersymmetric boundary state onto a conformal supersymmetric boundary state.

Starting from one of the maximally symmetric boundary states constructed above, we get the following coset boundary states :
\begin{equation}
\label{CosetBoundState}
\ket{B_{\mbox{\tiny coset}},\mu, x, \sigma} = W^\mathfrak{a}_\sigma \ket{B,\mu,x} = W^\mathfrak{a}_\sigma W_\mu^{\mathfrak{g}} W_x^\mathfrak{so} \ket{B,0,t} \;.
\end{equation}

In particular, for $\sigma = 0$, $W^\mathfrak{a}_\sigma$ acts like the identity on the state space $\mathcal{H}$, and the boundary states $\ket{B_{\mbox{\tiny coset}},\mu, x, 0}$ are the maximally symmetric boundary states. Because they are built from maximally symmetric boundary states, in even dimension these states will have a non-zero component in the Ramond-Ramond sectors for $d = 0 \mod 4$ if we choose the 0B GSO projection, and for $d = 2 \mod 4$ if we choose the 0A GSO projection. The fermionic modes belonging to the preserved vertex algebra, in particular those associated with the Cartan subalgebra of $\g$, still satisfy \eqref{FermCond}. 

It is not yet obvious that the boundary state \eqref{CosetBoundState} really satisfies Cardy's consistency condition. However, one can check (see \cite{Alekseev:2007in}) that \eqref{CosetBoundState} is exactly the boundary state constructed in \cite{Quella:2002ct}, if we choose the bosonic WZW model based on $\gk \oplus \sod$, and if we choose the embedding of $\hat{\mathfrak{a}}_{\kappa}$ defined in \eqref{PartCurrents}. The computations in \cite{Quella:2002ct} establish that the open string stretching between such branes fall into representations of the preserved algebra $\hat{\mathfrak{a}}_{\kappa} \oplus (\gk \oplus \sod)/\hat{\mathfrak{a}}_{\kappa}$, and therefore that Cardy's condition holds.

Note that one can generalize the construction in the case when we have a sequence of embeddings $\mathfrak{a}_1 \subset ... \subset \mathfrak{a}_p \subset \g$ of semi-simple Lie subalgebras \cite{Quella:2002ns}. The corresponding boundary state can again be constructed from a single maximally symmetric boundary state, by the action of $p+2$ Wilson operators associated to the the $p$ subalgebras $\{\mathfrak{a}_i\}$, $\g$ and $\mathfrak{so}(d)$.

\subsection{Twisted coset D-branes}

\label{SSTwistCosBranes}

The coset D-branes can also be twisted by outer automorphisms either of $\g$ or of the subalgebra $\mathfrak{a}$ \cite{Maldacena:2001ky, Quella:2002ct, Quella:2002ns}. To construct their supersymmetric equivalent, one needs only to apply the results of \cite{Quella:2002ns}, provided the subalgebra is chosen carefully so that supersymmetry is preserved.

We keep the same notations as in the previous subsection. Suppose we are given an outer automorphism $\Omega^\mathfrak{a}$ of $\mathfrak{a}$ and an outer automorphism $\Omega$ of $\g$. They induce automorphisms of $\hat{\mathfrak{a}}_{\kappa}$ and $\gk$, respectively (that we denote by the same symbol). The action of the automorphism of $\hat{\mathfrak{a}}_{\kappa}$ can be extended to $\gk$, by choosing its action to be trivial on the complement of $\hat{\mathfrak{a}}_{\kappa}$. The resulting map is not an automorphism of $\gk$ : it preserves only $\ak \oplus \gk/\ak$. We also extend both automorphisms on the fermionic modes by the adjoint action.  

Just as for ordinary twisted boundary states, $\Omega^\mathfrak{a}\Omega$ is an orthogonal transformation with respect to the Killing form. Therefore it induces an automorphism on $\sod$ by the adjoint action.

We can therefore consider the vertex algebra $\frac{\gk \oplus \sod}{\ak} \oplus \ak$, and apply the construction of \cite{Quella:2002ns} to get twisted coset states :
$$
\ket{B_{\tilde{\Omega}}, \dot{\mu}, \dot{\sigma}, x}\;.
$$
Cardy's condition follows from the results in this paper. The whole analysis carried out for twisted boundary states can be repeated to determine which is the necessary GSO projection for these states to have a non-zero component in the Ramond-Ramond sector. The dependence on the determinant of $\Omega^\mathfrak{a}\Omega$ is exactly the same.

To see that these states are indeed supersymmetric, one just has to show that the operator $G$ is left invariant by $\tilde{\Omega}$. $G$ is invariant under $\Omega$ because it is an automorphism of $\gk$.
$G^\mathfrak{a}$ is invariant under $\Omega^\mathfrak{a}$ because it is an automorphism of $\ak$. To see that $G^{\mathfrak{g}/\mathfrak{a}}$ is also invariant one should use the fact that it commutes with both $\pJ$ and $\psi|_{\mathfrak{a}}$. Therefore it belongs to the coset $\hat{\mathfrak{c}} / (\ak \oplus \hat{\mathfrak{f}}_\mathfrak{a})$, where $\Omega^\mathfrak{a}$ acts trivially. ($\hat{\mathfrak{f}}_\mathfrak{a}$ is the vertex algebra generated by $\psi|_{\mathfrak{a}}$.)

As for the coset boundary state, any product of Wilson loop operators of the form $W^\mathfrak{a}_\sigma W_\mu^{\mathfrak{g}} W_x^\mathfrak{so}$ acts on the twisted coset state to yield another supersymmetric consistent boundary state.

All of the boundary states constructed in this section are encompassed in the family of twisted coset boundary states. Maximally symmetric and twisted boundary states are twisted coset states with $\mathfrak{a} = \mathfrak{g}$, and the twisted coset state with trivial automorphism $\Omega^\mathfrak{a} = \Omega = \mbox{id}$ are the coset (or maximally symmetric) boundary states.

As they stand, our type 0 D-branes are unstable : one easily check that the spectrum of open string on the brane contains a tachyon, while the spectrum of strings stretching between a brane and an antibrane is free of instabilities. A D-brane supporting a tachyon cannot carry any non-trivial flow invariant, as a perturbation by the tachyon field makes it decay entirely into closed string radiation (see for instance section 3.3 of \cite{Moore:2003vf}). However, generic arguments \cite{Lerche:1986cx, Billo:2000yb} show that the inclusion of the ghosts reverses this situation by exchanging the brane/brane and brane/antibrane spectrum, so that in this setting the type 0 D-branes are stable.

\section{Supersymmetric Kondo perturbations}

\label{SupSymKondPert}

We will now make more precise the link between Kondo perturbations in the open string picture and Wilson operators in the closed string picture, in the case of super Wess-Zumino-Witten models. This argument is a slightly generalized version of the one appearing in \cite{Moore:2003vf}, pp.30-31.

Consider a \,sWZW model on a cylindrical worldsheet $S^1 \times [0,1]$, with worldsheet time along the periodic direction, and let $C$ be the boundary $S^1\times \{0\}$. After having tensored the state space with $\mathbb{C}^n$ (which corresponds to stacking $n$ D-branes of the same type at one boundary), one can perturb the supercharge $G$ as follows \cite{Hikida:2001py} :
\begin{equation}
\label{PertSupCharge}
\Delta G = -l \sqrt{\tilde{k}} \int_C d\sigma \psi^a(\sigma)A^a\;,
\end{equation}
where the normalization of the coupling $l$ is chosen for later convenience. $\sigma$ is a parametrization of $C$ and $A^a$ a set of $n\times n$ matrices forming a representation of a subalgebra  $\mathfrak{a} \subset \mathfrak{g}$ with highest weight $\tau$, and acting on the $\mathbb{C}^n$ factor of the state space. The maximally symmetric case is included, when $\mathfrak{a} = \mathfrak{g}$. 

We choose the same convention on the Lie algebra indices as in the previous section. Capital indices $A,B,C,...$ run over $1,...,d(\mathfrak{a})$, and indices $a,b,c...$ still run over $1,...,d$. The matrices $A^{B}$ form a representation of $\mathfrak{a}$ while $A|_{\mathfrak{a}^\perp} = 0$.
 
The perturbation \eqref{PertSupCharge} induces a perturbation on $L_0$, by imposing $(G_0)^2 = L_0 - \frac{c}{24}$ in the Ramond sector and $\{G_{\frac{1}{2}},G_{-\frac{1}{2}}\} = L_0$ in the Neveu-Schwarz sector. It reads :
$$
\Delta L_0 = l \sum_n J_n^B A^B + l^2 \tilde{k} \sum_{r,s} \psi^B_r A^B\psi^C_s A^C \;.
$$
Note that up to a term of order $l^2$, the right hand side coincides with the bosonic Kondo perturbation (\ref{KondoPert}). Using the definitions \eqref{PartCurrents} of the partial current $\pJ$  and \eqref{PartStressSupCh} of partial Virasoro generator $L^\mathfrak{a}$, in the NS sector, we can rewrite the zero mode of the perturbed stress tensor \nolinebreak:
\begin{align*}
L_0 + \Delta L_0 \; =& \; L^{\mathfrak{a}}_0 \; + \; L^{\mathfrak{g}/\mathfrak{a}}_0 \; + \; l \sum_n J_n^B A^B \; + \; l^2 \tilde{k} \sum_{r,s} \psi^B_r A^B\psi^C_s A^C \\ 
=& \; \sum_n \left ( \frac{1}{2\tilde{k}} : \pJ^A_{-n} \pJ^A_n : + \; l \pJ_n^B A^B +  \frac{1}{2}l^2 \tilde{k} A^B A^B \right ) + \; \frac{1}{2} \sum_r \left ( r :\psi^A_{-r} \psi^A_r: + \right. \\
& \; \left. + l(l\tilde{k}-1)f_{ABC} \sum_{s} \psi^A_r \psi^B_s A^C \right ) +  L^{\mathfrak{g}/\mathfrak{a}}_0 \\
=& \; \frac{1}{2\tilde{k}} \sum_n : (\pJ^B_{-n} + l\tilde{k}A^B) (\pJ^B_n + l\tilde{k}A^B) : \; + \; \frac{1}{2} \, \sum_r r :\psi^A_{-r} \psi^A_r: + \\
& \; + \frac{1}{2}l(l\tilde{k}-1)f_{ABC} \sum_{r,s} \psi^A_r \psi^B_s A^C +  L^{\mathfrak{g}/\mathfrak{a}}_0 \;.
\end{align*}
We wrote $L^{\mathfrak{a}}_0$ explicitly, expressed $J$ in term of $\pJ$, developed the last term and rearranged the terms. In the R sector the same computation holds verbatim, except for proper inclusions of the central terms $\frac{d}{16}$ and $\frac{d_{\mathfrak{a}}}{16}$.

It is not known how to treat this perturbation for generic $l$ beyond perturbation theory. But at the special coupling value $l = \tilde{k}^{-1}$, the operators $\pJ^B_n + l\tilde{k}A^B$ satisfy the commutation relations of the Kac-Moody algebra $\ak$. The perturbed Hamiltonian is therefore quadratic in the fields of $\ak \oplus \hat{\mathfrak{f}}_\mathfrak{a} \oplus \hat{\mathfrak{c}}/(\ak \oplus \hat{\mathfrak{f}}_\mathfrak{a})$. We expect the states of the perturbed theory to fall into modules for the bosonic part of this chiral algebra $\ak \oplus (\gk \oplus \sod)/\ak$. We will see now how the Hamiltonian can be diagonalized.

Consider a highest weight module $H_\mu \otimes H^{\mathfrak{so}}_x$ for $\gk \oplus \sod$, where $\mu$ and $x$ are integrable highest weights for $\gk$ and $\sod$, respectively. This module breaks into submodules for $\ak \oplus (\gk \oplus \sod)/\ak$ :
$$
H_\mu \otimes H^{\mathfrak{so}}_x = \bigoplus_{\sigma} \bigoplus_{[\mu,x,\sigma]} 
H^{\mathfrak{a}}_\sigma \otimes H^{cs}_{[\mu,x,\sigma]} \;,
$$
where $\sigma \in P^+_{\mathfrak{a},\kappa}$ is an integrable weight of $\ak$, $[\mu,x,\sigma]$ labels coset primary fields and $H^{cs}_{[\mu,x,\sigma]}$ is the corresponding module for the coset algebra $(\gk \oplus \sod)/\ak$. 

The fields in the coset $(\gk \oplus \sod)/\ak$ are left invariant under the perturbation \eqref{PertSupCharge}, but we saw that the components of the current generating $\ak$ are modified as $\pJ^B_n \mapsto \tilde{\pJ}^B_{-n} = \pJ^B_{-n} + A^B$. If we manage to decompose $H^{\mathfrak{a}}_\sigma \otimes \mathbb{C}^n$ into highest weight modules for the modified current, we will have diagonalized the perturbed Hamiltonian. It will then be easy to identify the perturbed theory from its spectrum.

$H^{\mathfrak{a}}_\sigma \otimes \mathbb{C}^n$ does not contain any highest weight vector for the perturbed currents. However, one can look for highest weight states in a completion $C_{\sigma,n}$ of $H^{\mathfrak{a}}_\sigma \otimes \mathbb{C}^n$, namely the full linear span of $H^{\mathfrak{a}}_\sigma \otimes \mathbb{C}^n$ (see \cite{Alekseev:2007in} for a more detailed version of the following argument). The vertex operators of the WZW model \cite{Tsuchiya:1988fy, Etingof:1998ru} intertwine the action of $\pJ$ and $\tilde{\pJ}$ in $C_{\sigma,n}$. The number of linearly independent vertex operators intertwining the action of $\tilde{\pJ}$ on $H^{\mathfrak{a}}_\sigma \otimes \mathbb{C}^n$ and the action of $\pJ$ on $H^{\mathfrak{a}}_\upsilon \subset C_{\sigma,n}$ is exactly given by the fusion rules $\mathcal{N}_{\sigma\tau}^{\mathfrak{a}\;\upsilon}$ of $\ak$. We therefore have an embedding :
\begin{equation}
\label{SubSpDiag}
\bigoplus_{\upsilon \in P^+_{\mathfrak{a},\kappa}} \mathcal{N}_{\sigma\tau}^{\mathfrak{a}\;\upsilon}H^{\mathfrak{a}}_\upsilon \subset C_{\sigma,n} \;,
\end{equation}
where we use a multiplicative notation to denote the direct sum of several copies of the same module. Note that the vectors of $H^{\mathfrak{a}}_\upsilon \subset C_{\sigma,n}$ are not normalizable with respect to the original norm on $H^{\mathfrak{a}}_\sigma \otimes \mathbb{C}^n$, but that one has naturally a ``renormalized'' norm on the subspace \eqref{SubSpDiag} of $C_{\sigma,n}$ induced by the inclusion.

Suppose for definiteness that the original open string theory described strings stretching between maximally symmetric D-branes of label $(\mu,x)$ and $(\lambda,y)$. The state space of the original theory is given by :
$$
\bigoplus_{\nu \in P^+_{k}} \mathcal{N}_{\mu\lambda}^{\;\;\nu} H_\nu \otimes \mathcal{N}_{xy}^{\;\;z} H^{\mathfrak{so}}_z \;,
$$
$\mathcal{N}_{\mu\lambda}^{\;\;\nu}$ and $\mathcal{N}_{xy}^{\;\;z}$ being the fusion rules of $\gk$ and $\sod$, respectively. The discussion above allows us to identify the state space of the perturbed theory as :
$$
\bigoplus_{\nu \in P^+_{k}} \bigoplus_{\sigma,\upsilon} \bigoplus_{[\nu,z,\sigma]}  \mathcal{N}_{\mu\lambda}^{\;\;\nu} \mathcal{N}_{\sigma\tau}^{\mathfrak{a}\;\upsilon}H^{\mathfrak{a}}_\upsilon \otimes \mathcal{N}_{xy}^{\;\;z} H^{cs}_{[\nu,z,\sigma]} \;,
$$
where the sum over $\sigma$ and $\upsilon$ runs over integrable $\ak$-weights and the sum on $[\nu,z,\sigma]$ runs over coset primary fields. The corresponding partition function reads :
$$
Z_{pert}(q) = \sum_{\nu \in P^+_{k}} \sum_{\sigma,\upsilon} \sum_{[\nu,z,\sigma]}  \mathcal{N}_{\mu\lambda}^{\;\;\nu} \mathcal{N}_{\sigma\tau}^{\mathfrak{a}\;\upsilon}\mathcal{N}_{xy}^{\;\;z} \chi^{\mathfrak{a}}_\upsilon(q) \chi^{cs}_{[\nu,z,\sigma]}(q) \;,
$$
where $\chi^{\mathfrak{a}}_\upsilon$ and $\chi^{cs}_{[\mu,z,\sigma]}$ are the specialized characters of $\ak$ and coset modules. Comparing with \cite{Quella:2002ct}, we see that this partition function describe open strings stretching between the D-branes corresponding to $\ket{B,\mu,x}$ and $\ket{B_{\mbox{\tiny coset}},\lambda, y, \tau} = W^\mathfrak{a}_\tau \ket{B,\lambda,y}$.

Therefore we see that there is a fixed point of the flow induced by the perturbation \eqref{PertSupCharge}. In the closed string sector, this fixed point is obtained from the unperturbed CFT by the insertion of the Wilson operator $W^\mathfrak{a}_\tau$ into the cylinder amplitude. We see in particular that the IR fixed point theory is superconformal, due to the fact that $W^\mathfrak{a}_\tau$ commutes with the generators of the superconformal algebra (see the discussion in section \ref{ConstrCosDbranes}). Note that in the maximally symmetric case : $\mathfrak{a} = \g$, the corresponding Wilson operator is $W^\mathfrak{g}_\tau$, which is constructed using the bosonic current $\bJ$. 

Hence the generalized Kondo flows take the form $d_\tau\ket{B} \mapsto W^\mathfrak{a}_\tau\ket{B}$. This result will allow us to check explicitly that the charges that we will assign to supersymmetric boundary states are invariants of these flows.

\section{Test states : the cohomology of the supercharge}

\label{TS-CSupC}

As explained in section \ref{SecOverview}, our aim is to measure the charges of the D-branes by formally computing their coupling to massless Ramond-Ramond states. We also remarked that such states do not exist in the state space of the sWZW model, so we will have to look for them in a generic highest weight module for the chiral algebra $\hat{\mathfrak{c}}$. Of course, a non-trivial part of our procedure for measuring the charges will be to complete the boundary states in the virtual sector containing the massless Ramond-Ramond states in a consistent way, in order to be able to compute the charges as overlaps, but this will be not be undertaken before the next section.
 
\subsection{Test states}

Let us write $D_\pm := \pm G_0 + i\bar{G}_0$. We have $D_+^\dagger = D_-$. For definiteness, let us decide that we are interested in measuring the charges of supersymmetric boundary states satisfying $D_-\ket{B} = 0$. The treatment of the boundary states supersymmetric with respect to the other supercharge is completely similar.

The first task is to identify a minimal set of test states. We will restrict to test states which are supersymmetric with respect to $D_+$, hence they should satisfy :
\begin{equation}
\label{SupSymCondRR}
D_+\ket{RR} = 0 \;.
\end{equation}
Non-supersymmetric test states do not seem to extract invariants of boundary RG flows from the boundary states. 

But we can restrict this set further : any test state of the form $\ket{RR} = D_+\ket{RR'}$ would have zero scalar product with a supersymmetric boundary state $\ket{B}$ satisfying $D_- \ket{B} = 0$. Hence our test states are in bijection with the elements of $\ker D_+$ modulo $\mbox{Im}\, D_+$, that is with the elements of the cohomology of $D_+$. Note that $(D_+)^2 = (G_0)^2 - (\bar{G}_0)^2 = L_0 - \bar{L}_0$ always vanishes on states satisfying the level matching condition, so $\mbox{Im}\, D_+ \subset \ker D_+$.

Let us examine now the condition that the Ramond-Ramond states should be massless. They are linear combinations of states of the form $\ket{RRb} \otimes \ket{RRf}$ where $\ket{RRb} \in \mathcal{H}^{\mathfrak{g}}$ and $\ket{RRf}$ lies in the Ramond-Ramond part of the $\sod$ state space $\mathcal{H}_X^{\mathfrak{so}}$. We can readily see that there is no massless RR field in the physical state space of the model. Indeed, their masses are given by :
\begin{equation}
\label{ImpossRRSupSym}
 \left ( L_0 + \bar{L}_0 - \frac{c}{12} \right ) \ket{RR} 
=
\left ( h_{\ket{RRb}} + h_{\ket{RRf}} - \frac{c}{12}\right ) \ket{RR} \;,
\end{equation}
where $h$ denotes the conformal dimension of the states. The nonexistence of massless Ramond-Ramond fields follows from the inequalities $h_{\ket{RRb}} \geq 0$, $h_{\ket{RRf}} \geq \frac{d}{8}$ and $c =  \frac{3k+h^\vee}{2(k+h^\vee)}d < \frac{3}{2}d$ \cite{Fuchs:1988gm}. 

If we want to pursue this approach, we have to look for states outside the physical state space of the model. So let us study the cohomology of $D_+$ on any highest weight module of the form :
\begin{equation}
\label{VirtStateSpace}
V_\lambda = H^\mathfrak{g}_\lambda \otimes \bar{H}^\mathfrak{g}_{\lambda^\ast} \otimes F^{2d}_{R}\;,
\end{equation}
for $\lambda$ an arbitrary weight of $\hat{\mathfrak{g}}$ at level $k$. $H^\mathfrak{g}_\lambda$ is the unique irreducible highest weight module with highest weight $\lambda$, and $F^{2d}_{R}$ is the Fock module for $2d$ real fermions in the Ramond sector. $\bar{H}^\mathfrak{g}_{\lambda^\ast}$ is the conjugated module to $H^\mathfrak{g}_\lambda$. It is generated by the action of $\gk$ on a state $\ket{\lambda^\ast}$ of weight $-\lambda$, annihilated by the positive modes of the current $\bar{J}$, as well as the zero modes of the generators associated to \emph{negative} roots. The component $(\bar{H}^\mathfrak{g}_{\lambda^\ast})_n$ of $\bar{H}^\mathfrak{g}_{\lambda^\ast}$ at a fixed grade $n \leq 0$ is therefore a lowest weight module for the horizontal Lie algebra $\g$\footnote{Integrable modules decompose into finite dimensional $\g$-modules at each grade, which makes this distinction between highest weight and lowest weight modules irrelevant. The situation is different when considering non-integrable modules, which generically decompose into infinite dimensional Verma $\g$-modules. This is the reason why we now have to distinguish in notation $\bar{H}^\mathfrak{g}_{\lambda}$ from $H^\mathfrak{g}_{\lambda}$.}.

Using \eqref{ImpossRRSupSym}, it is possible to guess for which choice of $\lambda$ we may obtain massless Ramond-Ramond states. Restricting our test states to be states of grade zero in $V_\lambda$, \eqref{ImpossRRSupSym} yields the equation :
\begin{equation}
\label{ExistMassLessRRStV-Rho}
0 = h_{\ket{RRb}} + h_{\ket{RRf}} - \frac{c}{12} = \frac{(\lambda,\lambda + 2\rho)}{k+h^\vee} + \frac{d}{8} - \frac{3k+h^\vee}{24(k+h^\vee)}d
\end{equation}
$$
\Rightarrow (\lambda,\lambda + 2\rho) + \frac{h^\vee d}{12} = 0 \;.
$$
By the Freudenthal-de Vries strange formula, this equality is satisfied when $\lambda = -\rho$, where $\rho$ denotes the Weyl vector of $\mathfrak{g}$ (half the sum of the positive roots). Therefore Ramond-Ramond ground states are massless on the  grade zero subspace of $V_{-\rho}$. Note that the module $V_{-\rho}$ is the only one among all $V_\lambda$ satisfying this condition, because $-\rho$ is the global minimum of $(\lambda,\lambda + 2\rho)$. We will now compute the cohomology of $D_+$ on this subspace.

\subsection{The cohomology of the supercharge}

\label{SectCohomSupCharge}

As we saw above, in $V_{-\rho}$, the kernel of $D_+$ is contained in the grade zero subspace $(V_{-\rho})_0$ (with respect to $L_0$). On this subspace, $G_0$ takes the simpler form :
$$
G_0|_{(V_{-\rho})_0} =  -\frac{1}{\sqrt{\tilde{k}}} \left ( \bJ^a_0 \psi^a_0 - \frac{1}{6}f_{abc}  \psi^a_0 \psi^b_0 \psi^c_0 \right ) \;.
$$
This operator appeared in the mathematical literature in \cite{Alekseev:math9903052} as a differential on the non-commutative Weil algebra, and in a more general form in \cite{MR1985723} as the ``cubic Dirac operator''. Note also that the deformation of $G_0$ induced by \eqref{PertSupCharge} was studied in \cite{freed-2005}, in the context of the computation of the twisted equivariant K-theory of compact Lie groups.

To compute the cohomology of $D_+$ on $(V_{-\rho})_0$, we will use a homotopy operator. The goal is to find $H$ such that $\{D_+,H\} = P$, where $P$ is an operator which commutes with $D_+$, and which is invertible on a subspace $V_P$ complementary to $\ker P$. The existence of such an operator implies that the cohomology of $D_+$ is trivial on $V_P$. Indeed, suppose that we have some state $\ket{RR} \in V_P$ such that $D_+ \ket{RR} = 0$. Then :
$$
\ket{RR} = P P^{-1} \ket{RR} = (D_+ H + HD_+ ) P^{-1} \ket{RR} = D_+ ( H P^{-1} \ket{RR}) \;,
$$
so that $\ket{RR}$ is cohomologically trivial. Moreover, as $D_+$ preserves $\ker P$, the cohomology of $D_+$ on $(V_{-\rho})_0$ is isomorphic to its cohomology on $\ker P$.

In our case, a homotopy operator is provided by (see \eqref{DefFermPol}) $H = \sqrt{\tilde{k}}\psi^\rho_{0+} := \sqrt{\tilde{k}} \rho^i \psi^i_{0+}$, where $i = 1,...,\mbox{rank }\mathfrak{g}$ runs over an orthonormal basis of the Cartan subalgebra $\mathfrak{h}$ of $\mathfrak{g}$, and $\rho$ is the Weyl vector. Then $P$ is given by $P = -J_0^\rho + \bar{J}_0^\rho = -\rho^i (J_0^i - \bar{J}_0^i)$, and one checks that $[D_+,P] = 0$ indeed. We should now identify $\ker P$. To this end, we rewrite the currents $J_0^\rho$ and $\bar{J}_0^\rho$ as :
$$
J_0^\rho = \bJ_0^\rho -\frac{1}{2}\sum_i \rho^i \sum_{\alpha \in \Delta} (\alpha^\vee)^i \psi_0^{-\alpha}\psi_0^\alpha = \bJ_0^\rho + (\rho,\rho) - \sum_{\alpha \in \Delta_+} \psi^{-\alpha}_0 \psi^\alpha_0 \;,
$$
$$
\bar{J}_0^\rho = \bar{\bJ}_0^\rho -\frac{1}{2}\sum_i \rho^i \sum_{\alpha \in \Delta} (\alpha^\vee)^i \bar{\psi}_0^{-\alpha}\bar{\psi}_0^\alpha = \bar{\bJ}_0^\rho - (\rho,\rho) + \sum_{\alpha \in \Delta_+} \bar{\psi}^\alpha_0 \bar{\psi}^{-\alpha}_0 \;,
$$
where $\Delta$ and $\Delta_+$ are the set of all roots and positive roots, respectively. We used here the non-orthonormal Cartan-Weyl basis induced from the root space decomposition of $\mathfrak{g}$, and wrote the structure constants of the Lie bracket explicitly in this basis. $\psi_0^\alpha$ is the zero mode of the fermion associated with $e^\alpha$, the generator of the root space $\mathfrak{g}_\alpha$. We can therefore rewrite :
$$
P = - \bJ_0^\rho + \bar{\bJ}_0^\rho - 2(\rho,\rho) + \sum_{\alpha \in \Delta_+} (\psi^{-\alpha}_0 \psi^\alpha_0 + \bar{\psi}^\alpha_0 \bar{\psi}^{-\alpha}_0) \;.
$$
Note that the last term is just the fermion number operator for the zero modes of the fermions associated with the roots of $\g$. 

It is now easy to read off $\ker P \subset (V_{-\rho})_0$ : it is given by the vectors with weight $-\rho$ ($\rho$) with respect to the holomorphic (antiholomorphic) bosonic currents and annihilated by the holomorphic (antiholomorphic) fermions associated with positive (negative) roots. Let us denote highest weight of state $H^{\mathfrak{g}}_{-\rho} \otimes \bar{H}^{\mathfrak{g}}_{-\rho}$ 
by $\ket{-\rho}$. Using the notation \eqref{BaseCliff}, we have :
\begin{equation}
\label{kerP}
\ker P = \left \{ \ket{-\rho} \otimes \sigma \ket{e \wedge  e_{\mathfrak{n}_+}} | e \in \bigwedge \mathfrak{h} \right \}
\end{equation}
where $e_{\mathfrak{n}_+}$ denotes the volume form on $\mathfrak{n}_+$, the subalgebra formed by the positive root spaces. $\sigma$ is the automorphism of $F^{2d}_R$ induced by $\bar{\psi}^\alpha_n \leftrightarrow \bar{\psi}^{-\alpha}_n$. The action of $\sigma$ ensures that the states above are annihilated by $\psi^\alpha_0$ and $\bar{\psi}^{-\alpha}_0$, rather than $\psi^\alpha_0$ and $\bar{\psi}^{\alpha}_0$. So we know that any state having no component on $\ker P$ is cohomologically trivial, and we reduced the problem to the study of the cohomology of $D_+$ on $\ker P$.

To proceed, one can remark that on $\ker P$, $G_0$ takes the form :
$$
G_0|_{\ker P} = -\frac{1}{\sqrt{\tilde{k}}} \left. \left ( \bJ^a_0 \psi^a_0 - \frac{1}{6}f_{abc}  \psi^a_0 \psi^b_0 \psi^c_0 \right ) \right |_{\ker P} = -\frac{1}{\sqrt{\tilde{k}}} \left ( -\rho^i \psi^i_0 + \frac{1}{2} \sum_{\alpha \in \Delta_+} \alpha^i \psi^i_0 \right ) = 0 \;,
$$
where we used the fact that $\bJ^\alpha_0$ and $\psi^\alpha_0$ vanish on $\ker P$, and made explicit the structure constants in the Cartan-Weyl basis. Therefore :
$$
D_+ |_{\ker P} = 0 \;.
$$
Finally, as $D_+$ commutes with $P$, we see that it preserves $\ker P$, so it is impossible to have $\ket{RR} = D_+ \ket{RR'}$ if $\ket{RR} \in \ker P$. Hence the cohomology is isomorphic to $\ker P$, and has dimension $2^r$, where $r$ denotes the rank of $\mathfrak{g}$. We have a representative for each class, given by \eqref{kerP} above.

Note that the restriction to massless Ramond-Ramond ground states which lead us to consider only $V_{-\rho}$ among all $V_{\lambda}$'s is not necessary. Indeed, using the homotopy operator $H' = \frac{1}{2}D_-$, we get $P' = L_0 + \bar{L}_0 - \frac{c}{12}$. As we saw in \eqref{ImpossRRSupSym}, $P'$ exactly computes the mass of the Ramond-Ramond states, so the fact that the cohomology can be non-trivial only in the kernel of $P'$ coincides with the condition for massless states. This shows that the cohomology is trivial in $V_{\lambda}$ for $\lambda \neq -\rho$.

\vspace{.5cm}

To compute the cohomology of $D_-$, which is used to probe the charges of the branes preserving the opposite supersymmetry, we can use the homotopy operator $H_- = \sqrt{\tilde{k}} \rho^i \psi^i_{0-}$. Then $P$ is identical, and we get in $\ker P$ a representative for each of the cohomology classes of $D_-$.

Let us adopt the following notation for our test states :
$$
\ket{RR, e} = \ket{-\rho} \otimes \sigma\ket{e \wedge  e_{\mathfrak{n}_+}} \;,
$$
with $e \in \bigwedge \mathfrak{h}$.

We take the opportunity of introducing here a grading that will be of much use in the remaining of this paper. $\mbox{ad}_P$ defines a grading $\mbox{gr}_P$ of $\hat{\mathfrak{c}} \oplus \hat{\mathfrak{c}}$ : $\mbox{gr}_P (J^\alpha_n) = (\rho,\alpha)$ and $\mbox{gr}_P (\psi^\alpha_n) = \pm \frac{2}{|\alpha|^2}$ depending if $\alpha$ is a positive or negative root. The anti-holomorphic partners have opposite grade. We define the \emph{total} grading $\mbox{gr}_T = h^\vee\mbox{gr}_{L_0} + \mbox{gr}_P$, where $\mbox{gr}_{L_0}$ is the grading induced by the zero mode $L_0$ of the stress tensor. As the scalar product of the highest root $\theta$ with the Weyl vector $\rho$ is given by $h^\vee-1$, we see that the only operators with zero total grade are the components $J_0|_{\mathfrak{h}}$ of the current and $\psi_0|_{\mathfrak{h}}$ along the Cartan subalgebra $\mathfrak{h}$ of $\g$. This grading induces a non-negative grading of $V_{-\rho}$, and $\ker P$ is the grade zero subspace.

\subsection{The generic test states}

\label{GenTestStates}

The test states that we found are not the most general ones, however. This comes from the fact that the $\gk$-module $H^\mathfrak{g}_{-\rho}$ does not carry a representation of the Lie group $G$. It is a member of a continuous family of non-isomorphic modules. 

As a $\g$-module, $H^\mathfrak{g}_{-\rho}$ is a direct sum of (infinite dimensional) highest weight modules with respect to the principal graduation defined by $\mbox{ad}\bJ^{\rho}_0$. That is, modules generated by a highest weight vector annihilated by all of the operators having a positive eigenvalue under the adjoint action of $\bJ^{\rho}_0$. Recalling that the Lie group $G$ acts by the adjoint action on the Kac-Moody algebra, we can choose another graduation $\mbox{ad}(g\bJ^{\rho}_0g^{-1})$, $g \in G$. From this graduation, we can construct a $\gk$-module $gH^\mathfrak{g}_{-\rho}$. $gH^\mathfrak{g}_{-\rho}$ is generated by $\gk$ from a highest weight state $\ket{g,-\rho}$ such that $g\bJ^{\alpha}_0g^{-1}\ket{g,-\rho} = 0$ for every $\alpha \in \Delta_+$, and $g\bJ^i_0g^{-1}\ket{g,-\rho} = -\rho^i \ket{g,-\rho}$ for every $\bJ^i_0$ in the Cartan subalgebra of the horizontal algebra $\g$. $\ket{g,-\rho}$ is also annihilated by the positive modes $\bJ^a_n$, $n > 0$. The modules $gH^\mathfrak{g}_{-\rho}$ form a continuous family of non-isomorphic modules. Put differently, the definition of a highest weight module requires the choice of a triangular decomposition of $\mathfrak{g} = \mathfrak{n}_- \oplus \mathfrak{h} \oplus \mathfrak{n}_+$, and the generators of the subalgebra $\mathfrak{n}_+$ are defined to annihilate the highest weight vector. Considering distinct triangular decompositions of $\mathfrak{g}$ yields non-isomorphic highest weight modules (except in the case of finite dimensional $\mathfrak{g}$-modules). Finally, note that the action of $\gk$ on $gH^\mathfrak{g}_{-\rho}$ is isomorphic to its action on $H^\mathfrak{g}_{-\rho}$ twisted by the inner automorphism $\mbox{ad}_{g^{-1}}$.

Similarly, one can construct a family $g\bar{H}^\mathfrak{g}_{-\rho}$ of highest weight module for the antiholomorphic copy of the Kac-Moody algebra. As a $\g$-module, $g\bar{H}^\mathfrak{g}_{-\rho}$ is a direct sum of highest weight $\g$-module with respect to the graduation provided by $-\mbox{ad}(g\bar{\bJ}^\rho_0g^{-1})$.

Define the $\hat{\mathfrak{c}} \oplus \hat{\mathfrak{c}}$-module :
$$
(g,g')V_{-\rho} =  gH^\mathfrak{g}_{-\rho} \otimes g'\bar{H}^\mathfrak{g}_{-\rho} \otimes F^{2d}_{R} \;.
$$
As mentioned above, the action of $\gk$ on $gH^\mathfrak{g}_{-\rho}$ is isomorphic to its $\mbox{ad}_{g^{-1}}$-twisted action on $H^\mathfrak{g}_{-\rho}$. As the superconformal generators are invariant under the adjoint action of the group $G$, $(g,g')V_{-\rho}$ also contains massless Ramond-Ramond states in its grade zero component.

The computation of the cohomology of $D_+$ on $(g,g')V_{-\rho}$ can be performed similarly as above. All the holomorphic operators should be conjugated by $g$, while antiholomorphic ones should be conjugated by $g'$. The cohomology is therefore supported on the kernel of the operator :
$$
P_{(g,g')} = - g J_0^\rho g^{-1} + g'\bar{J}_0^\rho g'^{-1} \;.
$$

\vspace{.5cm}

Note that the centralizer of $\bJ^{\rho}_0$ in $G$ is the Cartan torus $H$, so the set of modules $(g,g')V_{-\rho}$, $g,g' \in G$ can be tied together into a bundle $GV_{-\rho}$ over $G/H \times G/H$, if we define the fiber above the element $(gH, g'H)$ to be $(g,g')V_{-\rho}$. To be more precise, we take the trivial bundle $G \times G \times V_{-\rho}$, with the action of $\mathfrak{\hat{c}} \oplus \mathfrak{\hat{c}}$ on the fiber above $(g,g')$ to be the one realized in $(g,g')V_{-\rho}$, and we quotient it by the right actions of $H$ on each $G$ factors to get $GV_{-\rho}$.

In order to keep the notation as simple as possible, we will not be very careful in distinguishing the cosets $gH$ from representatives $g$. This is justified by the fact that all of the modules $(g,g')V_{-\rho}$ for $(g,g') \in g_0H \times g'_0H$ are isomorphic.


Finally, there is a natural map $\phi_g : gH_{-\rho} \rightarrow H_{-\rho}$. Denote by $\ket{g, -\rho}$ the highest weight vector of $gH_{-\rho}$. Then we define $\phi_g$ so that it sends $\ket{g, -\rho}$ on $\ket{1,-\rho} = \ket{-\rho}$, and $X\ket{g, -\rho}$ on $g^{-1}Xg \ket{-\rho}$, where $X$ is any product of the operators $\{\bJ^a_n\}$. This map intertwines the action of $\hat{\mathfrak{g}}$ on $gH_{-\rho}$ and its twisted action on $H_{-\rho}$ (the twist being the inner automorphism ad$_{g^{-1}}$). There is a similar intertwining map $\bar{\phi}_g$ between the antiholomorphic modules. On the fermionic side, the transformations $\psi^a_n \mapsto g^{-1}\psi^a_n g$ and $\bar{\psi}^a_n \mapsto g'^{-1}\bar{\psi}^a_n g'$ induces an automorphism $\phi^F_{(g,g')}$ of $F^{2d}_{R}$. More precisely, $\phi^F_{(g,g')}$ is defined by :
\begin{equation}
\label{DefPhiF}
\phi^F_{(g,g')} = g_\psi \circ g'_{\bar{\psi}} \;,
\end{equation}
where $g_\psi$ and $g'_{\bar{\psi}}$ denote the images of the elements $g,g' \in G$ by the holomorphic and antiholomorphic representations of $G$ on $F^{2d}_{R}$. (Infinitesimally, these two representations are generated by the zero modes of the fermionic currents \eqref{FermCurrent} and of their antiholomorphic counterpart.)
We can therefore define the following intertwining map :
\begin{equation}
\label{DefPhi}
\phi_{(g,g')} = \phi_g \otimes \bar{\phi}_{g'} \otimes \phi^F_{(g,g')} \;,
\end{equation}
which identifies $(g,g')V_{-\rho}$ and $V_{-\rho}$ as vector spaces.

\subsection{Gluing conditions and shifts on the group}

\label{SubSGlueCondShiGr}


The problem we will be interested in in the next section will be to solve in $GV_{-\rho}$ the gluing conditions satisfied by the boundary states. Such gluing conditions take the form of a set of equations of the type :
\begin{equation}
\label{GlueCondSample}
(X + \bar{X}')\ket{B} = 0 \;,
\end{equation}
where $X$ and $\bar{X}'$ are operators belonging respectively to the holomorphic and antiholomorphic copy of the chiral algebra $\hat{\mathfrak{c}}$. Suppose we are trying to solve \eqref{GlueCondSample} in the fiber $(g,g')V_{-\rho}$. The action of $\hat{\mathfrak{g}}$ in $(g,g')V_{-\rho}$ is only the action in $V_{-\rho}$ twisted by the inner automorphisms ad$_{g^{-1}}$ and ad$_{g'^{-1}}$ on the holomorphic and antiholomorphic side, respectively. Hence we readily see that solving \eqref{GlueCondSample} in $(g,g')V_{-\rho}$ is equivalent to solving the following gluing conditions in $V_{-\rho}$ :
\begin{equation}
\label{GlueCondSampleShift}
(g^{-1}Xg + (g')^{-1}\bar{X}'g')\ket{B} = 0 \;.
\end{equation}
More precisely, if we denote by $\ket{B_{(g,g')}}$ the solution of \eqref{GlueCondSample} in $(g,g')V_{-\rho}$, and $\ket{B'_{(1,1)}}$ the solution of \eqref{GlueCondSampleShift} in $V_{-\rho}$, we have $\phi_{(g,g')}\ket{B_{(g,g')}} = \ket{B'_{(1,1)}}$.


\subsection{Action of quantized Wilson loop operators}

Given a Kac-Moody algebra, one can construct quantized Wilson loop operators. These operators are series in the Kac-Moody current, and have a well-defined action on any highest weight module for the given Kac-Moody algebra. In particular, they will have a well-defined action on $V_{-\rho}$ and on the representatives of the cohomology in $\ker P$.

The computation of this action is straightforward for maximally symmetric Wilson operators. The relevant Kac-Moody algebra is $\gk$ (recall that this is the Kac-Moody algebra generated by the bosonic current $\bJ$), and the Wilson operators are the quantum equivalent of the classical Wilson loops \eqref{ClassWilsLoop} with the matrices $A^a$ forming a representation of $\mathfrak{g}$. From the general formula \eqref{EigValWilsLoop} giving the eigenvalue of $W_\mu$ on a Verma module of arbitrary highest weight $\eta$ at level $k$, we see that replacing $\eta = - \rho$, we get :
\begin{equation}
\label{EigValWOVRho}
W_\mu = d_\mu \mathbbm{1}_{V_{-\rho}} \qquad \mbox{on }V_{-\rho} \;,
\end{equation}
where $d_\mu$ is the dimension of the representation $A$ of $\mathfrak{g}$ of highest weight $\mu$.

The symmetry breaking Wilson operator is associated with a subalgebra $\hat{\mathfrak{a}}_\kappa$ generated by the partial current $\pJ$ (see \eqref{PartCurrents}). It acts by scalar multiplication on the highest weight $\hat{\mathfrak{a}}_\kappa$-modules, with eigenvalue given by \eqref{SymBreakWilLoop}. So we have to find which $\hat{\mathfrak{a}}_\kappa$-modules intersect $\ker P$, to find the action on the representatives of the cohomology. A little algebra yields :
\begin{equation}
\label{EigValPartCurr}
\begin{split}
\pJ^i_0 \ket{RR, e} =&\; \left ( J^i_0 + \frac{1}{2}f_{iAB} \psi_0^A \psi_0^B \right ) \ket{RR, e} \\
=& \;\left ( 0 - \frac{1}{2} \sum_{\alpha \in \Delta^+_{\mathfrak{a}}} \alpha^i  \right ) \ket{RR, e} = -\rho_{\mathfrak{a}}^i \ket{RR, e} \;,
\end{split}
\end{equation}
where $\Delta^+_{\mathfrak{a}}$ is the set of positive roots of $\mathfrak{a}$, and $\rho_{\mathfrak{a}}$ its Weyl vector. So all the states in $\ker P$ belong to $\hat{\mathfrak{a}}_\kappa$-modules of highest weight $-\rho_\mathfrak{a}$. Using the formula \eqref{SymBreakWilLoop} giving the spectrum of the symmetry breaking Wilson operator $W^\mathfrak{a}_\tau$, we get again :
$$
W^\mathfrak{a}_\tau = d_\tau \mathbbm{1}_{\ker P} \qquad \mbox{on }\ker P \;,
$$
where $d_\tau$ is the dimension of the representation of $\mathfrak{a}$ used to build the Wilson loop.

Note that all of these eigenvalues are integers : it will be a crucial fact to ensure that the charges are quantized.

\section{The charges of sWZW boundary states}

\label{SecChargeBoundSt}

In this section, we put the elements gathered in the last three sections together, and show how to associate to a given supersymmetric boundary state a quantity invariant under the generalized Kondo renormalization group flows. Then we compute the charges of the various branes constructed in section \ref{sWZWBranes}.

\subsection{The general prescription}

In the previous section, we had to look for Ramond-Ramond test states in the bundle of modules $GV_{-\rho}$. Because these modules are not part of the state space of the sWZW model, we cannot directly take a scalar product between the boundary state and a test state to compute their coupling. A prescription is therefore needed to measure the charge of the boundary state.

Obviously, one should add to the boundary state $\ket{B}$ a component $\ket{B_{-\rho}}$ in $\ker P$, to get a completed boundary state $\ket{\tilde{B}}$. The charge associated with the test state $\ket{RR,e}$ carried by $\ket{B}$ will then be given by $\braket{\tilde{B}}{RR, e}$. 

To do this in a natural way, we first solve the gluing conditions defining the boundary state in each of the fibers of $GV_{-\rho}$. In general, these gluing conditions cannot be solved in an arbitrary fiber, but only in the fibers above a submanifold $M \subset G/H \times G/H$. What we would like to do is to look for the component of the solution on the cohomology in each fiber, map it onto $\ker P$ with $\phi_{(g,g')}$ (defined in \eqref{DefPhi}), and then average it over $M$. We do not know yet how to do this precisely given arbitrary gluing conditions. However, this program can be realized for the boundary states of section \ref{sWZWBranes}.

We will see that these boundary states satisfy well-defined gluing conditions on $\psi|_\mathfrak{h}$. This property ensures that whenever we can solve the gluing conditions in some fiber $(m,m')V_{-\rho}$, $(m,m') \in M$, the solutions determine a one-dimensional subspace of the cohomology of the supercharge. Moreover, solving the original gluing conditions and mapping them back into $\ker P$ is equivalent to solving effective gluing conditions directly in $\ker P$, with automorphism $\Omega^{(m,m')} = \mbox{ad}_{m'^{-1}} \circ \Omega \circ \mbox{ad}_m$. After showing that $M$ is a disjoint union of connected components on which the effective automorphism is constant, we will produce a formula performing an average of the solutions on the finite set of connected components. This fixes a component of the boundary state in $\ker P$ up to a normalization constant $c$.

Finally, $c$ can be determined\footnote{A slight indeterminacy is left, see the remarks in section \ref{SecSomeRemarks}.} by considering the action of Wilson operators on the completed boundary state. This gives a component $\ket{B_{-\rho}}$ to the boundary state $\ket{B}$ on $\ker P$.

This procedure is consistent with the symmetries of the theory, in the sense that two boundary states differing by a shift on the group manifold are assigned the same component $\ket{B_{-\rho}}$, so they carry the same charges. This would not be the case if we restricted ourselves to a particular fiber of $GV_{-\rho}$. We will also check later that the charges obtained in this way are consistent with the intuition one may get from geometrical considerations.

We will now elaborate on the steps occurring in the procedure described above and justify our statements.

\subsection{Solving the gluing conditions}

Any supersymmetric boundary state satisfies a set of gluing conditions. Indeed, arbitrary correlators in the presence of this boundary condition should be invariant under superconformal transformations leaving the boundary fixed. This requirement translates into the gluing conditions \eqref{ConfCond} and \eqref{GlobSupSymCond} imposed on the corresponding boundary state. For all the known boundary states, however, the algebra preserved is much larger than the superconformal algebra, and additional gluing conditions are satisfied, as we noticed in section \ref{sWZWBranes}. Whenever the partition function for open strings stretching between a given brane $\ket{B}$ and a maximally symmetric brane decomposes into characters of a subalgebra $\hat{\mathfrak{s}} \subset \hat{\mathfrak{c}}$, $\ket{B}$ satisfy the gluing conditions :
\begin{equation}
\label{GlueCondGenSymAlg}
(X - (-i)^{2h_X}\Omega(\bar{X}))\ket{B} = 0
\end{equation}
for every generator $X$ of conformal dimension $h_X$ in $\hat{\mathfrak{s}}$. Here $\Omega$ is a map of $\hat{\mathfrak{c}}$ into $\hat{\mathfrak{c}}$ satisfying the intertwining relation $\Omega([X,Y]) = [\Omega(X),\Omega(Y)]$ for $X,Y \in \hat{\mathfrak{s}}$, but it does not necessarily preserve $\hat{\mathfrak{s}}$, and it can send it to another isomorphic subalgebra of $\hat{\mathfrak{c}}$. 

According to the prescription formulated above, one should start by solving the gluing conditions in $GV_{-\rho}$, that is in each of the fibers $(g,g')V_{-\rho}$, with $(g,g') \in G/H \times G/H$. By the remark of section \ref{SubSGlueCondShiGr}, solving \eqref{GlueCondGenSymAlg} in $(g,g')V_{-\rho}$ is equivalent to solving :
$$
(g^{-1}Xg - (-i)^{2h_X} g'^{-1}\Omega(\bar{X})g')\ket{B} = 0
$$ 
in $V_{-\rho}$.  After relabeling the subalgebra $g^{-1}\hat{\mathfrak{s}}g \rightarrow \hat{\mathfrak{s}}$ and the automorphism $\mbox{ad}_{g'^{-1}} \circ \Omega \circ \mbox{ad}_{g} \rightarrow \Omega$, the gluing condition can be rewritten in the same form as \eqref{GlueCondGenSymAlg}. So we can restrict our discussion to solutions of \eqref{GlueCondGenSymAlg} in $V_{-\rho}$. 

\vspace{.5cm}

So consider the chiral algebra $\hat{\mathfrak{c}}$ of the sWZW model, equipped with the total grading $\mbox{gr}_T$ defined in section \ref{SectCohomSupCharge}. The subalgebras $\hat{\mathfrak{s}}$ which can occur as symmetry algebras preserved by boundary states are conformal vertex subalgebras of $\hat{\mathfrak{c}}$. Their generators are series of normal ordered products of generators of $\hat{\mathfrak{c}}$ with well-defined grade. Therefore they decompose into subalgebras of negative, null and positive grade : $\hat{\mathfrak{s}} = \hat{\mathfrak{s}}_- \oplus \hat{\mathfrak{s}}_0 \oplus \hat{\mathfrak{s}}_+$.

We first extract a necessary condition for \eqref{GlueCondGenSymAlg} to have solutions. Any solution $\ket{B}$ to the full set of gluing conditions restricts onto a solution $\ket{B}^b$ of the bosonic gluing conditions after restriction on the even part $(V_{-\rho})^b$. This even part is given by $H_{-\rho} \otimes H^{\so}  \otimes \bar{H}_{-\rho} \otimes H^{\so}$, where $H^{\so} = H^{\so}_s \oplus H^{\so}_{s'}$ if $d$ is even and by $H^{\so} = H^{\so}_s$ if $d$ is odd (see section \ref{ChirAlg}). It is well known \cite{Behrend:1999bn} that after reinterpreting $\bar{H}_{-\rho} \otimes H^{\so}$ as the restricted dual of $H_{-\rho} \otimes H^{\so}$, the solution $\ket{B}^b$ becomes an operator $B : H_{-\rho} \otimes H^{\so} \rightarrow H_{-\rho} \otimes H^{\so}$, and the bosonic gluing conditions state that $B$ intertwines the action of $\hat{\mathfrak{s}}^b$ and the action of $\Omega\hat{\mathfrak{s}}^b$, the exponent $b$ denoting the bosonic part. A necessary condition for the existence of such an intertwiner is that $\Omega$ maps the generators of $\hat{\mathfrak{s}}^b$ represented by pronilpotent operators in $H_{-\rho} \otimes H^{\so}$ onto elements represented by pronilpotent operators. Pronilpotent means that for each vector in $H_{-\rho} \otimes H^{\so}$, there is a power of the pronilpotent operator sending it to zero. This is obviously the case for all the operators in $\mathfrak{c}_-$, and never the case for any operator in $\mathfrak{c}_+$\footnote{To be precise, this is true only for $L_0$-grade zero operators, because there is no null vector at grade zero in $H_{-\rho}$. However, $\Omega$ always preserves the $L_0$-grading, it is sufficient to consider operators at $L_0$-grade zero.}. Therefore it is mandatory that \nolinebreak:
\begin{equation}
\label{CondNec1}
\Omega(\hat{\mathfrak{s}}^b_-) \subset \hat{\mathfrak{c}}_- \qquad \mbox{and} \qquad \Omega(\hat{\mathfrak{s}}^b_+) \subset \hat{\mathfrak{c}}_+ \;.
\end{equation}
If this is not the case, there is no solution of \eqref{GlueCondGenSymAlg} in $V_{-\rho}$. If this condition is verified, we can proceed.

The irreducible highest weight $\hat{\mathfrak{s}}$-modules appearing in the decomposition of $\hat{\mathfrak{c}}$-modules are labeled by a set $P_\mathfrak{s}$. The modules for the subalgebra $\Omega\hat{\mathfrak{s}}$ are also classified by $P_\mathfrak{s}$, because $\hat{\mathfrak{s}}$ and $\Omega\hat{\mathfrak{s}}$ are isomorphic. One can define an action of $\Omega$ on $P_\mathfrak{s}$. Indeed, one can see each irreducible highest weight $\Omega\hat{\mathfrak{s}}$-module $H^{\Omega\mathfrak{s}}_\zeta$, $\zeta \in P_\mathfrak{s}$, as a $\hat{\mathfrak{s}}$-module, by composing the representation map with $\Omega$. $H^{\Omega\mathfrak{s}}_\zeta$ equipped with the action of $\hat{\mathfrak{s}}$ is isomorphic to some irreducible highest weight $\hat{\mathfrak{s}}$-module $H^{\mathfrak{s}}_{\Omega\zeta}$.

To solve the gluing condition \eqref{GlueCondGenSymAlg}, $V_{-\rho}$ should be decomposed into irreducible highest weight modules for $\hat{\mathfrak{s}} \oplus \Omega\hat{\mathfrak{s}}$, of the form $H^{\mathfrak{s}}_\zeta \otimes H^{\Omega\mathfrak{s}}_{\zeta'}$, and the gluing conditions \eqref{GlueCondGenSymAlg} can be solved in the summand $H^{\mathfrak{s}}_\zeta \otimes H^{\Omega\mathfrak{s}}_{\zeta'}$ if and only if $\Omega\zeta' = \zeta$. When this relation is satisfied, we can see $H^{\Omega\mathfrak{s}}_{\zeta'}$ equipped with the action of $\hat{\mathfrak{s}}$ as $H^{\mathfrak{s}}_\zeta$. Choosing an orthonormal basis $\{v_n\}$ of $H^{\mathfrak{s}}_\zeta$, the coherent state $\sum_n \ket{v_n} \otimes \ket{v_n}$ solves \eqref{GlueCondGenSymAlg}. See \cite{Behrend:1999bn}, section 2.1 for a detailed derivation.

Remark that once the polarization $\hat{\mathfrak{s}} = \hat{\mathfrak{s}}_- \oplus \hat{\mathfrak{s}}_0 \oplus \hat{\mathfrak{s}}_+$ has been fixed, the isomorphism class of an irreducible highest weight module $H^{\mathfrak{s}}_\zeta$, is completely defined by the action of the operators generating $\hat{\mathfrak{s}}_0$ on the grade zero subspace $(H^{\mathfrak{s}}_\zeta)_0$. Indeed, consider the module $M^{\mathfrak{s}}_\zeta$ freely generated from $(H^{\mathfrak{s}}_\zeta)_0$ by $\hat{\mathfrak{s}}_+$, the subalgebra of operators of positive grade. $H^{\mathfrak{s}}_\zeta$ is obtained by the quotient of $M^{\mathfrak{s}}_\zeta$ by its maximal submodule, which is generated by the action of $\hat{\mathfrak{s}}_+$ on all the null vectors of $M^{\mathfrak{s}}_\zeta$. But the condition that a vector is null involves only the action of $\hat{\mathfrak{s}}_0$ on $(H^{\mathfrak{s}}_\zeta)_0$. We see therefore that if the gluing conditions $\eqref{GlueCondGenSymAlg}$ for the elements of $\hat{\mathfrak{s}}_0$ can be solved on the highest weight subspace of $H^{\mathfrak{s}}_\zeta \otimes H^{\mathfrak{s}}_{\zeta'}$, we have $\zeta' = \Omega\zeta$ and there exists a solution to the whole set of gluing conditions.

As we eventually aim at taking a scalar product between $\ket{B}_{-\rho}$ and a test state $\ket{RR} \in \ker P$, we are only interested in the component of the solution of the gluing conditions on $\ker P$. By the discussion above, to find this component it is sufficient to solve on $\ker P$ the gluing conditions involving operators of total grade zero of $\hat{\mathfrak{s}}$ (provided $\Omega$ satisfies \eqref{CondNec1}, of course).

\vspace{.5cm}
Let us now be more specific. Recall that all of the boundary state constructed in section \ref{sWZWBranes} are specific instances of twisted coset boundary states. It is instructive to study the chiral algebra they preserve : $\hat{\mathfrak{s}} = \ak \oplus \hat{\mathfrak{f}}_{\mathfrak{a}} \oplus \hat{\mathfrak{c}}/(\ak \oplus \hat{\mathfrak{f}}_{\mathfrak{a}})$. $\ak$ is generated by the partial current $\pJ$ \eqref{PartCurrents} and $\hat{\mathfrak{f}}_{\mathfrak{a}}$ by $\psi|_\mathfrak{a}$. The coset algebra $\hat{\mathfrak{c}}/(\ak \oplus \hat{\mathfrak{f}}_{\mathfrak{a}})$ is formed by all the normal-ordered products of operators of $\hat{\mathfrak{c}}$ which (anti)commute with $\ak \oplus \hat{\mathfrak{f}}_{\mathfrak{a}}$. First, note that the full current $J|_\mathfrak{a}$ restricted to $\mathfrak{a}$ belongs to $\ak \oplus \hat{\mathfrak{f}}_{\mathfrak{a}}$. Consider now the fermionic fields $\psi|_{\mathfrak{h}^\perp}$ associated with the orthogonal complement $\mathfrak{h}^\perp$ of $\mathfrak{a}$ in the Cartan subalgebra $\mathfrak{h}$ of $\mathfrak{g}$. Obviously, $\psi|_{\mathfrak{h}^\perp}$ commutes with any fermionic field associated to $\mathfrak{a}$, because of orthogonality. It also commutes with the full current $J|_{\mathfrak{a}}$, again because of orthogonality. Therefore $\psi|_{\mathfrak{h}^\perp}$ belongs to $\hat{\mathfrak{c}}/(\ak \oplus \hat{\mathfrak{f}}_{\mathfrak{a}})$. Finally, using the same argument, we see that $J|_{\mathfrak{h}^\perp} \in \hat{\mathfrak{c}}/(\ak \oplus \hat{\mathfrak{f}}_{\mathfrak{a}})$. So $J|_{\mathfrak{h}}$ and $\psi|_{\mathfrak{h}}$ belong to the preserved chiral algebra. Using the notation of the previous paragraph, we proved that $\hat{\mathfrak{s}}_0 = \hat{\mathfrak{c}}_0$ for generic twisted coset boundary states. 

The gluing conditions satisfied on this subspace are given by  :
\begin{equation}
\label{ExplGlueCond}
\begin{split}
(\psi^j_0 + i\Omega(\bar{\psi}^j_0))\ket{B} &= 0 \;, \\
(J^j_0 + \Omega(\bar{J}^j_0))\ket{B} &= 0 \;,
\end{split}
\end{equation}
where $j$ is an index running on the Cartan subalgebra of $\mathfrak{g}$ and $\Omega$ is the product of an automorphism of $\mathfrak{g}$ and an automorphism of $\mathfrak{a}$. As it preserves the Killing form, it is an orthogonal transformation of $\g$. In general it does not preserve the Lie bracket, however.

We can now derive another necessary condition for the existence of a solution to the gluing conditions. By definition, $\Omega$ preserves the $L_0$-grading, and also preserves the real form of the Kac-Moody algebra $\hat{\mathfrak{g}}$. Remark that the only real grade zero generators with a nontrivial kernel in $V_{-\rho}$ are $J^j_0$, the generators of the Cartan subalgebra, which vanish on $\ker P$. The other real generators of grade zero have a trivial kernel in $V_{-\rho}$ (they are basically sums of positive and negative ladder operators). Therefore the set $\{J^j_0\}$ is necessarily preserved by the intertwining map $B$, so $\Omega$ has to preserve the Cartan subalgebra of $\mathfrak{g}$.


The necessary condition \eqref{CondNec1} applied to coset fields forces $\Omega$ to preserve the whole triangular decomposition of $\mathfrak{g}$, and not only the Cartan subalgebra. This fact can be seen as follows. Consider the coset generators with non-trivial $\mathfrak{g}$-weights and zero $L_0$-grade. As the coset generators are normal ordered products of generators of $\hat{\mathfrak{c}}$, their weights belong to the root lattice of $\mathfrak{g}$. (They are also necessarily orthogonal to the weight space of $\mathfrak{a}$, else it would be impossible for such operators to commute with the Cartan generators of $\mathfrak{a}$.) Any operator of this type which weight is not a linear combination of the simple roots with negative integer coefficients is necessarily pronilpotent. The fact that such operators have to be mapped among themselves by $\Omega$ forces the latter to preserve the triangular decomposition of $\mathfrak{g}$.

To summarize, the gluing conditions are solvable in the fiber above $(m,m')$ only if $\mbox{ad}_{m'^{-1}}\circ \Omega \circ \mbox{ad}_{m}$ preserves the triangular decomposition of $\mathfrak{g}$. Remark that the gluing conditions imposed on the bosonic fields are trivial when we consider them in $\ker P$, because the full currents in the Cartan subalgebra vanishes there. So the component of the solution in $\ker P$ is determined by the gluing condition on the fermionic zero modes.

If $\Omega$ is an involution, these fermionic gluing conditions force the component of the solution to \eqref{ExplGlueCond} in $\ker P$ to be (up to normalization) :
$$
\ket{RR,e_{(m,m')}} = \ket{-\rho} \otimes \sigma \ket{e_\Omega e_{\mathfrak{n}_+}} = \ket{RR,e_\Omega} \;,
$$
where we chose a basis $\{h^j\}$ of $\mathfrak{h}$ which diagonalizes $\Omega$ and defined the volume form $e_\Omega$ on the eigenspace of eigenvalue $-1$ : $e_\Omega = \bigwedge_{j|\Omega(h^j)=-h^j} h^j$. 

In general, we have to choose a lift $\tilde{\Omega}$ of $\Omega \in O(r)$ into $Pin_+(r)$, where $r$ is the rank of $\mathfrak{g}$ and we see $\Omega$ as an orthogonal transformation of the Cartan subalgebra. Such an element of $Pin_+(r)$ can be expressed as a polynomial in the fermionic antiholomorphic generators $\{\bar{\psi}^j_0\}$. 
By definition, $\tilde{\Omega}$ satisfies $\Omega (\bar{\psi}^j_0) = \alpha(\tilde{\Omega}) \bar{\psi}^j_0 \tilde{\Omega}^{-1}$, where $\alpha$ is the involution $\mbox{ad}_{(-1)^{\bar{\mathcal{F}}}}$. ($\alpha$ multiplies odd Clifford elements by $-1$ and leaves invariant even elements). The solution to the gluing conditions \eqref{ExplGlueCond} therefore reads :
$$
\ket{RR,e_{(m,m')}} = \tilde{\Omega}\ket{RR,1} \, .
$$

Let us remark that in the case of interest to us $\hat{\mathfrak{s}} = \ak \oplus \hat{\mathfrak{f}}_{\mathfrak{a}} \oplus \hat{\mathfrak{c}}/(\ak \oplus \hat{\mathfrak{f}}_{\mathfrak{a}})$, whenever we can solve the gluing conditions, the solution determines a one-dimensional subspace of $\ker P$.

\subsection{Averaging}

\label{SubSecAveraging}

We describe now how to average over $M$ the component of the solution on the cohomology. We must first study the structure of the manifold $M$. In the previous section, we defined $M$ as the set points $(g,g') \in G/H \times G/H$ such that $\Omega^{(g,g')} := \mbox{ad}_{g'^{-1}} \circ \Omega \circ \mbox{ad}_g$ preserves the triangular decomposition of $\mathfrak{g}$. 

Consider the subalgebra $\mathfrak{a}$ which Lie bracket is preserved by $\Omega$, and its corresponding Lie group $A \subset G$. $\Omega$ maps $\mathfrak{a}$ on another subalgebra $\mathfrak{a}^\Omega \subset \mathfrak{g}$, isomorphic to $\mathfrak{a}$. After exponentiation, $\Omega$ can be seen as a group isomorphism from $A$ to $A^\Omega := \exp \mathfrak{a}^\Omega$. Note however that usually this map does not extend naturally to a map from $G$ to $G$. Consider now pairs of elements $(a, \Omega a)$, $a \in A$, and $\Omega^{(a, \Omega a)} = \mbox{ad}_{(\Omega a)^{-1}} \circ \Omega \circ \mbox{ad}_a$. $\Omega^{(a, \Omega a)}$ coincides with $\Omega$ on $\hat{\mathfrak{s}}$. Indeed, if we write $a = e^x$, $x \in \mathfrak{a}$, then $\Omega a = \exp \Omega x$ and we have for a generic $X \in \hat{\mathfrak{s}}$ :
\begin{equation}
\label{OmInvOnTwistDiag}
\Omega^{(a, \Omega a)}(X) = e^{-\Omega x} \Omega(e^x X e^{-x}) e^{\Omega x} = \Omega(X) \;.
\end{equation}
If $X$ is a coset field, the last equality is true because $X$ commutes with $e^x$ and $\Omega(X)$ commutes with $e^{\Omega x}$. If $X$ is a field in $\hat{\mathfrak{a}} \oplus \hat{\mathfrak{f}}_\mathfrak{a}$, then $\Omega$ satisfies the intertwining property $\Omega(e^x X e^{-x}) = e^{\Omega x} \Omega(X) e^{-\Omega x}$, so \eqref{OmInvOnTwistDiag} is true as well. 
This shows that for any $(m,m') \in M$, $(am, \Omega am') \in M$. Moreover, the images in $\ker P$ of the solutions to the gluing conditions above these two points are the same, because the effective automorphism in $V_{-\rho}$ is the same.

Let us now fix an arbitrary point $(m,m') \in M$, so that $\Omega^{(m,m')}$ preserves the triangular decomposition of $\g$. Another class of solutions are provided by pairs of elements $(w,w')$ of the Weyl group of $\mathfrak{g}$ such that $w'^{-1} \circ \Omega^{(m,m')} \circ w$ still preserves the triangular decomposition of $\mathfrak{g}$. Such a pair $(w,w')$ can be written as $(\mbox{ad}_{g_w},\mbox{ad}_{g_{w'}})$ with $g_w,g_{w'} \in G$, so $(mg_w ,m'g_{w'}) \in M$. Let us call $\mathcal{W}$ the set of pairs $(w,w')$ satisfying the condition above, modulo the left action of elements of the form $(\mbox{ad}_{m^{-1}am}, \mbox{ad}_{m'^{-1} (\Omega a) m'})$.

We see that $M$ contains a submanifold of the form $\bigsqcup_i M_i$, where each of the $M_i$ is generated from a point $(m,m')$ by the left action of the elements $(a,\Omega a)$, and the $M_i$'s are the images of $M_0 \ni (m,m')$ under the right action of $(g_w,g_{w'})$, $(w,w') \in \mathcal{W}$. It seems that this exhausts all of the solutions to the gluing conditions, so that $M = \bigsqcup_i M_i$, but we do not have a proof of this statement.

\vspace{.5cm}

Suppose that $M = \bigsqcup_i M_i$ as described above. Then $\mathcal{W}$ parametrizes the components $M_i$ of $M$. We saw that the automorphisms associated to points in a given component $M_i$ of $M$ coincide. The corresponding gluing conditions have therefore the same line of solutions in $\ker P$ and we should be able to reduce the average to a sum over the connected components $M_i$. This can be done as follows. 

Any element $w$ of the Weyl group can be seen as an orthogonal transformation of $\h$. The transformation $\epsilon(w) w$ has determinant 1 for any Weyl group element $w$. Hence it can be lifted to an element $\tilde{w}$ of $Pin$. Pairs of such $Pin$ elements naturally act on $\ker P$ by :
\begin{equation}
\label{ActWeylKerP}
(\tilde{w}, \tilde{w}') \cdot \ket{RR,e} = \tilde{w}'^{-1}_{\bar{\psi}} \tilde{w}^{-1}_\psi  \ket{RR,e} = (-1)^{\epsilon(w)|\Omega|} \tilde{w}'^{-1}_{\bar{\psi}} \tilde{\Omega} \, \tilde{w}_{\bar{\psi}}  \ket{RR,1} \;,
\end{equation}
where $\tilde{w}_\psi$ is the Clifford element $\tilde{w}$ expressed in term of the Clifford generators $\{\psi^i_0\}$. We also expressed $\ket{RR,e}$ as $\tilde{\Omega}\ket{RR,1}$ for $\tilde{\Omega}$ some polynomial in $\{\bar{\psi}^i_0\}$ .

As $\tilde{w}'^{-1}_{\bar{\psi}} \tilde{\Omega} \tilde{w}_{\bar{\psi}}$ is a lift into $Pin$ of $w'^{-1} \circ \Omega \circ w$, $\tilde{w}'^{-1}_{\bar{\psi}} \tilde{\Omega} \tilde{w}_{\bar{\psi}} \ket{RR,1}$ solves the gluing conditions for $\Omega^{(mg_w,m'g_{w'})}$ if and only if $\tilde{\Omega}\ket{RR,1}$ solves the gluing conditions for $\Omega^{(m,m')}$. We can therefore define the charge of the boundary state as :
\begin{equation}
\label{PrescrMesChargeNDiag}
c\ket{B_{-\rho}} = \sum_{(w,w') \in \mathcal{W}} (-1)^{\epsilon(w)|\Omega|}  \tilde{w}'^{-1}_{\bar{\psi}} \tilde{\Omega} \, \tilde{w}_{\bar{\psi}}  \ket{RR,1} \;,
\end{equation}
where $\tilde{\Omega} \ket{RR,1}$ solves the gluing conditions for $\Omega^{(m,m')}$.
If each class in $\mathcal{W}$ does contain a diagonal element $(w,w)$, $w \in W$, then \eqref{PrescrMesChargeNDiag} coincides with the simpler formula :
\begin{equation}
\label{PrescrMesChargeDiag}
c\ket{B_{-\rho}} = \sum_{(w,w) \in \mathcal{W}} \ket{RR,w(e)} \;,
\end{equation}
where $\ket{RR,e}$ solves the gluing conditions for $\Omega^{(m,m')}$ and $w(e)$ denotes the action of the Weyl group on $\h$ extended on $\bigwedge \h$.

The subspace defined by \eqref{PrescrMesChargeNDiag} and \eqref{PrescrMesChargeDiag} is invariant upon a left shift by $g_L$ and a right shift by $g_R$ on the group manifold. Indeed, upon such a transformation, $\Omega \mapsto \mbox{ad}_{g_R} \circ \Omega \circ \mbox{ad}_{{g_L}^{-1}}$, so that $M$ is shifted in $G/H \times G/H$ and $\mathcal{W}$ is invariant. This feature is crucial to ensure that the charges are invariant under the truly marginal boundary RG flows corresponding to these transformations.

Remark that \eqref{PrescrMesChargeNDiag} is ambiguous. Indeed, $c\ket{B_{-\rho}}$ depends on the choice of the lifts $\tilde{w}$ and $\tilde{w}'$ into $Pin$ of the Weyl group elements $w$ and $w'$. In most of the examples, we will be able to use \eqref{PrescrMesChargeDiag}. In the other examples, namely the D-branes of $SU(4)$ which are twisted by the product of an automorphism of a subalgebra and an outer automorphism of $SU(4)$, the natural choice of the lifts produces a physically sensible answer, but we were not able to define a canonical choice that would make \eqref{PrescrMesChargeNDiag} completely well-defined.

\subsection{Normalization and periodicity}

\label{SubSecNormPer}

It still remains to fix the normalization of the element $\ket{B_{-\rho}}\in \ker P$ obtained. Let $\{\ket{B_m}\}$, $m \in \mathcal{I}$ be the set of elementary boundary states satisfying the same gluing conditions as our boundary state $\ket{B}$, for $\mathcal{I}$ a set indexing the elementary boundary states (for instance in the case of maximally symmetric boundary states, $\Omega = \mathbbm{1}$ and $\mathcal{I} = P^+_k \times \{t,f\}$, the set of integrable highest weight of $\gk$ times the two weights $t$ and $f$ of $\sod$ associated with boundary states preserving the supercharge $D_-$).

The completed boundary states read $\ket{\tilde{B}_m} = \ket{B_m} + q_m \ket{B_{-\rho}}$, where $q_m$ are a set of a priori arbitrary real numbers. Now the action of Wilson operators on boundary states can be used to constrain the numbers $q_m$. Indeed, we saw that Wilson operators have a well defined action \eqref{EigValWOVRho} in $V_{-\rho}$. We also saw that they act on boundary states : whenever $\ket{B}$ is a consistent boundary state, $\ket{B'} = W^\mathfrak{g}_\mu \ket{B}$ is also a consistent boundary state. Demanding this property to be preserved when Wilson operators act on completed boundary states $\ket{\tilde{B}} = \ket{B} + q_B \ket{B_{-\rho}}$ and $\ket{\tilde{B}'} = \ket{B'} + q_{B'} \ket{B_{-\rho}}$ yields $q_{B'} = d_\mu q_B$. These relations can be used to express all of the $q_m$ in term of one of them. The value of the remaining free parameter is arbitrary and has no physical meaning\footnote{Indeed, the Ramond-Ramond test states do not exist as physical states. Their couplings to D-branes have a meaning only as invariants of the renormalization group flows, and not as an interaction that could be tested with a physical probe. As the Kondo flows send stacks of branes onto stacks of branes, only ratios of charges are relevant.}. Note that the charges of all the branes preserving the given gluing conditions are integer linear combinations of the charges of the elementary boundary states, so the numbers $q$ can be chosen to be all integers. 

There is an extra subtlety that will account for the periodicity of the charges. Recall from section \ref{WilsLoopKondRGF} that we defined an equivalence relation on Wilson operators such that two Wilson operators are in the same equivalence class if and only if they have an identical action on the physical state space $\mathcal{H}$. In particular, we had $[W^{\textfrak{g}}_\mu] = [\epsilon(w)W^{\textfrak{g}}_{w(\mu)}]$ where $w$ is an element of the affine Weyl group of $\gk$ and $\epsilon(w)$ its sign. As these two operators produce the same boundary state when applied to $\ket{B}$, we see that the completed boundary states should be identified as :
\begin{equation*}
\begin{split}
W^{\textfrak{g}}_\mu \ket{\tilde{B}} \sim &\; \epsilon(w)W^{\textfrak{g}}_{w(\mu)} \ket{\tilde{B}} \\
\Rightarrow W^{\textfrak{g}}_\mu \ket{B} + d_\mu q_B \ket{B_{-\rho}} \sim &\; W^{\textfrak{g}}_\mu \ket{B} + \epsilon(w) d_{w(\mu)} q_B \ket{B_{-\rho}} \;.
\end{split}
\end{equation*} 
Hence the the charges should be identified according to :
\begin{equation}
\label{CondPerCharges}
d_\mu q_B \sim \epsilon(w) d_{w(\mu)} q_B \;,
\end{equation}
which means that only the class of $q_B$ modulo an integer $M$ is unambiguous and well-defined. As the classes of maximally symmetric Wilson operators form an algebra isomorphic to the fusion ring of $\gk$ \eqref{RingClassWilsOp}, \eqref{CondPerCharges} is actually equivalent to :
\begin{equation}
\label{FusRingRepMod}
d_\lambda d_\mu = \mathcal{N}_{\lambda\mu}^{\;\;\nu} d_\nu \mod M \;.
\end{equation}
This equation is exactly the one considered in \cite{Fredenhagen:2000ei, Bouwknegt:2002bq}. Therefore the periodicity $M$ of the charge coincides with the one expected from K-theory computations \cite{Braun:2003rd}. $M$ is an integer which depends both on the level $k$ and on the Lie algebra $\g$. But as Wilson operators act on every boundary state, \eqref{FusRingRepMod} holds independently of which boundary state we are considering, so the integer $M$ is the same for all of the types of charges. Note that \eqref{CondPerCharges} for $\mu = 0$ was also used in \cite{Alekseev:2000jx, Maldacena:2001ky} to compute $M$.

We see now that $\ket{B}$ carries $q_B \!\mod M$ units of the charge associated to the element $\ket{B_{-\rho}}$, while the other (orthogonal) possible charges of $\ket{B}$ vanish. In particular, the charges are quantized. 

\subsection{Invariance under the generalized Kondo flows}

\label{SubSecInvKondFlows}

We also immediately see that the charges defined in this way are the same between the UV and IR fixed points of the generalized Kondo flows discussed in section \ref{SupSymKondPert}. Indeed, if the initial D-brane of the flow is a stack of $n$ D-branes with boundary state $\ket{B}$, then the IR fixed point of the flow is given by the action of the Wilson loop operator $W^\mathfrak{a}_\tau$, where $\tau$ labels the $n$-dimensional representation of $\mathfrak{a}$ used in the perturbation \eqref{KondoPert} : $n\ket{B} \mapsto W^\mathfrak{a}_\tau \ket{B}$. 

The Wilson operators commute with both $J|_\mathfrak{h}$ and $\psi|_\mathfrak{h}$, so neither the action of these operators nor the generalized Kondo flows modify the gluing conditions \eqref{ExplGlueCond}. Moreover, for the completed boundary states, we have :
$$
n(\ket{B} + \ket{B_{-\rho}}) \mapsto W^\mathfrak{a}_\tau(\ket{B} + \ket{B_{-\rho}}) = W^\mathfrak{a}_\tau\ket{B} + n\ket{B_{-\rho}} \;,
$$
so the charge of $W^\mathfrak{a}_\tau\ket{B}$ is the same as the charge of the stack $n\ket{B}$.

Note also that a class of flows which are not of the Kondo type was uncovered in \cite{Alekseev:2002rj}. These flows link the following twisted D-brane configurations :
$$
d_{\dot{\mu}}\ket{B_\Omega, \dot{0}, x} \rightarrow \ket{B_\Omega, \dot{\mu}, x}
$$
Our charges are also invariant under these flows, because the charges of the states $\ket{B_\Omega, \dot{\mu}, x}$ are proportional to $d_{\dot{\mu}}$ (see section \ref{ChTwStates}).

\subsection{Maximally symmetric D-branes}

\label{SubSecChMaxSymSt}

Now we measure the charges of the supersymmetric boundary states constructed in section \ref{sWZWBranes}. We start with the maximally symmetric D-branes.  Let us recall that the maximally symmetric boundary states preserving the supercharge $D_-$ satisfy the following gluing conditions :
\begin{equation}
\label{RepGlueCondMaxSym}
\begin{array}{rl}
(\psi^a_r + i \bar{\psi}^a_{-r}) \ket{B} &= 0 \;,\\
(J^a_n  + \bar{J}^a_{-n})\ket{B} &= 0 \;.
\end{array}
\end{equation}
Consider these equations in $(g,g')V_{-\rho}$. The second condition simply states that the operator $B$ should intertwine the action of $\hat{\mathfrak{g}}$ in $gH_{-\rho}$ and $g'H_{-\rho}$. Of course, this is possible only if $gH = g'H$. Therefore, it is possible to solve these gluing conditions only in the fibers above the diagonal $M = G/H \subset G/H \times G/H$. The equations \eqref{RepGlueCondMaxSym} in $(g,g)V_{-\rho}$ are equivalent to :
\begin{equation}
\label{RepGlueCondMaxSymVR}
\begin{array}{rl}
(g^{-1}\psi^a_rg + i g^{-1}\bar{\psi}^a_{-r}g) \ket{B} &= 0 \;,\\
(g^{-1}J^a_n g + g^{-1} \bar{J}^a_{-n}g)\ket{B} &= 0 \;.
\end{array}
\end{equation}
in $V_{-\rho}$. But the system \eqref{RepGlueCondMaxSymVR} is equivalent to \eqref{RepGlueCondMaxSym}. The solution in $\ker P$ to these gluing conditions is $\ket{RR,1}$ and this solution is constant over $M$. Therefore, maximally symmetric boundary states carry a charge along $\ket{RR,1}$. We learn that $\ket{RR,1}$ is the Ramond-Ramond test state probing the so-called ``D0-brane charge''.

Remark that if we had rather considered maximally symmetric boundary states shifted on the group by an element $\hat{g} \in G$, which satisfy :
$$
\begin{array}{rl}
(\psi^a_r + i \hat{g}\bar{\psi}^a_{-r}\hat{g}^{-1}) \ket{B} &= 0 \;,\\
(J^a_n  + \hat{g}\bar{J}^a_{-n}\hat{g}^{-1})\ket{B} &= 0 \;.
\end{array}
$$
then we would have been able to solve the gluing conditions in all of the fibers above $M' = \{(g,g') \in G/H \times G/H \; | g' = \hat{g}g \}$. The image in $\ker P$ of the component of solution on the cohomology would still be $\ket{RR,1}$, because of the intertwining property of the map $\phi^F$. This is equally true for any type of boundary states : our prescription assigns identical charges to D-branes differing by a mere shift on the group manifold. This is what we expect from invariants of the boundary renormalization group flows, as shifted branes are linked by marginal deformations of the open string CFT.

To determine the charges, recall that the maximally symmetric boundary states can be expressed as :
\begin{equation}
\label{MaxSymBoundStatesFromOne}
\ket{B,\mu,x} = W_\mu^{\mathfrak{g}} W_x^\mathfrak{so} \ket{B,0,t} \;.
\end{equation}
Let us complete $\ket{B,0,t}$ as follows : $\ket{\tilde{B},0,t} = \ket{B,0,t} + \ket{RR,1}$. As we mentioned previously, the global normalization of the charges is not physically relevant, and can be fixed so that all the charges are integers. Then from \eqref{MaxSymBoundStatesFromOne}, we obtain :
$$
\ket{\tilde{B},\mu,x} = W_\mu^{\mathfrak{g}} W_x^\mathfrak{so} \ket{\tilde{B},0,t} = \ket{B,\mu,x} \pm d_\mu \ket{RR,1} \;,
$$
where the sign is $+$ when $x = t$ and $-$ when $x=f$. Indeed, $W_f^\mathfrak{so}$ acts by $-1$ in the whole Ramond-Ramond sector. The charges are then :
$$
\braket{\tilde{B},\mu,x}{RR,e} = \left \{ \begin{array}{lll} \pm d_\mu & \!\!\!\!\! \mod M & \quad \mbox{if }e = 1 \\ 0 &\!\!\!\!\! \mod M & \quad \mbox{if } e \perp 1 \;. \end{array} \right.
$$
Therefore we get exactly the results expected from \cite{Fredenhagen:2000ei}. The fact that one gets the opposite charge for boundary states differing only by the label $x$ confirms that these states form brane-antibrane pairs.

\subsection{Coset D-branes}

\label{ChCosStates}

The gluing conditions satisfied by coset boundary state are given by :
\begin{equation}
\label{RepGlueCondCoset}
\begin{array}{rl}
(\psi^A_r + i \bar{\psi}^A_{-r}) \ket{B} &= 0 \;,\\
(J^A_n  + \bar{J}^A_{-n})\ket{B} &= 0 \;.
\end{array}
\end{equation}
where $A$ is an index running on a basis of the subalgebra $\mathfrak{a}$, and :
\begin{equation}
(X - (-i)^{2h_X} \bar{X}) \ket{B} = 0
\end{equation}
for each of the fields $X$ in the coset vertex algebra ($h_X$ is the conformal dimension of $X$).

We are interested in solving these gluing conditions in $(g,g')V_{-\rho}$, and this is equivalent solving in $V_{-\rho}$ the gluing conditions where the holomorphic fields have been conjugated with $g^{-1}$ and the antiholomorphic fields by $g'^{-1}$. Choosing now the index $A$ to run over a basis of the subalgebra $g^{-1}\mathfrak{a}g$, we get from \eqref{RepGlueCondCoset} :
\begin{equation}
\begin{array}{rl}
(\psi^A_r + i g'^{-1}g \bar{\psi}^A_{-r}g^{-1}g') \ket{B} &= 0 \;\\
(J^A_n  + g'^{-1}g \bar{J}^A_{-n}g^{-1}g')\ket{B} &= 0 \;,
\end{array}
\end{equation}
and :
\begin{equation}
(X - (-i)^{2h_X} g'^{-1}g\bar{X}g^{-1}g') \ket{B} = 0 \;.
\end{equation}
The gluing conditions on the operators of total grade zero are given by :
\begin{equation}
\begin{array}{rl}
(\psi^j_0 + i g'^{-1}g \bar{\psi}^j_0g^{-1}g') \ket{B} &= 0 \;,\\
(J^j_0  + g'^{-1}g \bar{J}^j_0g^{-1}g')\ket{B} &= 0 \;.
\end{array}
\end{equation}
According to the discussion after equation \eqref{ExplGlueCond}, such gluing conditions can be solved on $\ker P$ only if the automorphism ad$_{g'^{-1}g}$ preserves the Cartan subalgebra of $\mathfrak{g}$, as well as the positive and negative root spaces. This forces $g'H = gH$, so $M$ is again given by the diagonal $G/H \subset G/H \times G/H$. The gluing conditions on the total grade zero subspace are the same as for maximally symmetric boundary states, so the solution on $\ker P$ is $\ket{RR,1}$ for each $m \in M$. Note however that the full solution to the gluing conditions is not constant, only its component on $\ker P$ is. So the coset states carry the same type of charge as the maximally symmetric ones, the D0-brane charge.

The coset D-branes can be obtained by the action of symmetry breaking Wilson operators on the maximally symmetric boundary states, according to \eqref{CosetBoundState} :
$$
\ket{B_{\mbox{\tiny coset}},\mu, x, \sigma} = W^\mathfrak{a}_\sigma W_\mu^{\mathfrak{g}} W_x^\mathfrak{so} \ket{B,0,t} \;.
$$
So, keeping the convention $\ket{\tilde{B},0,t,0} = \ket{B,0,t,0} + \ket{RR,1}$ adopted in the maximally symmetric case, the completed boundary state reads :
$$
\ket{\tilde{B}_{\mbox{\tiny coset}},\mu, x, \sigma} = \ket{B_{\mbox{\tiny coset}},\mu, x, \sigma} \pm d_\sigma d_\mu \ket{RR,1} \;,
$$
where again, the sign depends on $x$. The charges of coset states are then given by :
$$
\braket{\tilde{B}_{\mbox{\tiny coset}},\mu, x, \sigma}{RR,e} = \left \{ \begin{array}{lll} \pm d_\sigma d_\mu & \!\!\!\!\! \mod M & \quad \mbox{if }e = 1 \\ 0 & \!\!\!\!\! \mod M & \quad \mbox{if } e \perp 1 \;. \end{array} \right.
$$

\subsection{Twisted D-branes}

\label{ChTwStates}

To find the charge of the twisted supersymmetric boundary states, we should solve the gluing conditions \eqref{TwistCond} and \eqref{TwistCondFerm} in $(g,g')V_{-\rho}$, which amounts to solving :
\begin{equation}
\label{GlueCondTwistRep}
\begin{array}{rl}
(g^{-1}J^a_ng + g'^{-1}\Omega(\bar{J}^a_{-n})g')\ket{B} &= 0 \;,\\
(g^{-1}\psi^a_ng + ig'^{-1}\Omega(\bar{\psi}^a_{-n})g') \ket{B} &= 0
\end{array}
\end{equation}
in $V_{-\rho}$, where $\Omega$ is an outer automorphism of $\mathfrak{g}$. Any such outer automorphism of $\mathfrak{g}$ is conjugate to the automorphism $\Omega_D$ coming from the symmetry of the Dynkin diagram of $\mathfrak{g}$ : $\Omega = \mbox{ad}_{\tilde{g}} \circ \Omega_D$. Recall also that $\Omega$ can be lifted to an automorphism of $G$, and that we have $\mbox{ad}_g \circ \Omega = \Omega \circ \mbox{ad}_{\Omega g}$ as operators on $\mathfrak{g}$. The system \eqref{GlueCondTwistRep} can then be rewritten :
\begin{equation}
\label{GlueCondTwistRep2}
\begin{array}{rl}
(J^a_n + \hat{g}\Omega_D(\bar{J}^a_{-n})\hat{g}^{-1})\ket{B} &= 0 \;,\\
(\psi^a_n + i\hat{g}\Omega_D(\bar{\psi}^a_{-n})\hat{g}^{-1}) \ket{B} &= 0 \;,
\end{array}
\end{equation}
where $\hat{g} = g'^{-1}(\Omega g)\tilde{g}$. As $\Omega_D$ preserves the total grading, $\mbox{ad}_{\hat{g}}$ should also preserve it, so $g'H = (\Omega g)\tilde{g}^{-1}H$. $M$ is therefore the ``twisted diagonal'' in $G/H \times G/H$ defined by the previous equation. Again, the solution to the gluing conditions is constant on this diagonal, and we have \nolinebreak:
$$
\ket{B_{-\rho}}|_{\ker P} = \ket{RR,e_{\Omega_D}} \;,
$$
where $e_{\Omega_D} = \prod_{e^j|\Omega_D e^j=-e^j} e^j$, for a suitable basis $\{e^j\}$ of the Cartan subalgebra diagonalizing $\Omega_D$. In particular, we see that branes twisted by distinct conjugated outer automorphisms carry the same charge.

In general, we cannot express every twisted boundary state from one of them using Wilson operators, as was done in the maximally symmetric case \eqref{MaxSymBoundStatesFromOne}. However, the action of Wilson operators satisfies :
$$
W_\lambda^{\mathfrak{g}} \ket{B_\Omega, \dot{\mu}, x} = \sum_{\dot{\nu} \in P^+_{\Omega,k}} \mathcal{N}^{\Omega\;\dot{\nu}}_{\lambda\dot{\mu}} \ket{B_\Omega, \dot{\nu}, x} \;,
$$
where $\mathcal{N}^{\Omega\;\dot{\nu}}_{\lambda\dot{\mu}}$ are fusion rules for twisted representations \cite{Gaberdiel:2002qa}. As these relations must still hold after the completion of boundary states, we should have :
\begin{equation}
\label{EqChargeTwist}
d_{\lambda} q_{\dot{\mu}} = \sum_{\dot{\nu} \in P^+_{\Omega,k}} \mathcal{N}^{\Omega\;\dot{\nu}}_{\lambda\dot{\mu}} q_{\dot{\nu}} \;.
\end{equation}
It was shown in \cite{Gaberdiel:2003kv} that a solution $q_{\dot{\mu}}$ valued in $\mathbb{Z}/M\mathbb{Z}$ to these equations is provided by $q_{\dot{\mu}} = d_{\dot{\mu}} \mod M$. $d_{\dot{\mu}}$ denotes dimension of the grade zero subspace of the $\Omega$-twisted representation of $\gk$ labeled by $\dot{\mu}$. This grade zero subspace is a representation of the finite Lie subalgebra of $\g$ left fixed by $\Omega$.

We learn that the charges of $\Omega$-twisted boundary states are given by :
$$
\braket{\tilde{B}_\Omega, \dot{\mu}, x}{RR,e} = \left \{ \begin{array}{lll} \pm d_{\dot{\mu}} & \!\!\!\!\! \mod M & \quad \mbox{if }e = e_{\Omega_D} \\ 0 & \!\!\!\!\! \mod M & \quad \mbox{if } e \perp e_{\Omega_D} \;.\end{array} \right.
$$
Again, the $\pm$ sign depends on the label $x$. We remark that as $\braket{RR,e_{\Omega_D}}{RR,1} = 0$, the twisted D-branes do not carry any D0-brane charge.

\subsection{Twisted coset D-branes}

\label{SubSectChTwistCosSt}

The twisted coset states break the symmetry algebra down to $\ak \oplus \hat{\mathfrak{f}}_{\mathfrak{a}} \oplus \hat{\mathfrak{c}}/(\ak \oplus \hat{\mathfrak{f}}_{\mathfrak{a}})$, and the twist $\tilde{\Omega} = \Omega^\mathfrak{a}\Omega$ is a product of an automorphism $\Omega$ of $\hat{\mathfrak{c}}$ and an automorphism $\Omega^\mathfrak{a}$ of $\ak \oplus \hat{\mathfrak{f}}_{\mathfrak{a}}$, extended trivially on the coset fields. 

This case is much more complicated than the three previous ones. Except for the general principles described in section \ref{SubSecAveraging}, we do not know any method other than case by case inspection to identify the manifold $M$ and compute $\ket{B_{-\rho}}$. We will study boundary states of this type in detail below, in the case of $G = SU(4)$. For now, we suppose that we obtained a component \nolinebreak:
$$
\ket{B_{-\rho}} = \ket{RR,e_{\tilde{\Omega}}}\;,
$$
by solving the gluing conditions in $GV_{-\rho}$ and averaging the solutions over $M$.

Like for twisted states, it is in general not possible to express every twisted coset boundary state from one of them by the action of Wilson operators. However, from the explicit expressions of the twisted boundary states \cite{Quella:2002ns}, we have the relations :
$$
W_\lambda^{\mathfrak{g}} \ket{B_{\tilde{\Omega}}, \dot{\mu}, \dot{\sigma}, x} = \sum_{\dot{\nu} \in P^+_{\Omega,k}} \mathcal{N}^{\Omega\;\dot{\nu}}_{\lambda\dot{\mu}} \ket{B_{\tilde{\Omega}}, \dot{\nu}, \dot{\sigma}, x} \;,
$$
$$
W_\tau^{\mathfrak{a}} \ket{B_{\tilde{\Omega}}, \dot{\mu}, \dot{\sigma}, x} = \sum_{\dot{\upsilon} \in P^+_{\Omega^\mathfrak{a},\kappa}} \mathcal{N}^{\Omega^\mathfrak{a}\;\dot{\upsilon}}_{\tau\dot{\sigma}} \ket{B_{\tilde{\Omega}}, \dot{\mu}, \dot{\upsilon}, x} \;,
$$
where $P^+_{\Omega^\mathfrak{a},\kappa}$ labels the $\Omega^\mathfrak{a}$-twisted representations of $\ak$ and $\mathcal{N}^{\Omega^\mathfrak{a}\;\dot{\upsilon}}_{\tau\dot{\sigma}}$ are the corresponding twisted fusion rules. Imposing the same equations on the completed boundary states yields constraints on the charges, which are solved in the same way as for twisted states.

The charges of the twisted coset boundary states are therefore given by :
$$
\braket{B_{\tilde{\Omega}}, \dot{\mu}, \dot{\sigma}, x}{RR,e} = \left \{ \begin{array}{lll} \pm d_{\dot{\mu}}d_{\dot{\sigma}} & \!\!\!\!\! \mod M & \quad \mbox{if }e = e_{\tilde{\Omega}} \\ 0 & \!\!\!\!\! \mod M & \quad \mbox{if } e \perp e_{\tilde{\Omega}} \;. \end{array} \right.
$$
As always, the $\pm$ sign depends on the label $x$, and $d_{\dot{\mu}}$ and $d_{\dot{\sigma}}$ are the dimensions of the representations of the subalgebras of $\g$ and $\mathfrak{a}$ left fixed by $\Omega$ and $\Omega^\mathfrak{a}$, respectively.

\subsection{Some remarks}

\label{SecSomeRemarks}

\begin{itemize}
  \item As technicalities may have obscured our point, let us state it again. We found the most general set of massless states in the Ramond-Ramond sector, and remarked that these states live in a continuous family of modules that can be tied together into a bundle over $G/H \times G/H$. As this sector does not belong to the physical spectrum of the theory, we were forced to add to the boundary states a component in this sector. The most natural way of doing so is to solve the gluing conditions defining the boundary state in this virtual sector. Because we have not a single module, but a family of them, and that there is no reason to distinguish one of them from the others, we had to find all of the solutions to the gluing conditions, and average them. This determined a component on the cohomology for each boundary state. To fix the relative normalization of this component, we used the action of Wilson operators. The charges are then given by the overlaps between the completed boundary states and the test states.
  \item Note that the eigenvalues of Wilson operators on highest weight $\gk$-modules are given by an analytic function \eqref{EigValWilsLoop} on the space of weights. Therefore, using Wilson operators to fix the normalization of the boundary state in the virtual sector can be seen as the appropriate continuation of the components of the boundary state from the physical sector of the theory to $V_{-\rho}$. This continuation is not unique, and the ambiguity induces the periodicity of the charges.
  \item Our analysis of the solutions to the gluing conditions in $GV_{-\rho}$ is not completely rigorous. We are not certain that the disjoint union $\sqcup_i M_i$ described in section \ref{SubSecAveraging} contains all of $M$. Moreover, for some types of boundary states, the averaging procedure may contain ambiguities. While these questions definitely deserve further study, the complete agreement with geometry that we will find in section \ref{SecRelHom} seems to support our analysis.
	\item This procedure may not succeed when applied to hypothetical elementary boundary states that would not satisfy well-defined gluing conditions involving $\psi_0|_\mathfrak{h}$. Indeed, given a set of gluing conditions \eqref{ExplGlueCond}, the element of $\ker P$ solving them is uniquely determined, up to normalization. This normalization is then fixed by the action of Wilson operators. Now if a boundary state did not satisfy gluing conditions of the type of \eqref{ExplGlueCond}, there could exist several linearly independent solutions to its gluing conditions with a non-trivial component on $\ker P$. Then the action of the Wilson operators would clearly not be sufficient to fix the charge of this boundary state. Supersymmetric versions of the boundary states recently constructed in \cite{Blakeley:2007gu} may have this property. 
	\item Remark an interesting coincidence. Massless Ramond-Ramond states and a non-trivial cohomology of the supercharge appear only in the module $V_{-\rho}$. But it is also in this module only that the eigenvalue of the Wilson operator does not get renormalized by the generalized Kondo flow $d_\tau \mathbbm{1} \mapsto W^\mathfrak{a}_\tau$. This crucial fact ensures that the charges we constructed are invariant under the Kondo flows. Note also that the states in the cohomology are supersymmetric with respect to both $G_0$ and $\bar{G}_0$, so this phenomenon looks like one more instance of the general fact that supersymmetric quantities are often protected from renormalization. It may be interesting to understand it better from the point of view of representation theory.
	\item The equations \eqref{RingClassWilsOp} \eqref{CondPerCharges} \eqref{FusRingRepMod} \eqref{EqChargeTwist} obtained from the action of Wilson operators to determine the normalization of the charges of a set of elementary boundary states are similar to the ones appearing in the earlier works \cite{Fredenhagen:2000ei, Alekseev:2000jx, Maldacena:2001ky, Bouwknegt:2002bq, Gaberdiel:2003kv}, and we used the results of these authors to solve them. But we would like to point out here a fundamental difference in the way they were derived. In all the papers above, the structure of the flows or the action of marginal deformations of the WZW model was used to derive constraints on the charges, and these constraints took the form of the equations above. Here, we were forced to consider the action of Wilson operators as the only sensible way of fixing the normalization of the component of the completed boundary state in $\ker P$. We did not make any assumption on the structure of the renormalization group flows. The quantization of the charges and their invariance under the RG flows is a consequence of our procedure, rather than a postulate.
	\item Actually, except in some specific cases, these equations do not allow to determine completely the charge of the boundary states. Indeed, given a boundary state, they determine an element in the appropriate $\mathbbm{Z}/M\mathbbm{Z}$ factor, but only up to automorphisms of this cyclic group. This means that the charges assigned to members of a given family of state linked by Kondo flows are determined only up to automorphisms of $\mathbbm{Z}/M\mathbbm{Z}$. As long as we work with a single family, this is irrelevant, but when we consider two distinct families of boundary states which carry charges in the same factor $\mathbbm{Z}/M\mathbbm{Z}$, then this indeterminacy is relevant. Given two boundary states belonging to distinct Kondo families but carrying charges in the same $\mathbbm{Z}/M\mathbbm{Z}$ factor, we cannot compare their charges. This constitutes a shortcoming of our construction.
	\item We saw that the cohomology of the supercharge leads to $2^r$ independent charges. However, in all the examples we will construct (see the next section), the boundary states couple only to $2^{r-1}$ of them. Indeed, the known boundary states couple only to test states $\ket{RR,e}$, with $e \in \bigwedge (\mathfrak{h}/\mathbbm{C}h^\rho)$, where $h^\rho$ is the generator in the Cartan subalgebra associated to the Weyl vector of $\mathfrak{g}$. The fact that D-branes cannot couple to elements in $\bigwedge (\mathfrak{h}/\mathbbm{C}h^\rho) \wedge h^\rho$ will be confirmed by the geometric analysis of the charges to be undertaken in section \ref{SecRelHom}. Basically, a brane coupling to such a state would wrap non-trivially the homology class of the 3-sphere in the Lie group, which is impossible because there is a non-trivial NS-NS flux through this cycle. 
	
The set of possible charges for the boundary state is therefore given by :
$$
(\mathbb{Z}/M\mathbb{Z})^{(2^{r-1})} \;.
$$
This group coincides with the twisted K-theory of the target space Lie group $G$ \cite{Braun:2003rd}. This confirms the intimate link between twisted K-theory, invariants of the renormalization group flows and (a generalization of ) Ramond-Ramond charges. We have not found yet a direct interpretation of our algebraic construction in term of K-theory, though.
\end{itemize}

\section{Examples}

\label{SecExamples}

We now turn to some examples. We are mostly interested in determining what type of charges each of the Kondo families of boundary states carry. The magnitude of the charge of each state in a family is determined (up to an automorphism of $\mathbb{Z}/M\mathbb{Z}$) as described in sections \ref{SubSecChMaxSymSt} to \ref{SubSectChTwistCosSt}.

\subsection{$SU(2)$}

There are only two families of boundary states in our classification that can be realized in $SU(2)$. The first are the maximally symmetric boundary states, and we already know that they carry a charge along $\ket{RR,1}$.

Another type of boundary states, the ``B-type'' branes, were constructed in \cite{Maldacena:2001ky}, using what can now be identified as a twisted coset construction. The subalgebra $\mathfrak{a}$ is a $\mathfrak{u}(1)$ sitting in $\mathfrak{su}(2)$, and the automorphism is the sign reversal on $\mathfrak{u}(1) \simeq \mathbb{R}$. Choosing $e^3$ as the generator of $\mathfrak{u}(1)$, the gluing conditions on the grade zero subspace of the preserved algebra satisfied by such states are therefore :
$$
(J_0^3 - \bar{J}_0^3)\ket{B} = 0 \;,\qquad (\psi_0^3 - i\bar{\psi}_0^3)\ket{B} = 0 \;,
$$
as well as gluing conditions on the coset fields (which in this case form the vertex algebra of parafermions). These gluing conditions can be solved in $V_{-\rho}$, and the corresponding element in $\ker P$ is $\ket{RR,e^3}$. According to the results of section \ref{SubSecAveraging}, one should also consider gluing conditions of the form :
$$
(g^{-1} J_0^3 g - g'^{-1} \bar{J}_0^3 g')\ket{B} = 0 \;, \qquad (g^{-1} \psi_0^3 g - i g'^{-1} \bar{\psi}_0^3 g')\ket{B} = 0
$$
in $V_{-\rho}$. They can also be solved for $g = g' = \hat{g}$, such that $\mbox{ad}_{\hat{g}}$ is the unique nontrivial element $w$ in the Weyl group. The corresponding element of $\ker P$ is $\ket{RR, we^3} = -\ket{RR,e^3}$. 
Other couples $(g,g')$ do not preserve the triangular decomposition of $\mathfrak{g}$.

The charge therefore vanishes, according to the expectations of \cite{Maldacena:2001ky}. Remark that the test state $\ket{RR,e^3}$ associated with the Weyl vector of $\mathfrak{su}(2)$ does not seem to couple to any boundary state.

\subsection{$SU(3)$}

In $SU(3)$, we have the maximally symmetric boundary states which carry D0-brane charge along $\ket{RR,1}$, but also boundary states twisted with respect to an outer automorphism $\Omega$ of $SU(3)$. We are free to choose any outer automorphism, because they differ from each other by conjugations by elements of $SU(3)$. So let us pick $\Omega$ to be the automorphism generated by the symmetry of the Dynkin diagram of $\mathfrak{su}(3)$, which preserves the triangular decomposition. According to the derivation above, twisted states carry a charge along the element of the exterior algebra $\bigwedge \mathfrak{h}$ corresponding to the volume form of the eigenspace of eigenvalue $-1$ of $\Omega$. This eigenspace is one-dimensional, given by the difference of the two simple roots $\alpha_1$ and $\alpha_2$, so the twisted states carry a charge $\ket{RR,h^{\alpha_1-\alpha_2}}$, where $h^{\lambda}$ is the element of $\mathfrak{h}$ dual to $\lambda \in \mathfrak{h}^\ast$.

Note again that the test states in the subspace generated by the elements $h^{\rho}$ and $h^{\rho} \wedge h^{\alpha_1-\alpha_2}$ of $\bigwedge \mathfrak{h}$ do not couple to any known boundary state ($\rho = \alpha_1 + \alpha_2$ in $\mathfrak{su}(3)$). 

\subsection{$SU(4)$}

We have again maximally symmetric boundary states which carry a charge along $\ket{RR,1}$, and twisted states which carry a charge along $\ket{RR, h^{\alpha_1-\alpha_3}}$. (The outer automorphism of $SU(4)$ induced from the symmetry of the Dynkin diagram exchanges the simple roots $\alpha_1$ and $\alpha_3$.) Let us now examine some twisted coset boundary states.

\subsubsection*{Twisted coset states from the embedding of $SU(3)$}

In \cite{Gaberdiel:2004za}, the authors constructed\footnote{We specialize here their construction in $SU(n)$ to the case of $SU(4)$.} twisted coset boundary states associated with the embedding $SU(3) \subset SU(4)$, and conjectured that they should carry all the possible K-theory charges of $SU(4)$. 

So let us follow \cite{Gaberdiel:2004za} and consider the complex conjugation automorphism $\Omega_{C3}$ of $SU(3)$, extended trivially on the coset fields. It does not preserve the triangular decomposition of $SU(4)$, but it can be expressed as $\mbox{ad}_{\hat{g}} \circ \Omega_{D3}$, where $\Omega_{D3}$ coincides on $SU(3)$ with the automorphism generated by the symmetry of the Dynkin diagram of $SU(3)$. $\hat{g}$ is given explicitly by :
\begin{equation}
\label{GroupElConjDynSu3}
\hat{g} = \left (
\begin{array}{cccc}
0 & 0 & 1 & 0 \\
0 & -1 & 0 & 0 \\
1 & 0 & 0 & 0 \\
0 & 0 & 0 & 1
\end{array}
\right )\;.
\end{equation}
$\Omega_{D3}$ preserves the triangular decomposition and acts trivially on the coset fields. Therefore the gluing conditions can be solved in $(1,\hat{g})V_{-\rho}$, and using our result for $SU(3)$, we see that the resulting element of the cohomology is $\ket{RR,h^{\alpha_1 -\alpha_2}}$. According to section \ref{SubSecAveraging}, we can also solve the gluing conditions in $(a,\Omega_{C3}(a)\hat{g})$, $a \in SU(3)$, and these elements constitute $M_0$. One can check that the image in $\ker P$ of the solution to the gluing conditions is constant over $M_0$. There are also some $M_i$, $i>0$. To identify them, one should look for pairs of elements $(w_i,w_i')$ of the Weyl group of $\mathfrak{su}(4)$ such that $w'_i \circ \Omega_D \circ w_i^{-1}$ preserves the triangular decomposition of $\mathfrak{su}(4)$. We should also identify pairs which differ from each other by the left action of an element of the twisted diagonal $(a,\Omega_{C3}(a))$, $a\in SU(3)$, because each of them map $M_0$ onto the same $M_i$. 
Explicit numerical computations with the Weyl group of $SU(4)$ exhibit eight such classes. Representatives can be taken to be all of the form $(w,w)$, and are given explicitly by \nolinebreak:
\begin{equation}
\label{WelGroupElSu3Su4}
\begin{array}{ll}
w & \mbox{Action on }(\alpha_1,\alpha_2) \\
\hline \mbox{id} &: (\alpha_1,\alpha_2) \mapsto (\alpha_1,\alpha_2)  \\
r_{\alpha_3} &: (\alpha_1,\alpha_2) \mapsto (\alpha_1, \alpha_2 + \alpha_3) \\
r_{\alpha_1}\circ r_{\alpha_2} \circ r_{\alpha_3}&: (\alpha_1,\alpha_2) \mapsto (\alpha_2, \alpha_3) \\
r_{\alpha_2}\circ r_{\alpha_3} &: (\alpha_1,\alpha_2) \mapsto (\alpha_1 + \alpha_2, \alpha_3)\\
\end{array} \;,
\end{equation}
where $r_{\alpha_i}$ is the reflexion with respect to the plane orthogonal to $\alpha_i$.
The charge is therefore given by :
$$
\ket{B_{-\rho}} \sim \ket{RR, h^{\alpha_1-\alpha_2} + h^{\alpha_1-\alpha_2-\alpha_3} + h^{\alpha_2 - \alpha_3} + h^{\alpha_1 + \alpha_2 - \alpha_3}} \sim \ket{RR,h^{\alpha_1 - \alpha_3}}\;,
$$
so these boundary states carry the same charge as the boundary states twisted by the outer automorphism of $SU(4)$. This already rules out the conjecture of \cite{Gaberdiel:2004za}, according to which all the possible K-theory charges can be realized by such boundary states. We will see in the next section that this result could have been guessed from purely geometric considerations.

\vspace{.5cm}

The remaining family of boundary states appearing in the construction of \cite{Gaberdiel:2004za} are states twisted by the composition of the complex conjugation automorphism $\Omega_{C4}$ of $SU(4)$ with the complex conjugation automorphism $\Omega_{C3}$ of $SU(3) \subset SU(4)$. This product of automorphism does not preserve the triangular decomposition of $\mathfrak{g}$, so the corresponding gluing conditions cannot be solved in $V_{-\rho}$. According to the principles exposed in section \ref{SecChargeBoundSt}, it is possible to solve them in another fiber $(g,g')V_{-\rho}$ of $GV_{-\rho}$ only if $\mbox{ad}_{g'^{-1}} \circ \Omega_{C4} \circ \Omega_{C3} \circ \mbox{ad}_g$ preserves the triangular decomposition. This happens for $g = \hat{g}$, $g' = \hat{g}'$ for $\hat{g}$ given by \eqref{GroupElConjDynSu3} and :
\begin{equation}
\label{ElGrpDiagOm4}
\hat{g}' = \left (
\begin{array}{cccc}
0 & 0 & 0 & 1 \\
0 & 0 & -1 & 0 \\
0 & 1 & 0 & 0 \\
-1 & 0 & 0 & 0
\end{array}
\right )\;.
\end{equation}
because we then have $\mbox{ad}_{\hat{g}'^{-1}} \circ \Omega_{C4} = \Omega_{D4}$ and $\Omega_{C3} \circ \mbox{ad}_{\hat{g}} = \Omega_{D3}$, where $\Omega_{D4}$ and $\Omega_{D3}$ are the automorphisms of $\mathfrak{su}(4)$ and $\mathfrak{su}(3)$ generated by the Dynkin diagram symmetries. So we can solve the gluing conditions in $(\hat{g}, \hat{g}')V_{-\rho}$. The pairs of Weyl group elements are now of the form $(w, \Omega_{D4} w)$, with $w$ any element in the list \eqref{WelGroupElSu3Su4}. To express the corresponding solutions to the gluing conditions, we use the principal basis (see section \ref{SecRelHom}) :
\begin{equation}
\label{BasePrinSu4}
\begin{array}{ccl}
(\hat{h}^1)^\ast & = & \frac{1}{\sqrt{20}}(3\alpha_1 + 4\alpha_2 + 3\alpha_3) \\
(\hat{h}^2)^\ast  & = & \frac{1}{2}(\alpha_1 - \alpha_3) \\
(\hat{h}^3)^\ast  & = & \frac{1}{\sqrt{20}}(\alpha_1 - 2\alpha_2 + \alpha_3) \\
\end{array}
\end{equation}
and decompose $\bar{\psi}_0|_\mathfrak{h}$ on this basis to get generators $\bar{\psi}^j_0$.
We also have to pick a lift $\tilde{\Omega}$ of $\Omega_{D4} \circ \Omega_{D3}$ into Pin$(r)$  ($r$ being the rank $\mathfrak{g}$). We choose :
$$
\tilde{\Omega} = \frac{1}{\sqrt{6}} \left ( 1-\sqrt{5}\, \bar{\psi}^2_0\bar{\psi}^3_0 \right ) \;.
$$
One can check that the adjoint action of $\tilde{\Omega}$ on the basis element $\bar{\psi}^j_0$ reproduces the action of $\Omega_{D4} \circ \Omega_{D3}$. Now (minus) the elementary Weyl reflections are implemented into Pin$(r)$ by the adjoint action of the following elements :
$$
\begin{array}{rcl}
w = r_{\alpha_1} &:& (\tilde{w})_{\bar{\psi}} = \frac{1}{\sqrt{2}} \left ( \frac{1}{\sqrt{5}} \bar{\psi}^1_0 - \bar{\psi}^2_0 + \frac{2}{\sqrt{5}} \bar{\psi}^3_0 \right ) \\
w = r_{\alpha_2} &:& (\tilde{w})_{\bar{\psi}} = \frac{1}{\sqrt{10}} \left (\bar{\psi}^1_0 - 3 \bar{\psi}^3_0 \right ) \\
w = r_{\alpha_3} &:& (\tilde{w})_{\bar{\psi}} = \frac{1}{\sqrt{2}} \left ( \frac{1}{\sqrt{5}} \bar{\psi}^1_0 + \bar{\psi}^2_0 + \frac{2}{\sqrt{5}} \bar{\psi}^3_0 \right ) \\
\end{array} \;.
$$
These are just the expression of the unit vectors along the simple roots in the basis \eqref{BasePrinSu4}. To apply the prescription \eqref{PrescrMesChargeNDiag}, we have to compute\footnote{These computations are most efficiently performed with a program dealing with Clifford algebras. We used the Mathematica package Clifford.m \cite{ARAGON-CAMARASA} that can be found at http://www.fata.unam.mx/aragon/software/.} $(\tilde{w}')_{\bar{\psi}} \tilde{\Omega} (\tilde{w})^{-1}_{\bar{\psi}}$ for each $w$ in the list \eqref{WelGroupElSu3Su4} and for $w' = \Omega_{D4}w$. We get :
$$
\begin{array}{ll}
w & (\tilde{w}')_{\bar{\psi}} \tilde{\Omega} (\tilde{w})^{-1}_{\bar{\psi}} \\
\hline \mbox{id} & \frac{1}{\sqrt{6}} - \sqrt{\frac{5}{6}} \bar{\psi}_0^2 \bar{\psi}_0^3 \\
r_{\alpha_3} & \sqrt{\frac{2}{3}}-\frac{1}{\sqrt{30}}\bar{\psi}_0^1 \bar{\psi}_0^2-\sqrt{\frac{3}{10}} \bar{\psi}_0^2 \bar{\psi}_0^3 \\
 r_{\alpha_1}\circ r_{\alpha_2} \circ r_{\alpha_3}& \frac{1}{\sqrt{6}}+\sqrt{\frac{5}{6}} \bar{\psi}_0^2 \bar{\psi}_0^3\\
r_{\alpha_2}\circ r_{\alpha_3} & \sqrt{\frac{2}{3}}+\frac{1}{\sqrt{30}}\bar{\psi}_0^1 \bar{\psi}_0^2+\sqrt{\frac{3}{10}} \bar{\psi}_0^2 \bar{\psi}_0^3\\
\end{array} \;.
$$
Summing these four terms gives $\sqrt{6}$ (as a Clifford element), which shows that these boundary states carry only D0-brane charge along $\ket{RR,1}$.

Therefore in the case of $SU(4)$, the boundary states constructed in \cite{Gaberdiel:2004za} do not carry any new charge compared to maximally symmetric and twisted boundary states.

\subsubsection*{Twisted coset states from the embedding of the Cartan torus}

In \cite{Gaberdiel:2004hs}, the same authors presented another set of boundary states which may carry all of the possible K-theory charges. They used the twisted coset construction, using this time the Cartan torus as subgroup, and reflexions across perpendicular hyperplanes in the Cartan subalgebra as automorphisms. The specific orthonormal basis in the Cartan subalgebra of $SU(4)$ chosen in \cite{Gaberdiel:2004hs} is $(h^1,h^2,h^3)$, with :
\begin{equation}
\label{BasCartSubGab}
\begin{array}{rl}
(h^1)^\ast &= \frac{1}{\sqrt{2}}\alpha_2 \vspace{.1cm}\\
(h^2)^\ast &= \sqrt{\frac{2}{3}}(\alpha_1 + \frac{1}{2}\alpha_2) \vspace{.2cm}\\
(h^3)^\ast &= \frac{2}{\sqrt{3}} \omega_3 = \frac{1}{2\sqrt{3}}(\alpha_1 + 2 \alpha_2 + 3 \alpha_3) \;,
\end{array}
\end{equation}
where $\omega_3$ is the third fundamental weight. Let us define $\hat{\Omega}$ to be diag$(\pm 1, \pm 1, 1)$ in this basis, and the identity outside the Cartan subalgebra. Set $\Omega = \hat{\Omega} \circ (\Omega_{C4})^\epsilon$, where as above $\Omega_{C4}$ is complex conjugation on $\mathfrak{su}(4)$ and $\epsilon = 0,1$. We want to compute the charges carried by the twisted coset states constructed from $\Omega$. 

If $\hat{\Omega} = \mbox{id}$, we recover the maximally symmetric boundary states or the twisted ones, depending on the value of $\epsilon$.

If we have a non-trivial $\hat{\Omega}$ with $\epsilon = 0$, $\Omega$ preserves the Cartan subalgebra and the grading, so we can solve the gluing conditions defining the boundary state in $V_{-\rho}$. The component in the cohomology is given by $\ket{RR,e_{\Omega}}$, where $e_{\Omega} \in \bigwedge \mathfrak{h}$ is the volume form on the eigenspace of eigenvalue $-1$. We also have to look for pairs $(w,w')$ of elements of the Weyl group $W$ such that $w' \circ \Omega \circ w^{-1}$ preserves the Cartan subalgebra and the grading. But this is the case for any diagonal element $(w,w)$, $w \in W$. The component $\ket{B_{-\rho}}$ to be added to the boundary state in $\ker P$ therefore reads :
$$
\ket{B_{-\rho}} = \sum_{w \in W} \ket{RR, w e_{\Omega}}\;.
$$
If the negative eigenspace of $\Omega$ has dimension 1, $\ket{B_{-\rho}} = 0$, as we are averaging an element of the Cartan subalgebra over the Weyl group. If it has dimension 2, $e_{\Omega} = h^1 \wedge h^2$ and we see that $\ket{RR, e_{\Omega}} + \ket{RR, r_{\alpha_2} e_{\Omega}} = 0$, which implies that the average over the whole Weyl group also vanishes. So none of these states carry any charge. This canceling phenomenon similar to the one we noticed for parafermionic B-branes in $SU(2)$.

\vspace{.5cm}

If $\epsilon = 1$ and $\hat{\Omega}$ is non-trivial, the analysis is a bit more delicate, because $\Omega_{C4}$ does not preserve the triangular decomposition of $\mathfrak{g}$. However, we have the relation $\Omega_{C4} = \mbox{ad}_{\hat{g}'^{-1}} \circ \Omega_{D4}$, where again $\Omega_{D4}$ is the automorphism generated from the Dynkin diagram of $\mathfrak{su}(4)$, and $\hat{g}'$ is the group element \eqref{ElGrpDiagOm4}.
Therefore $\hat{\Omega} \circ \Omega_{C4} = \Omega_{C4} \circ \hat{\Omega} = \mbox{ad}_{\hat{g}'^{-1}} \circ \Omega_{D4} \circ \hat{\Omega}$. Now $\Omega' := \Omega_{D4} \circ \hat{\Omega}$ obviously preserves the triangular decomposition of $\mathfrak{g}$, so we can solve the gluing conditions in $(1,\hat{g}')V_{-\rho}$, and this is equivalent to solving the gluing conditions for $\Omega'$ in $V_{-\rho}$. Let us now try to find the other solutions. If $w' \circ \Omega' \circ w^{-1}$ preserves the triangular decomposition, for $w,w'$ in the Weyl group, then $w' \circ \Omega_{D4} \circ  w^{-1}$ also does, because $\hat{\Omega}$ acts non trivially only on the Cartan subalgebra. The condition that positive root spaces be mapped into positive root spaces forces $w' = \Omega_{D4} w$. We should now solve the gluing conditions $(\psi^j_0 + i\Omega'\bar{\psi}^j_0)\ket{B} = 0$ in $\ker P$, and average with the action of the (whole) Weyl group.

As this involves summing twenty-four Clifford elements, we will not write the computation explicitly here. The analysis is completely analogous to the one performed at the end of the previous section for the boundary states twisted by both outer automorphisms of $SU(3)$ and $SU(4)$. The resulting charges all vanish. 

This analysis rules out the conjecture of \cite{Gaberdiel:2004hs} as well. Interestingly, the problem of finding boundary states accounting for all of the possible charges in $SU(n)$, $n > 3$ is still open.

Let us however recall again that the formula used to perform the average in the case of the boundary states twisted by both the automorphism of $SU(4)$ and of the subalgebra may be dependent of the particular choice of lift into $Pin$ of the Weyl group elements we average on.

\section[Relation to homology]{Relation to homology\footnote{Thanks to Rudolf Rohr for his patience when answering our numerous questions on the Kostant conjecture. For some background on the algebraic models of Lie group cohomology, see for instance \cite{Greub1976, Guillemin1999}.}}

\label{SecRelHom}

In this section, we discuss how our construction can be interpreted in term of the familiar (co)homological classifications of D-branes. 

\subsection{The Kostant conjecture}

It is well known that the cohomology classes of Lie groups are in bijection with bi-invariant (that is left and right invariant) forms. Such a form $\omega$ is completely determined by its component $e_\omega$ in $(\bigwedge T^{\ast})_1 G \sim \bigwedge \mathfrak{g}^\ast$, the fiber at the identity element $1 \in G$ of the exterior algebra of the cotangent bundle. The requirement that the form is bi-invariant means that $e_\omega \in \bigwedge \mathfrak{g}^\ast$ is invariant under the coadjoint action of $\mathfrak{g}$ on $\bigwedge \mathfrak{g}^\ast$. This provides an algebraic model for the cohomology of Lie group :
$$
H^\bullet(G,\mathbbm{R}) = \left ( \bigwedge \mathfrak{g}^\ast \right )^\mathfrak{g} \;,
$$
where $(...)^\mathfrak{g}$ denotes the invariant part under the action of $\mathfrak{g}$.
Naturally, we also get a model for homology :
$$
H_\bullet(G,\mathbbm{R}) = \left ( \bigwedge \mathfrak{g} \right)^\mathfrak{g} \;,
$$
where $\mathfrak{g}$ acts now by the adjoint action. 

Recall also that the homology ring $(\bigwedge \mathfrak{g})^\mathfrak{g}$ is generated by the subspace of primitive elements $P \subset \bigwedge \mathfrak{g}$, so that $(\bigwedge \mathfrak{g})^\mathfrak{g} = \bigwedge P$. Their degrees are given by twice the exponents of the Lie algebra plus one, and their precise definition can be found in \cite{Greub1976}, section V.4. Integer valued homology and cohomology can also be described by the corresponding $\mathbbm{Z}$ valued rings generated by suitably normalized primitive elements.

These two rings have dimension $2^{r}$, where $r$ is the rank of the Lie algebra. Recall that our test states were classified by $\bigwedge \mathfrak{h}$, which is also $2^{r}$-dimensional. It turns out that there is a natural map between these two rings, which have received some attention in the Lie representation literature in the context of the so-called Kostant conjecture \cite{Bazlov2003}. Let us describe this map.

There is a (anti)symmetrization map from the exterior algebra to the Clifford algebra : 
$$
\begin{array}{lrcl}
s : & \bigwedge \mathfrak{g} & \rightarrow & \mbox{Cl}(\mathfrak{g}) \vspace{.2cm}\\
& e^1 \wedge e^2 \wedge ...\wedge e^m & \mapsto & \frac{1}{m!} \sum_{\sigma \in S_m} \mbox{sgn}(\sigma) e^{\sigma(1)} \cdot e^{\sigma(2)} \cdot ... \cdot e^{\sigma(m)} \;,
\end{array}
$$
where $S_m$ is the symmetric group of order $m$, $\mbox{sgn}$ is the sign map and $\cdot$ is the Clifford product. $s$ is an isomorphism of vector spaces. It restricts to a (vector space) isomorphism of the invariant parts :
$$
s : \left ( \bigwedge \mathfrak{g} \right )^\mathfrak{g}  \rightarrow  \mbox{Cl}(\mathfrak{g})^\mathfrak{g} \;.
$$

The Harish-Chandra map $hc$ sends $\mbox{Cl}(\mathfrak{g})^\mathfrak{g}$ to $\mbox{Cl}(\mathfrak{h})$. Its precise definition can be found in \cite{Bazlov2003}. Informally, it can be described as a normal ordering with respect to the triangular decomposition of $\mathfrak{g}$, followed by an orthogonal projection on $\mbox{Cl}(\mathfrak{h}) \subset \mbox{Cl}(\mathfrak{g})$.

Finally, there is a map $c$, the Chevalley map, which identifies $\mbox{Cl}(\mathfrak{h})$ and $\bigwedge \mathfrak{h}$ as vector spaces. Denote $\epsilon(h)$, $h \in \mathfrak{h}$ the endomorphism of $\bigwedge \mathfrak{h}$ consisting of left multiplication by the element $h$ : $\epsilon(h)(x) = h \wedge x$. Denote by $\iota(h)$ the endomorphism of $\bigwedge \mathfrak{h}$ consisting of the contraction of elements of $\bigwedge \mathfrak{h}$ with $h$. Let $\gamma(h) = \epsilon(h) + \iota(h)$. It can be shown that $\gamma$ extends to a homomorphism $\mbox{Cl}(\mathfrak{h}) \rightarrow \mbox{End}\bigwedge \mathfrak{h}$. Define finally $c(h) = \gamma(h)(1)$, for an arbitrary $h \in \mbox{Cl}(\mathfrak{h})$ (that is, apply $\gamma(h) \in \mbox{End}\bigwedge \mathfrak{h}$ to $1 \in \bigwedge \mathfrak{h}$).

A non trivial result is that the sequence of maps $\Phi = c \circ hc \circ s : (\bigwedge \mathfrak{g})^\mathfrak{g} \rightarrow \bigwedge \mathfrak{h}$ is an isomorphism of algebras \cite{Bazlov2003}. Indeed this is not obvious, because the maps $c$ and $s$ are only isomorphisms of vector spaces.

To state the Kostant conjecture, we need one more ingredient, the principal basis of the Cartan subalgebra \cite{Kostant1959, Rohr}. Consider the principal $\mathfrak{sl}(2)$ subalgebra of $\mathfrak{g}$, that is, the $\mathfrak{sl}(2)$ subalgebra containing $\rho^\vee \in \mathfrak{h}$ and $e^+ = \sum_{\alpha} e^{\alpha}$, where the sum runs over the simple roots of $\mathfrak{g}$. $e^\alpha$ is the generator of the corresponding root space, and $\rho^\vee$ is half the sum of the positive coroots. These two elements determine a unique $e^-$ such that $\{e^+, \rho^\vee, e^-\}$ generate a subalgebra isomorphic to $\mathfrak{sl}(2)$, the principal subalgebra. $\mathfrak{g}$ decomposes into a direct sum of module for the principal subalgebra, and restricting this decomposition to the Cartan subalgebra $\mathfrak{h}$, we get an orthogonal decomposition of $\mathfrak{h}$. We therefore get a canonical orthogonal basis of $\mathfrak{h}$, the principal basis. For instance, the adjoint module of the principal subalgebra always determines a subspace $\mathbbm{C}\rho^\vee$, and $\rho^\vee$ can be taken as an element of the orthogonal basis.

\vspace{.5cm}

The Kostant conjecture states that the image by $\Phi$ of the primitive elements $\{p_i\}$ are monomials of degree one in $\bigwedge \mathfrak{h}$. Moreover, $\{\Phi(p_i)\}$ form an orthonormal basis of $\mathfrak{h}$, which coincides with the principal basis induced from the principal decomposition of the Langlands dual $\mathfrak{g}^\vee$ of $\mathfrak{g}$. (Recall that the roots of Langlands dual algebra coincide with the coroots of the original Lie algebra.) We denote this basis by $\{\hat{h}^i\}$. 
This conjecture is actually a theorem for all of the infinite series $A$, $B$, $C$ and $D$ and for $F_2$ \cite{Bazlov2003}. It remains a conjecture only for the other exceptional Lie algebras.

The isomorphism $\Phi$ gives us the possibility of interpreting the abstract charges provided by our procedure as homology charges. If a brane has been assigned a charge $\ket{B_{-\rho}} \in \ker P \sim \bigwedge \mathfrak{h}$, one can decompose it into a (exterior) polynomial in the elements $\{\hat{h}^i\}$ of the principal basis of the Langlands dual. Each element of the principal basis corresponds to a primitive element, which itself corresponds to an elementary homology class.

Let us note that the homology class of degree 3, present in all compact simple Lie groups, always corresponds to $h^{\rho} \in \bigwedge \mathfrak{h}$ under the Kostant conjecture. It is well known that D-branes cannot wrap the homology class of the 3-sphere in Lie groups, because there is a non-trivial NS-NS flux through it. This flux makes it impossible to find a globally defined potential for the NS-NS 3-form on the brane worldvolume, what makes the Wess-Zumino term in the open string WZW action ill-defined. We conjecture that D-branes will couple only to states in $\bigwedge (\mathfrak{h}/\mathbbm{C}e^\rho)$, a property that we checked in the various examples. Moreover, we expect integer overlaps between any D-brane state and the test states labeled by integer polynomials in the generators of the principal basis (after a suitable normalization of these generators). This property was also satisfied in the examples, but definitely needs to be tested further. As we already mentioned in the remarks above, if we discard test states outside $\bigwedge (\mathfrak{h}/\mathbbm{C}e^\rho)$, we get a $2^{r-1}$ dimensional space of test states, and the possible charges of the branes live in :
$$
(\mathbbm{Z}/M\mathbbm{Z})^{(2^{r-1})}\;.
$$
This charge group precisely coincides with the twisted K-theory of the Lie group.


\subsection{Back to the examples}

Let us now return to our examples and compare the algebraic charges we computed in section \ref{SecExamples} with homology. We are here implicitly taking a limit where the level $k$ of the sWZW model is sent to infinity, so that we can see the D-branes as submanifolds of the Lie group $G$ and compute their homology charges. While this geometrical picture breaks down for finite $k$, the algebraic approach developed above for the computation of the charges is well-defined for any $k$.

As they couple to $\ket{RR,1}$ only, we immediately see that maximally symmetric D-brane never carry any homology charge. This phenomenon is well-known \cite{Bachas:2000ik,Pawelczyk:2000ah} : they are stabilized by a $U(1)$ flux on their worldvolume.

The homology of $SU(2)$ contains only the class $h_3$ corresponding to the 3-sphere forming the group manifold $SU(2)$. The parafermionic B-branes of $SU(2)$ can be seen as thickened D1-strings \cite{Maldacena:2001ky}, so they are not expected to carry any charge, as has been confirmed by our computation.

\vspace{.5cm}

The homology of $SU(3)$ is given by $\bigwedge P$, where $P$ is generated by primitive elements $p_1$ of degree 3 and $p_2$ of degree 5 which correspond respectively to the homology class of the 3-sphere and of the 5-sphere (homologically, $SU(3) \sim S^3 \times S^5$). $SU(3)$ (as well as $SU(n)$) is its own Langlands dual. Under the principal decomposition, the adjoint representation of $\mathfrak{su}(3)$ decomposes into $(2) \oplus (4)$. We use here Dynkin index notation for the representation of the principal $\mathfrak{sl}(2)$. The intersection of $(2)$ with $\mathfrak{h}$ gives $h^\rho = h^{\alpha_1 +\alpha_2}$, which is the image under $\Phi$ of $p_1$, and the intersection of $(4)$ with $\mathfrak{h}$ gives $h^{\alpha_1 - \alpha_2}$, which is the image under $\Phi$ of $p_2$. We saw that the twisted branes couple to the Ramond-Ramond test state associated with $h^{\alpha_1 - \alpha_2}$, so we expect them to carry a non-trivial homology charge with respect to the class of the 5-sphere. This was shown from geometric considerations in \cite{Maldacena:2001xj}, indeed.

\vspace{.5cm}

The homology of $SU(4)$ is given by $\bigwedge P$, where $P$ is generated by primitive elements of degree 3, 5 and 7. They correspond respectively to the homology class of the 3-sphere, of the 5-sphere and of the 7-sphere sitting in $SU(4)$. The data coming from the principal decomposition of the Cartan subalgebra is summarized in the table below :
$$
\begin{array}{cccc}
\mathfrak{sl}(2)\mbox{-module} & \mbox{Basis element in }\mathfrak{h} & \mbox{Homology class} & \Omega\mbox{-parity}\\
(2) & h^{3\alpha_1 + 4\alpha_2 + 3\alpha_3}  & 3\mbox{-sphere} & + \\
(4) & h^{\alpha_1 - \alpha_3} & 5\mbox{-sphere} & - \\
(6) & h^{\alpha_1 - 2\alpha_2 + \alpha_3} & 7\mbox{-sphere} & +
\end{array}.
$$
The first column lists the Dynkin indices of the $\mathfrak{sl}(2)$-modules appearing in the decomposition of the adjoint representation of $\mathfrak{su}(4)$ under the action of the principal subalgebra. The second column shows the basis element determined by the module. The third column describes the corresponding homology class, and the last column shows the parity of this class under the action of the outer automorphism of $SU(4)$. 

Recall that we found that the D-branes twisted by the automorphism of $SU(4)$ carried a charge along $h^{\alpha_1 - \alpha_3}$, so they wrap around the 5-sphere in $SU(4)$. One can also check this geometrically. For instance, the D-brane passing through the identity element wraps the conjugacy class $C_\Omega = \{g\Omega_{D4}(g^{-1})|g\in SU(4)\}$. This means that the tangent space to the D-brane is the negative eigenspace of $\Omega_{D4}$, which is five-dimensional. Moreover, if $\Omega_{D4} x = -x$ and $g = \exp x$, then $\Omega_{D4} g^{-1} = g$. So we even have $C_\Omega = \{\exp x | x\in \mathfrak{su}(4), \Omega_{D4} x = -x\}$. One can therefore write explicitly the conjugacy class as :
$$
\left (
\begin{array}{cccc}
z_1 & z_2 & z_3 & 0 \\
-\bar{z}_2 & \bar{z}_1 & 0 & z_3 \\
-\bar{z}_3 & 0 & \bar{z}_1 & -z_2 \\
0 & -\bar{z}_3 & \bar{z}_2 & z_1
\end{array}
\right ) \;,
$$
with $|z_1|^2 + |z_2|^2 + |z_3|^2 = 1$, so it is topologically a 5-sphere. It remains to check that it carries a non-trivial homology charge. This can be done\footnote{Thanks to Pavol Severa for some help on this point.} using the fact that $C_\Omega$ defines a map $SU(4) \stackrel{f}{\rightarrow} S^5 \stackrel{i}{\hookrightarrow} SU(4)$, where $i$ is the inclusion. $i \circ f$ is given by the (Lie group) product of two maps : $i \circ f = \mbox{id} \cdot (\Omega_{D4} \circ \mbox{inv})$, where $\cdot$ is the product on the group and $\mbox{inv}$ the inversion map. To check that the homology is non-trivial is equivalent to checking that the cohomology class $[i^\ast \phi_5]$ of the pull-back of the bi-ivariant 5-form $\phi_5$ of $SU(4)$ onto $S^5$ is a non-zero multiple of the class $[\omega]$ of the volume form $\omega$ on $S^5$. There is no simple way of verifying this directly, but we can check that $[(i \circ f)^\ast\phi_5]$ is non-trivial, which implies that $[i^\ast\phi_5]$ is non-trivial either. To this end, we just use that $[(f_1\cdot f_2)^\ast \phi_5] = [f_1^\ast \phi_5] + [f_2^\ast \phi_5]$. $\phi_5$ is multiplied by $-1$ under inversion, as any odd form, but there is another $-1$ factor from the outer automorphism. Therefore $[(i \circ f)^\ast\phi_5] = 2[\phi_5]$, and $C_\Omega$ is non-trivial in homology indeed. 

Concerning the states constructed with the twisted coset construction from the subgroup $SU(3) \subset SU(4)$, geometry also leads us to expect them to carry a homology charge along $S^5$. If we consider again the twisted D-brane passing through the identity element of the group, we know from \cite{Maldacena:2001xj} that it is a 5-dimensional submanifold wrapping the homology class of $S^5$ in $SU(3)$. The image of this homology class under the push-forward induced by the inclusion $SU(3) \subset SU(4)$ is the homology class of $S^5$ of $SU(4)$, so we expect the boundary states twisted with respect to $SU(3)$ to wrap $S^5$, like the states twisted by the automorphism of $SU(4)$. 

This shows that already from purely geometrical considerations, one can guess that the D-branes twisted by the automorphisms of $SU(4)$ and $SU(3)$ carry the same type of charge, as was showed by our algebraic analysis.

We therefore find a complete agreement between the geometrical and CFT pictures of the D-brane charges. 

\section{Discussion and conclusion}

\label{SecConcl}

Despite its successes, our construction does have some shortcomings. We saw that we cannot compare the charges of boundary states belonging to two distinct Kondo families carrying charges in the same factor $\mathbbm{Z}/M\mathbbm{Z}$. In particular, we know that in $SU(4)$, the boundary states twisted by the automorphisms of $SU(4)$ and $SU(3)$ carry a charge along $\ket{RR,h^{\alpha_1-\alpha_3}}$, but we cannot compare the magnitude of the charges between boundary states of the two families. As we mentioned earlier, our procedure may not be applicable for some boundary states which break too many of the bulk symmetries, and the averaging procedure is not always completely well-defined, what constitutes a second shortcoming. Let us also emphasize that we restricted our discussion to simply connected Lie groups when we chose the charge conjugation modular invariant for the bosonic part of the sWZW model. The extension of these results to non simply connected Lie groups seems to be non trivial.

We believe that this construction should have a conceptual mathematical interpretation that we have been unable to find up to now. A proper mathematical formulation may overcome the two shortcomings described above. It could also help to establish a clear link between our algebraic charges and the twisted K-theory of the Lie group, beyond the mere observation of their isomorphism.

An interesting problem would be to identify more boundary RG flows that do not fall into the family of generalized Kondo flows, and check whether the charges we obtained are really invariant. We saw that this is true for the flows described in \cite{Alekseev:2002rj}. Interestingly, our results indicate that the problem of constructing boundary states carrying charges from each of the factors of \eqref{K-theoryG} is still open beyond $SU(3)$. We will not make here a conjecture about the form such boundary states may take, but in principle one should be able to test future proposals using our procedure. Let us finally mention that this construction can certainly be extended to generic $N=1$ supersymmetric coset models, with many potential applications in the study of boundary renormalization group flows in physically pertinent models. We will return to these issues in a future work.

\section*{\normalsize Acknowledgments}

I would like to thank Costas Bachas, Stefan Fredenhagen, Thomas Quella, Rudolf Rohr, Pavol Severa and Jan Troost for discussions. Special thanks to Anton Alekseev for so many crucial discussions, and for introducing me to the cohomology of the Dirac operator in the non-commutative Weil algebra, what triggered this work. This research is supported in part by the Swiss National Science Foundation.

{
\small
\providecommand{\href}[2]{#2}\begingroup\raggedright\endgroup
}

\end{document}